\documentclass[]{aastex631}
\NewPageAfterKeywords
\usepackage{amsmath}
\usepackage{CJKutf8}
\usepackage[counterclockwise, figuresright]{rotating}
\usepackage{xcolor}

\shorttitle{Physical GP SFHs}
\shortauthors{Iyer \& Speagle et al.}
\usepackage{fontawesome} % GitHub icon
\graphicspath{{./}}
\begin{document}

\title{Stochastic Modelling of Star Formation Histories III. Constraints from Physically-Motivated Gaussian Processes}

\author[0000-0001-9298-3523]{Kartheik G. Iyer}
\altaffiliation{Equal contribution}
\affiliation{Dunlap Institute for Astronomy \& Astrophysics, University of Toronto, 50 St George Street, Toronto, ON M5S 3H4, CA}
\email{kartheik.iyer@dunlap.utoronto.ca}

\author[0000-0003-2573-9832]{Joshua S. Speagle (\begin{CJK*}{UTF8}{gbsn}沈佳士\ignorespacesafterend\end{CJK*})}
\altaffiliation{Equal contribution}
\affiliation{Department of Statistical Sciences, University of Toronto, 9th Floor, Ontario Power Building, 700 University Ave, Toronto, ON M5G 1Z5, CA}
\affiliation{David A. Dunlap Department of Astronomy \& Astrophysics, University of Toronto, 50 St George Street, Toronto ON M5S 3H4, CA}
\affiliation{Dunlap Institute for Astronomy \& Astrophysics, University of Toronto, 50 St George Street, Toronto, ON M5S 3H4, CA}
\affiliation{Data Sciences Institute, University of Toronto, 17th Floor, Ontario Power Building, 700 University Ave, Toronto, ON M5G 1Z5, CA}
\email{j.speagle@utoronto.ca}

\author[0000-0003-3287-5250]{Neven Caplar}
\affiliation{Department of Astrophysical Sciences, Princeton University, 4 Ivy Lane, Princeton, NJ 08544, USA}

\author[0000-0002-1975-4449]{John C. Forbes}
\affiliation{Center for Computational Astrophysics, Flatiron Institute, 162 Fifth Avenue, New York, NY 10010, USA}

\author[0000-0003-1530-8713]{Eric Gawiser}
\affiliation{Department of Physics and Astronomy, Rutgers, the State University, Piscataway, NJ 08854, USA}

\author[0000-0001-6755-1315]{Joel Leja}
\affiliation{Department of Astronomy \& Astrophysics, The Pennsylvania
State University, University Park, PA 16802, USA}
\affiliation{Institute for Computational \& Data Sciences, The Pennsylvania State University, University Park, PA, USA}
\affiliation{Institute for Gravitation and the Cosmos, The Pennsylvania State University, University Park, PA 16802, USA}

\author[0000-0002-8224-4505]{Sandro Tacchella}
\affiliation{Kavli Institute for Cosmology, University of Cambridge, Madingley Road, Cambridge, CB3 0HA, UK}
\affiliation{Cavendish Laboratory, University of Cambridge, 19 JJ Thomson Avenue, Cambridge, CB3 0HE, UK}



\begin{abstract}
Galaxy formation and evolution involves a variety of effectively stochastic processes that operate over different timescales. The Extended Regulator model provides an analytic framework for the resulting variability (or `burstiness') in galaxy-wide star formation due to these processes. It does this by relating the variability in Fourier space to the effective timescales of stochastic gas inflow, equilibrium, and dynamical processes influencing GMC creation and destruction using the power spectral density (PSD) formalism. We use the connection between the PSD and auto-covariance function (ACF) for general stochastic processes to reformulate this model as an auto-covariance function, which we use to model variability in galaxy star formation histories (SFHs) using physically-motivated Gaussian Processes in log SFR space. Using stellar population synthesis models, we then explore how changes in model stochasticity can affect spectral signatures across galaxy populations with properties similar to the Milky Way and present-day dwarfs as well as at higher redshifts. We find that, even at fixed scatter, perturbations to the stochasticity model (changing timescales vs overall variability) leave unique spectral signatures across both idealized and more realistic galaxy populations. Distributions of spectral features including H$\alpha$ and UV-based SFR indicators, H$\delta$ and Ca-H,K absorption line strengths, D$_n$(4000) and broadband colors provide testable predictions for galaxy populations from present and upcoming surveys with Hubble, Webb \& Roman. The Gaussian process SFH framework provides a fast, flexible implementation of physical covariance models for the next generation of SED modeling tools. Code to reproduce our results can be found at \faGithub\href{https://github.com/kartheikiyer/GP-SFH}{kartheikiyer/GP-SFH}.
\end{abstract}

% from https://astrothesaurus.org/concept-select/


\section{Introduction} \label{sec:intro}

\defcitealias{2019MNRAS.487.3845C}{Paper I}
\defcitealias{2020MNRAS.497..698T}{TFC20}

Large galaxy surveys like SDSS \citep{2000AJ....120.1579Y, 2002AJ....124.1810S, 2009ApJS..182..543A}, GAMA \citep{2009A&G....50e..12D} and COSMOS \citep{2007ApJS..172....1S, 2007ApJS..172...70L,  2022ApJS..258...11W} reveal an enormous diversity in galaxy demographics across different epochs, and many studies in modern galaxy evolution have devoted considerable energy in attempting to explain this diversity from a physical standpoint, usually through analytical models and cosmological simulations \citep[see review by][]{2015ARA&A..53...51S}.

A part of this picture involves the stochasticity in star formation, which is regulated by a variety of physical processes acting over many orders of magnitude in spatial and temporal scales. This stochasticity or `burstiness' can be physically observed through short-timescale star formation rate (SFR) indicators such as H$\alpha$ or UV-based SFRs, which probe recent star formation averaged over the most recent $\sim 4-10$ and $\sim 10-100$ Myr respectively \citep{2014ARA&A..52..415M, 2021MNRAS.501.4812F,  2022MNRAS.513.2904T}. It can also be observed by studying resolved star formation in galaxies, where most star formation appears to to occur in discrete clumps traced by the rest-UV that are spatially offset from the bulges of galaxies where most older stellar populations live \citep{2014AAS...22314511G, 2020MNRAS.499..814H}, with the creation and destruction of these clumps due to ISM physics and feedback leading to stochasticity on the timescales of the clump lifetimes \citep{2018ApJ...861....4S, 2021ApJ...918...13S}. On a population level, this could leave signatures in the scatter of scaling relations like the SFR-M$_*$ correlation\footnote{Although building falsifiable tests to test these signatures can be challenging \citep{2014arXiv1406.5191K, 2016ApJ...832....7A}.} \citep{noeske2007star, daddi2007multiwavelength, elbaz2007reversal} and the mass-metallicity relation \citep{tremonti2004origin}, which show overall coherent behaviour for the `average' galaxy that is tied to the growth of their parent dark-matter halos, but significant variation (at the $\sim 0.3-0.5$ dex level) in the star formation rates of individual galaxies that seem to fluctuate around these average relations \citep{kauffmann2006gas, 2016MNRAS.457.2790T,  2019MNRAS.484..915M}.

These fluctuations in SFR are regulated by a variety of physical processes ranging from the local creation and destruction of stars in giant molecular clouds (GMCs), to dynamical processes like disk formation and bulge growth, to galaxy-wide processes that include mergers, galactic winds from stellar and AGN feedback, and baryon cycling that couples a galaxy to its surrounding circumgalactic medium (CGM) \citep{2020MNRAS.497..698T, 2020MNRAS.498..430I}. On the largest scales, however, a galaxy's growth is tied to the reservoir of fuel available to it to form stars, which are in turn tied to the accretion rates of their parent halos, galactic depletion-times and outflows, and the large-scale structure of the environment they live in \citep[see review by][]{2018ARA&A..56..435W}.

Analyzing the stochasticity of star formation across a range of timescales\footnote{Using population-level statistics in this work, since we currently lack the constraining power in our observations to perform this analysis for individual galaxies.} therefore provides us with a way to constrain the relative strengths of these physical processes. A particularly effective way to quantify and assess this is by quantifying the `burstiness' or stochasticity of a galaxy's past star formation as a function of timescale, and comparing these to theoretical predictions. Fourier space has proved instructive in this regard, with the power spectral density (PSD) of galaxy star formation over time being used to construct analytical models of galaxy stochasticity due to different physical processes \citep{2014MNRAS.443..168F, 2020MNRAS.497..698T}, study the regulation of star formation across different cosmological simulations \citep{2020MNRAS.498..430I}, and constrain simple models of stochasticity using observational data from large galaxy surveys \citep{2019MNRAS.487.3845C, 2020ApJ...895...25W}.

These studies hit on a fundamental aspect of galaxy evolution, that the growth of pure dark matter halos is essentially a scale-free process that leads to a power-law PSD \citep{guszejnov2018universal, kelson2020gravity}. This is the reason why (to first order) the halo mass is such a good predictor of galaxy properties and methods like sub-halo abundance matching have met with such remarkable success. Other aspects including stellar feedback, baryon cycling and multi-phase ISM astrophysics decouple star formation from this hierarchical build-up and add additional stochasticity to this on a range of (often interwoven) timescales, leading to an overall complex power spectrum that can be studied and understood with careful analysis. \citet[][Paper I in this series, hereafter CT19]{2019MNRAS.487.3845C} defines the PSD of a galaxy's star formation history (SFH) and assuming the shape of a simple broken-power law, linked it to SFR distributions averaged over different timescales. \citet[][Paper II in this series, hereafter TFC20]{2020MNRAS.497..698T} builds on this, using  the widely successful\footnote{A review by \citet{2020ARA&A..58..157T} finds that the model can reproduce the combined evolution of molecular gas fractions, star formation rates
% galactic winds,
and gas-phase metallicities.} gas regulator model \citep[][]{2013ApJ...772..119L} coupled with the stochastic inflow of gas (\citealt{kelson2020gravity}, \citetalias{2020MNRAS.497..698T}) to derive a more general form for the PSD of galaxy SFHs.

The PSD of the Extended Regulator model depends only on an overall level of stochasticity for gas inflow and GMCs, and characteristic timescales for effective gas inflow, equilibrium, and GMC lifetimes. \citetalias{2020MNRAS.497..698T} formulates this PSD and proposes the timescales for a few galaxy populations (Milky-way analogues, dwarf galaxies, galaxies at high redshift and massive galaxies at cosmic noon) given our current knowledge of burstiness in these galaxy populations. However, to verify these models and observationally probe these timescales, we need a framework in which the PSDs can be tested observationally.

To observationally measure the PSD (i.e. the variability in SFR over different timescales), we can leverage the fact that a range of spectral features measure changes in the star formation rate averaged over different timescales. Some of the commonly measured spectral features include the nebular H$\alpha$ line, rest-UV+IR SED that traces emission from young massive stars \& re-emitted radiation from dust, H$\delta$ absorption from the photospheres of smaller (mostly A-type) stars, and the 4000 \AA{} break from the accumulation of ionized metal absorption lines for older stellar populations \citep{2012ApJ...759...67G, 2021MNRAS.501.4812F}. Taken together, these indicators probe fluctuations in SFRs on timescales ranging from $\sim 5$ Myr to $\gtrapprox10$ Gyr, and their ratios have previously been used to obtain estimates of the burstiness of galaxy populations \citep{guo2016bursty, emami2018closer, broussard2019stochastic, 2019ApJ...884..133F,  2020ApJ...892...87W}.

Spectral energy distribution (SED) fitting methods go a step further and estimate the full star formation histories (SFHs) of individual galaxies using the full range of available multiwavelength spectral information, whether it is photometry, spectroscopy, or a combination of the two \citep{moped, vespa, dye, pacifici, pacifici2016timing, smith2015deriving, leja2017deriving, 2017ApJ...838..127I, carnall2018inferring, leja2018measure, 2019ApJ...879..116I}. A few of these methods that implement non-parametric SFHs \citep[][]{pacifici, leja2017deriving, 2019ApJ...879..116I} also allow users to implement priors on SFR burstiness on one or more timescales. The non-parametric Dense Basis method \citep{2017ApJ...838..127I, 2019ApJ...879..116I} for SFH reconstruction that allows us to incorporate physical priors on SFR stochastcity through a Gaussian process covariance function (also called the kernel). These kernels are related to the PSDs through the Weiner-Khinchin theorem\footnote{Assuming that galaxy SFHs, or their oscillation around a fiducial `main sequence' are stationary processes - see Section 5.2 for more discussion about this.} \citep[][]{wiener1930generalized, khinchin1938theory}, which allows us to relate the frequency-domain PSDs to the time-domain auto-covariance functions (ACFs).

For the first time, this opens up a way to explicitly incorporate an analytical framework for correlated SFRs over a range of timescales into an SED modeling framework. This is a crucial development for multiple reasons:
\begin{enumerate}
    \item It provides a physical explanation for how SFRs vary over time through the three timescales - the timescale over which stochastic gas inflow is correlated, the mass-loss or equilibrium timescale on which gas is consumed/removed from the reservoir, and and average GMC lifetime timescales. While a range of different physical processes are responsible for regulating SF in galaxies, these three timescales have been shown to capture a significant portion of the effective dynamics of galaxy populations\footnote{While we currently consider the \citetalias{2020MNRAS.497..698T} model in this work, other models for galaxy growth \citep[][]{2000AJ....120..728H, 2012MNRAS.421...98D, 2022arXiv220504273A} that can be reformulated as an autocorrelation function for a Gaussian process could also be tested in future studies.} (\citetalias{2020MNRAS.497..698T}, \citealt{2020ARA&A..58..157T}).
    \item It offers a framework to forward-model galaxy observations based on a stochasticity model to compare against existing models and to determine sensitive spectral features for future observations \citep[e.g.][]{2014ApJ...790..143W, 2021AAS...23734208P}.
    \item It allows us to explicitly compare against existing priors e.g., uncorrelated SFRs in adjacent timesteps, or the Dirichlet \& continuity type priors \citep[][]{leja2017deriving, 2019ApJ...879..116I} that assume an arbitrary amount of stochasticity and/or correlation between adjacent SFR bins in a model to test their efficacy and their relation to the effective timescales in the regulator model.
    \item It provides an intuitive framework for modeling SFH priors, or incorporating SFH priors from cosmological simulations by estimating the extended regulator model parameters directly from their SFHs.
\end{enumerate}

\begin{figure}
    \centering
    \includegraphics[width=\textwidth, page=1, trim=0.0cm 1.0cm 0.5cm 0.0cm, clip]{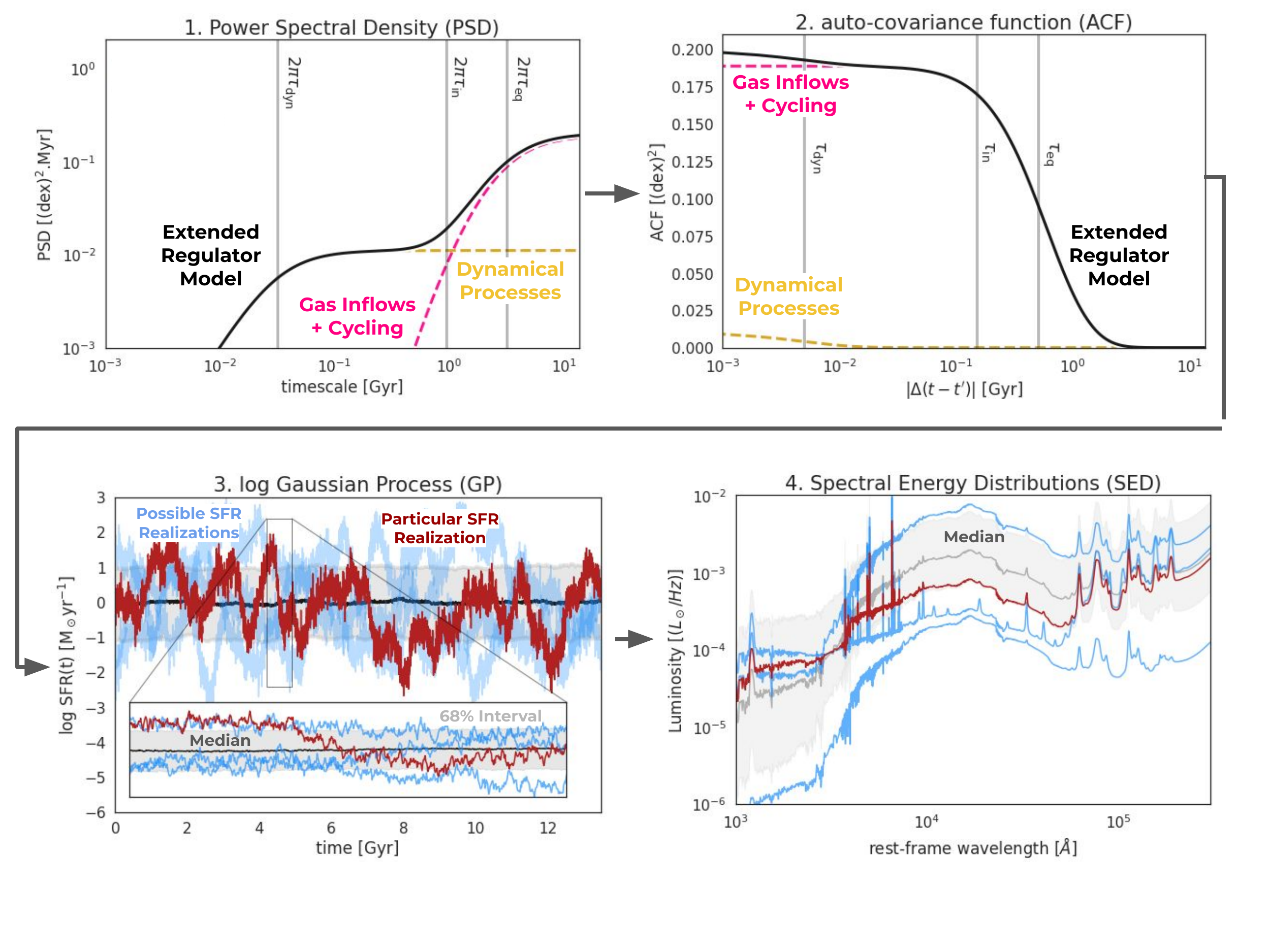}
    \caption{An illustration of the overall strategy used in this paper to test the impact that stochasticity in star formation histories (SFHs) can have on galaxy spectral energy distributions (SEDs). We start with power spectral density (\textit{PSD; top left}) of a dwarf-like galaxy population defined within the framework of the Extended Regulator model of \citetalias{2020MNRAS.497..698T} (see \S\ref{subsec:cases}), with individual components highlighted. We then convert this PSD to the corresponding auto-covariance function (\textit{ACF; top right}). Both the PSD and ACF highlight that variability on short timescales is dominated by giant molecular clouds (GMCs) while variability on longer timescales is dominated by a combination of gas inflows and cycling between atomic and molecular gas. Using the ACF, we define a Gaussian Process (\textit{GP; bottom left}) to quickly and easily sample many realizations of the $\log {\rm SFR}(t)$ over time relative to some mean $\log {\rm SFR}_{\rm base}(t)$, which here we take to be $1$M$_\odot/$yr. We again highlight both long-term and short-term (see inset) variability driven by gas inflows/cycling and dynamical processes, respectively. Finally, we generate SEDs at $z\sim 0$ by running the GP-sampled SFHs through a stellar population synthesis model (\textit{bottom right}). The variability in the SFHs leads to differences in the overall normalization (from differences in total stellar mass formed) as well as particular spectral features (from differences in relative contributions of various stellar populations). Properties apart from the SFH assumed while generating the spectra are detailed in Section \ref{subsec:fsps}, and are held constant to highlight differences from the SFHs. Understanding how the distributions of spectral features corresponding to various galaxy populations correspond to the properties of their PSDs will allow us to put constraints on the timescales on which gas inflows, baryon cycling, and GMC physics drive galaxy evolution.
    }
    \label{fig:overview}
\end{figure}

In this paper, we incorporate the physically motivated SFR stochasticity model proposed in \citetalias{2020MNRAS.497..698T} within the framework of a Gaussian process \citep[GP;][]{gp_book,2019ApJ...879..116I}. Using this, we then use the flexibility of the Extended Regulator model of \citetalias{2020MNRAS.497..698T} to define GPs corresponding to various regimes of stochasticity that we might find in galaxy populations - ranging from the bursty behaviour of galaxies at high redshifts to the long-timescale correlated behaviour of Milky Way analogues at lower redshifts. We then use these GPs to generate mock SFHs in a computationally inexpensive manner, which is crucial if these are to be used in SED fitting. By running these SFHs through a stellar population synthesis framework \citep[FSPS;][]{conroy2009propagation, conroy2010propagation, ben_johnson_2021_4737461}, we then model spectra corresponding to these SFHs and use them to identify observables that can be used to tell the models apart, laying the foundations of future work where this can be incorporated into SED fitting packages. A schematic representation of our main approach is shown in Figure \ref{fig:overview}, and the code used to implement the GPs is publicly available at \url{https://github.com/kartheikiyer/GP-SFH}.

This paper is structured as follows. In \S\ref{sec:model}, we go through the formalism and provide a description of the Extended Regulator model first presented in \citetalias{2020MNRAS.497..698T}, derive the associated auto-covariance function (ACF), and highlight specific cases likely to correspond to various galaxy populations of interest. In \S\ref{sec:gp}, we describe the implementation of the derived ACF as a ``physical kernel'' in a Gaussian process (GP), and describe how it can be used as a prior in SED fitting codes that have flexible models for galaxy SFHs. In \S\ref{sec:single}, we investigate how differences in the underlying ACFs can translate over to spectral signatures using stellar population synthesis (SPS) models, and investigate how these differences manifest in populations of simulated galaxies. We discuss our findings in \S\ref{sec:disc} and conclude in \S\ref{sec:conc}.

Note that while it is most natural to have a process using the base-$e$ logarithm $\ln {\rm SFR}(t)$, we convert to the base-10 logarithm $\log {\rm SFR}(t)$ in most plots to facilitate comparisons with other quantities and measurements in the literature. Throughout this paper magnitudes are in the AB system; we use a standard $\Lambda$CDM cosmology with $\Omega_m=0.3$, $\Omega_\Lambda=0.7$ and $H_0 = 70$ km s$^{-1}$ Mpc$^{-1}$ .

\section{Modelling Stochasticity in Star Formation Rates} \label{sec:model}

We start by building physical intuition of how different physical processes related to galaxies can affect stochasticity and correlations in the SFRs of individual galaxies across cosmic time, summarized through their power spectral densities (PSDs) and associated auto-covariance functions (ACFs). In \S\ref{subsec:psd_acf_intro}, we provide a brief set of definitions for the PSD and ACF and their relationship with each other. In \S\ref{subsec:regulator_ext_intro}, we derive results for the Extended Regulator model presented in \citetalias{2020MNRAS.497..698T}. Finally, in \S\ref{subsec:cases} we highlight the four special cases from \citetalias{2020MNRAS.497..698T} -- Milky Way analogues, dwarfs, galaxies at cosmic noon, and galaxies at high-redshift -- along with a description of their general expected behavior. A more detailed set of derivations are presented in \S\ref{ap:model}.

Following \citetalias{2020MNRAS.497..698T}, we will assume that the stochastic processes described by the Extended Regulator model correspond to variability in the natural log of the star formation rate, $\ln {\rm SFR}$ around a time-dependent mean $\mu(t)$ and are captured entirely by the corresponding PSD/ACF. This will allow us to model SFHs as log-Gaussian Processes, which we will return to in Section~\ref{sec:gp}.

\subsection{Terms and Definitions}
\label{subsec:psd_acf_intro}

We start by informally defining a \textit{stochastic process} as something that can generate infinite realizations of a \textit{time series}
$\{ x_1, x_2, \dots x_n \} \equiv \{x_t\}_{1}^{n} \equiv \mathbf{x}_{n}$ at any times $t=1,\dots,n$ (i.e. the $x_t$ values change every time we simulate from the process). The collection of $\mathbf{x}_{n}$ values will then follow some joint probability distribution $P(\mathbf{x}_{n})$ which is defined by the stochastic process.

The simplest way to explore the correlation structure in a given stochastic process is to compute the \textit{auto-covariance function} (ACF)\footnote{The prefix ``auto-'' is often used to emphasize that the calculation is done at two different times for the same process, rather than between two different processes.}
\begin{equation}
    \mathcal{C}(t, t') = \int_{-\infty}^{+\infty} \left[x_t - \mu(t)\right] \left[x_{t'} - \mu({t'})\right] \, P(x_t, x_{t'}) \, {\rm d}x_t {\rm d}x_{t'}
\end{equation}
between $x_t$ and $x_{t'}$ at two times $t$ and $t'$, where $\mu(t)$ is the time-dependent mean and $P(x_t, x_{t'})$ is the joint distribution of $x_t$ and $x_{t'}$ defined by the process. Assuming our process is \textit{stationary} so that the auto-covariance function only depends on the separation (i.e. time lag) between any two given times $\tau \equiv t - t'$ rather than the individual times $t$ and $t'$ themselves, we can instead write the auto-covariance function as
\begin{equation}
    \mathcal{C}(t,t') = \mathcal{C}(t-t') \equiv \mathcal{C}(\tau)
\end{equation}

In addition to defining correlation structure as a function of time $t$, we can also do the same as a function of frequency $f$. We first define a ``windowed'' version of $x(t)$
\begin{equation}
    x_T(t) \equiv x_t w_T(t) =
    \begin{cases}
    x_t & t - \frac{T}{2} < t < t + \frac{T}{2} \\
    0 & {\rm otherwise}
    \end{cases}
\end{equation}
for a window function $w_T(t)$ with some width (duration) $T$ centered around $t$. Taking its \textit{Fourier transform}
\begin{equation}
    \hat{x}_T(f) \equiv \int_{-\infty}^{+\infty} x_T(t) \, e^{-2\pi i f t} \, {\rm d}t
\end{equation}
the \textit{power spectral density} (PSD) is then
\begin{equation}
    \mathcal{S}(f) \equiv \lim_{T \rightarrow \infty}\frac{1}{T} |\hat{x}_T(f)|^2
\end{equation}
where the limit $T \rightarrow \infty$ assumes the stochastic process is not localized in time. We can interpret the PSD as the relative amount of variance as a function of frequency, where larger values indicate stronger correlations at particular frequencies.

While the ACF and PSD can be computed directly from a given stochastic process, they can also be directly computed from each other. Based on the Wiener-Khinchin theorem, in the continuous-time limit the PSD $\mathcal{S}(f)$ and ACF $\mathcal{C}(\tau)$ are Fourier pairs and we can convert between the two using:
\begin{equation}
    \mathcal{S}(f) = \int_{-\infty}^{+\infty} \mathcal{C}(\tau) \, e^{-2\pi i f \tau} \, {\rm d}\tau \quad\Longleftrightarrow\quad
    \mathcal{C}(\tau) = \int_{-\infty}^{+\infty} \mathcal{S}(f) \, e^{+2\pi i \tau f} \, {\rm d}f
\end{equation}
This property is extremely useful, as many stochastic processes (such as the Extended Regulator model in \S\ref{subsec:regulator_ext_intro}) can be much easier to describe in frequency rather than in time (and vice versa).

\subsection{Extended Regulator Model} \label{subsec:regulator_ext_intro}

\citetalias{2020MNRAS.497..698T} introduced the \textit{Extended Regulator model} as a way to characterize how stochastic processes that drive
\begin{enumerate}
    \item gas inflow rates,
    \item gas cycling (between atomic and molecular gas) in equilibrium, and
    \item the formation and disruption of giant molecular clouds (GMCs)
\end{enumerate}
relate to $\ln {\rm SFR}(t)$. Assuming that the behavior of each component follows a \textit{damped random walk} with some de-correlation timescale $\tau_{\rm dec}$ and variability $\sigma$, each PSD can be shown to have a broken power-law PSD of the form
\begin{equation}
    \mathcal{S}_{\rm DRW}(f) = \frac{s^2}{1 + (2\pi\tau_{\rm dec})^2 f^2}
\end{equation}
where $s^2$ is the absolute normalization (scatter squared) for $f=0$. Making the well-justified assumptions that
\begin{enumerate}
    \item the process describing the behavior of GMCs is largely independent of those describing gas inflow and cycling in equilibrium and
    \item the processes describing gas inflow and equilibrium gas cycling are coupled,
\end{enumerate}
the full PSD of the Extended Regulator model can be written as
\begin{align}
    \mathcal{S}_{\rm ExReg}(f) &= \mathcal{S}_{\rm in}(f) \times \mathcal{S}_{\rm eq}(f) + \mathcal{S}_{\rm dyn}(f) \nonumber \\
    &= \frac{s_{\rm in}^2}{1 + (2\pi\tau_{\rm in})^2 f^2} \times \frac{s_{\rm eq}^2}{1 + (2\pi\tau_{\rm eq})^2 f^2} + \frac{s_{\rm dyn}^2}{1 + (2\pi\tau_{\rm dyn})^2 f^2} \\
    &\boxed{= \frac{s_{\rm gas}^2}{1 + ((2\pi\tau_{\rm in})^2 + (2\pi\tau_{\rm eq})^2) f^2 + (2\pi\tau_{\rm in})^2 (2\pi\tau_{\rm eq})^2 f^4} + \frac{s_{\rm dyn}^2}{1 + (2\pi\tau_{\rm dyn})^2 f^2}} \nonumber \\
    &\equiv \mathcal{S}_{\rm gas}(f) + \mathcal{S}_{\rm dyn}(f) \nonumber
\end{align}
where $s_{\rm gas}^2 = s_{\rm in}^2 s_{\rm eq}^2$ is the total variability in gas inflows and equilibrium cycling, $s_{\rm dyn}^2$ is the variability in dynamical processes including the creation and destruction of GMCs, and $\tau_{\rm in}$, $\tau_{\rm eq}$, and $\tau_{\rm dyn}$ are the de-correlation timescales associated with gas inflows, cycling in equilibrium, and GMC formation/disruption respectively, and we use the values of $\beta_l = 0$, $\beta_h = 2$ for the power-law slopes of the gas inflow term as defined in Eqn. 23 and Table 2 of \citetalias{2020MNRAS.497..698T}.

Using the Wiener-Kninchin Theorem, the corresponding ACF is then
\begin{align}
    \mathcal{C}_{\rm ExReg}(\tau) &= \mathcal{C}_{\rm gas}(\tau) + \mathcal{C}_{\rm dyn}(\tau) \nonumber \\
    &\boxed{= \sigma_{\rm gas}^2 \times \frac{\tau_{\rm in} \, e^{-|\tau|/\tau_{\rm in}} - \tau_{\rm eq} \, e^{-|\tau|/\tau_{\rm eq}}}{\tau_{\rm in} - \tau_{\rm eq}} + \sigma_{\rm dyn}^2 \times e^{-|\tau|/\tau_{\rm dyn}}
    }
\end{align}
for all $\tau_{\rm in} \neq \tau_{\rm eq}$ and where we've replaced $s \rightarrow \sigma$ to emphasize that $\mathcal{S}(f=0) = s^2 \neq \sigma^2 = \mathcal{C}(\tau=0)$. See \S\ref{ap:model} for further details and more general results.

Compared to the PSD, the ACF offers different insights into the correlation structure. In particular, it shows that a damped random walk leads to correlations that decay exponentially with time ($\propto e^{-|\tau|/\tau_{\rm dec}}$)\footnote{This is proportional to the Mat{\'e}rn $\nu=1/2$ kernel sometimes used in more general applications of Gaussian Processes. See \S\ref{sec:gp} for additional details.}. Note also that the \textit{variance} of the Extended Regulator model now becomes
\begin{equation}
    \sigma_{\rm ExReg}^2 \equiv \mathcal{C}_{\rm ExReg}(\tau=0) = \sigma_{\rm gas}^2 + \sigma_{\rm dyn}^2
\end{equation}
since there are now multiple independent stochastic processes involved.

\begin{figure}
    \centering
    \includegraphics[width=\textwidth]{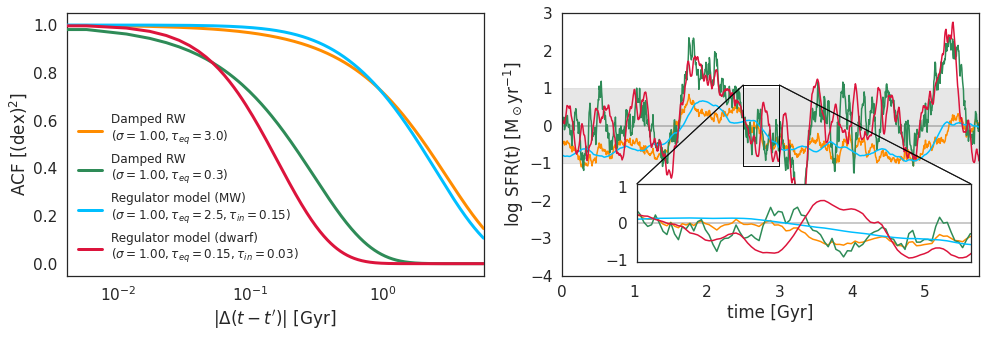}
    \caption{The effects of varying the gas inflow timescale $\tau_{\rm in}$ and equilibrium gas cycling time $\tau_{\rm eq}$ on galaxy star formation histories. We highlight four ACFs: two damped random walks with long and short correlation timescales $\tau_{\rm eq}$, respectively, and two Regulator model realizations (see \S\ref{subsec:regulator_ext}) corresponding to a Milky Way analogue and a dwarf galaxy (see \S\ref{subsec:cases}). To be comparable across different timescales, the four ACFs have all been scaled such that $\mathcal{C}(\tau=0) \approx 1.0$ dex (\textit{left panel}). We then generate a single realisation of the SFH $\log {\rm SFR}(t)$ using a GP with each of the four ACFs \textit{with a fixed random number seed} to highlight differences that arise from varying the ACF (\textit{right panel}). ACFs with longer-timescale correlations tend to spend larger stretches of time above and below the mean SFH, while short-timescale kernels tend to show more ``bursty'' behaviour that generate larger (but short-lived) excursions from the mean. The Milky Way analogue model in particular includes correlations at both long and short timescales from gas cycling and inflows, respectively, giving it the smoothest, most correlated SFH.
    }
    \label{fig:varying_kernel_params}
\end{figure}

We can quantify the extent and strength of the correlations using the \textit{auto-correlation time}
\begin{equation}
    \tau_A \equiv \frac{1}{\sigma^2}\int_{-\infty}^{+\infty} \mathcal{C}(\tau) \,{\rm d}\tau
\end{equation}
which is a measure of the relative correlation contributed by all possible time lags $\tau$. For the Extended Regulator model, evaluating this expression gives
\begin{equation}
    \tau_{A, {\rm ExReg}} = 2 \times \left(\frac{\sigma_{\rm gas}^2}{\sigma_{\rm ExReg}^2}\tau_{\rm gas} + \frac{\sigma_{\rm dyn}^2}{\sigma_{\rm ExReg}^2}\tau_{\rm dyn}\right) \equiv 2 \left(a_{\rm gas} \tau_{\rm gas} + (1-a_{\rm gas}) \tau_{\rm dyn} \right)
\end{equation}
where we've defined $a_{\rm gas}$ as the fractional contribution to the variance from the gas component. In the limit where $\sigma_{\rm gas} \ll \sigma_{\rm dyn}$ so GMC formation/disruption is the dominant source of variability, this gives
\begin{equation}
    \tau_{A, {\rm dyn}} = 2 \tau_{\rm dyn}
\end{equation}
which is just twice the GMC formation/disruption time (since $\tau$ can range from $-\infty$ to $+\infty$). In the limit where $\sigma_{\rm gas} \gg \sigma_{\rm dyn}$, we instead have
\begin{equation}
    \tau_{A, {\rm gas}} = 2 \tau_{\rm gas} = 2 (\tau_{\rm in} + \tau_{\rm eq})
\end{equation}
which is related instead to the combined timescales involved in gas inflows $\tau_{\rm in}$ and cycling $\tau_{\rm eq}$. This final case, where the GMC contribution to the variability is assumed to be negligible, is what \citetalias{2020MNRAS.497..698T} refer to as the \textit{Regulator model}.

The PSD and ACF for the Extended Regulator model are shown in Figure \ref{fig:overview} while comparisons between the Regulator model (two damped random walks) and a single damped random walk are shown in Figure \ref{fig:varying_kernel_params}.

\subsection{Special Cases} \label{subsec:cases}

Following \citetalias{2020MNRAS.497..698T}\footnote{which bases its assumptions on studies of galaxies across cosmic time \citep[e.g.][and references therein]{2020ARA&A..58..157T}}, we consider the following special illustrative cases to highlight the behavior of the Extended Regulator model in four different regimes:
\begin{enumerate}
    \item \textit{Milky Way Analogue (MWA)}: $(\tau_{\rm in}, \tau_{\rm eq}, \tau_{\rm dyn}) = (150, 2500, 25)\,{\rm Myr}$. Based on the long-term secular evolutionary trends seen in the MW, this includes an extremely large $\tau_{\rm eq}$ (2.5\,Gyr) along with approximate order of magnitude ($>5$) differences between various timescales, with $\tau_{\rm eq} \gg \tau_{\rm in} \gg \tau_{\rm dyn}$. SFHs will be dominated by the long-running equilibrium timescale, with small perturbations from changes to the inflow rate along with small amounts of additional white noise from GMCs on much shorter timescales.
    \item \textit{Dwarf}: $(\tau_{\rm in}, \tau_{\rm eq}, \tau_{\rm dyn}) = (150, 30, 10)\,{\rm Myr}$. Although it has the same $\tau_{\rm in}$ as MW, $\tau_{\rm eq}$ has substantially decreased to account for the much more rapid gas cycling (and $\tau_{\rm dyn}$ to a lesser extent for similar reasons) expected in these low-mass galaxies. This has the effect of making the SFHs substantially burstier on short timescales ($\lesssim 100 Myr$). It also includes smaller changes in scale ($\sim 5$), leading to somewhat larger impacts from $\tau_{\rm in} > \tau_{\rm eq} > \tau_{\rm dyn}$. SFHs will be dominated by the variability timescales associated with gas inflows, but with larger perturbations from equilibrium and white noise from GMCs compared to our Milky Way analogue.
    \item \textit{Cosmic Noon}: $(\tau_{\rm in}, \tau_{\rm eq}, \tau_{\rm dyn}) = (100, 200, 50)\,{\rm Myr}$. This case is designed to simulate a typical $10^9 \,M_\odot$ galaxy around $z \sim 2$. The equilibrium time is larger by an order of magnitude relative to the dwarf case due to the larger overall mass, with  smaller changes in scale ($\sim 2$) and longer-lived GMCs. As $\tau_{\rm eq} \gtrapprox \tau_{\rm eq} \gtrapprox \tau_{\rm dyn}$, all timescales remain quite relevant, leading to larger and more correlated fluctuations.
    \item \textit{High-$z$}: $(\tau_{\rm in}, \tau_{\rm eq}, \tau_{\rm dyn}) = (16, 15, 6)\,{\rm Myr}$. Our last case is designed to simulate the conditions for a galaxy at $z\sim 4-6$. Here, $\tau_{\rm eq} \approx \tau_{\rm in}$ with both only $\sim 2\tau_{\rm dyn}$, with extremely short timescales due to the lower masses involved along with the more disruptive environments that many galaxies find themselves in. Since the gas-related timescales are almost identical, we expect this case to have behavior most similar to a Matern32 kernel (see \S\ref{subsubsec:matern32}) but with additional perturbations caused by GMCs on somewhat similar timescales.
\end{enumerate}

We consider two classes of models when deciding on the values of the scatter $\sigma_{\rm gas}$ and $\sigma_{\rm dyn}$:
\begin{enumerate}
    \item \textit{Fixed (0.4 dex)}: We normalize our results such that $\sigma_{\rm gas} = 0.39\,{\rm dex}$ and $\sigma_{\rm dyn} = 0.07\,{\rm dex}$, so that the relative contribution from gas inflows/cycling versus GMC formation/disruption are always fixed. This allows us to fix the scatter in log SFR at 0.4 dex, and isolate the impact that relative changes in timescales may have on SFHs and associated observables. We choose 0.4 dex since it is close to the commonly measured value for the scatter in the SFR-M$_*$ correlation \citep{2016ApJ...820L...1K, 2018ApJ...866..120I, 2021arXiv211004314L}.
    \item \textit{Variable (\citetalias{2020MNRAS.497..698T})}: We normalize our results to the values reported in \citetalias{2020MNRAS.497..698T} (see their Figure 9) of $\sigma_{\rm MWA} = 0.17$ dex, $\sigma_{\rm Dwarf} = 0.53$ dex, $\sigma_{\rm Noon} = 0.24$ dex and $\sigma_{\rm High-z} = 0.27$ dex. This involves relative changes in both the overall scatter and the relative contributions from gas inflows/cycling versus GMC formation/disruption.
\end{enumerate}

\begin{figure}
    \centering
    \includegraphics[width=\textwidth, page=2, trim=0.0cm 7.5cm 0.0cm 0.0cm, clip]{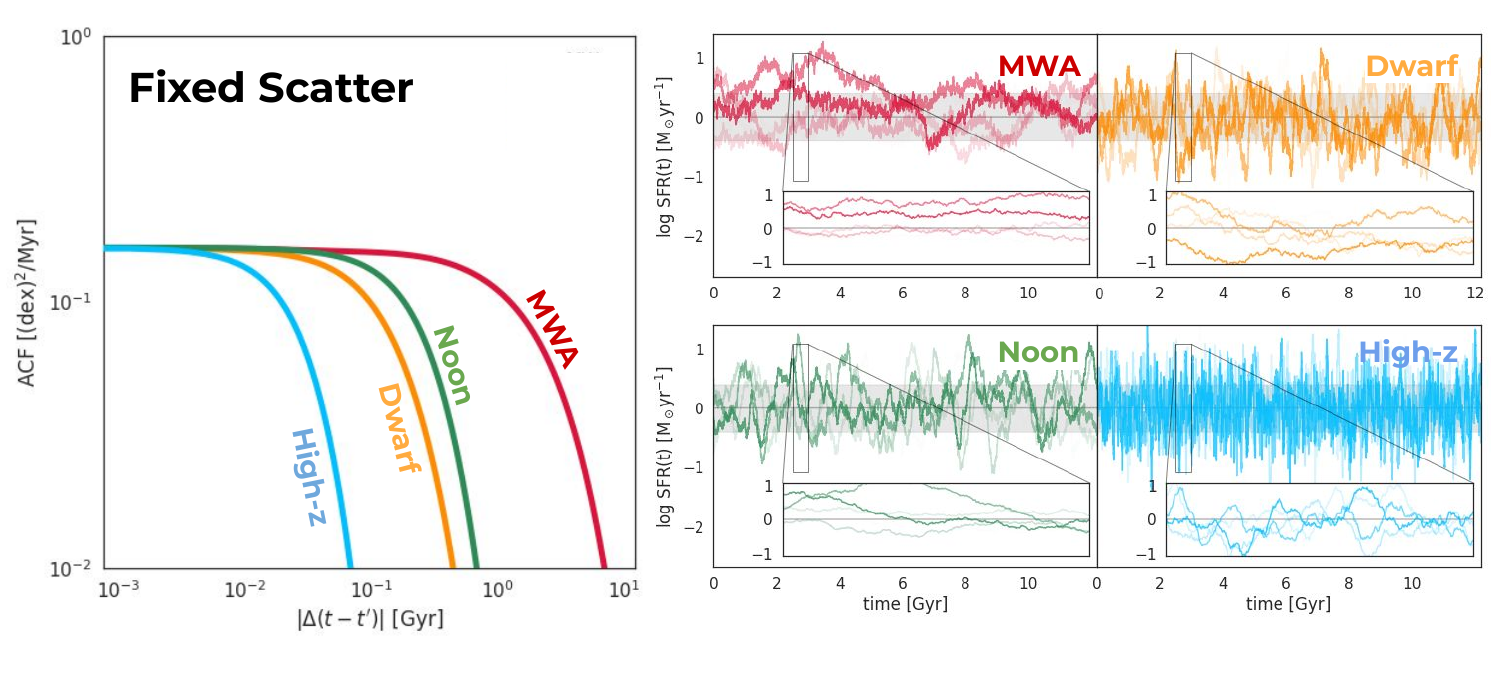}
    \caption{An illustration of the ACF (\textit{left panel}) and several corresponding realizations of SFHs (\textit{right panels}) for the four cases discussed in \S\ref{subsec:cases}: Milky Way analogues at $z=0.1$ (MWA; red), dwarf galaxies at $z=0.1$ (orange), massive galaxies at cosmic noon at $z=2$ (``noon''; green), and galaxies at high redshifts (high-$z$; blue). SFHs are generated assuming a base SFR of $1M_\odot/yr$ for the age of the universe (i.e. ignoring formation times) and with a fixed total scatter of $1\,{\rm dex}$ and a fixed relative contribution of GMC formation/disruption to the variance of $f_{\rm dyn} = 0.03$. These highlight the relative changes in behavior that arise only from varying correlation timescales $\tau_{\rm in}$, $\tau_{\rm eq}$, and $\tau_{\rm dyn}$. Within each panel, SFHs are plotted with various intensities to help the eye distinguish them.}
    \label{fig:extended_regulator_kernel_cases1}
\end{figure}

\begin{figure}
    \centering
    \includegraphics[width=\textwidth, page=3, trim=0.0cm 7.5cm 0.0cm 0.0cm, clip]{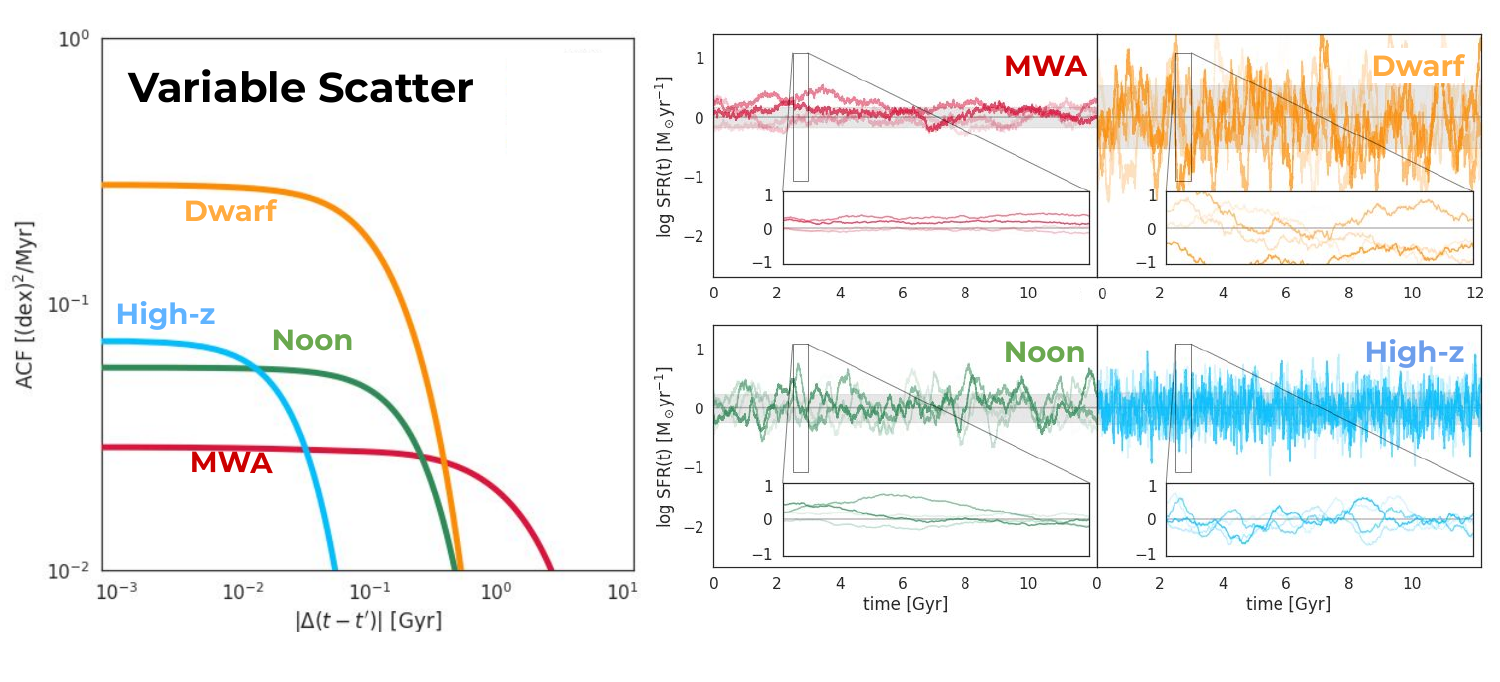}
    \caption{Similar to Figure \ref{fig:extended_regulator_kernel_cases1}, but the overall scatter for each case is set to the values used in \citetalias{2020MNRAS.497..698T}. These highlight overall changes in behavior that arise both from varying correlation timescales as well as changes in the overall burstiness of SFHs.}
    \label{fig:extended_regulator_kernel_cases2}
\end{figure}

The general behaviour of each case for fixed and variable scatter is highlighted in Figures \ref{fig:extended_regulator_kernel_cases1} and \ref{fig:extended_regulator_kernel_cases2}, respectively.

\section{Gaussian Process Implementation} \label{sec:gp}

We now describe how we can use the ACFs computed in \S\ref{sec:model} to quickly generate realizations of galaxy star formation histories (SFHs). In \S\ref{subsec:gp}, we provide a brief overview of Gaussian Processes (GPs). In \S\ref{subsec:fsps}, we describe how we use them to generate realizations of synthetic galaxy spectra. Our implementation is publicly available at \url{https://github.com/kartheikiyer/GP-SFH} and summarized in Figure \ref{fig:overview}.

\subsection{Brief Overview of Gaussian Processes} \label{subsec:gp}

A \textit{Gaussian Process (GP)} is a generalization of the Gaussian distribution to the space of functions \citep{gp_book}. Similar to our definition of a stochastic process, this means that our GP can generate an infinite set of values $\mathbf{y}_{n} \equiv \{y_t\}_{t=1}^{t=n}$ at any time $t=1,\dots,n$ whose joint probability distribution will always follow a multidimensional Gaussian distribution
\begin{equation}
    P(\mathbf{y}_{n}) = \mathcal{G}(\mathbf{y}_{n} | \boldsymbol{\mu}_{n}, \mathbf{C}_{n,n})
\end{equation}
where $\boldsymbol{\mu}_{n}$ is the $n$ vector of mean values generated from some mean function $\mu(t)$ at times $t=1,\dots,n$ and $\mathbf{C}_{n,n}$ is the $n \times n$ covariance matrix evaluated at each pair of times $t=1,\dots,n$ and $t'=1,\dots,n$. If we have some values $\mathbf{y}_{m}$ that are known and want to predict a set of new possible values at our given times $t=1,\dots,n$, this can be done by exploiting the fact that the joint distribution is
\begin{equation}
    P(\mathbf{y}_{n},\mathbf{y}_{m}) = \mathcal{G}\left(
    \begin{bmatrix} \mathbf{y}_{n} \\ \mathbf{y}_{m} \end{bmatrix} |
    \begin{bmatrix} \boldsymbol{\mu}_{n} \\ \boldsymbol{\mu}_{m} \end{bmatrix} ,
    \begin{bmatrix} \mathbf{C}_{n,n} & \mathbf{C}_{n,m} \\ \mathbf{C}_{m,n} & \mathbf{C}_{m,m} \end{bmatrix}\right)
\end{equation}
which gives a conditional distribution of
\begin{equation}
\label{eqn:gp_cov}
    P(\mathbf{y}_{n} | \mathbf{y}_{m}) = \mathcal{G}\left(\boldsymbol{\mu}_{n} + \mathbf{C}_{n,m} \mathbf{C}_{m,m}^{-1} (\mathbf{y}_{m} - \boldsymbol{\mu}_{m}), \mathbf{C}_{n,n} - \mathbf{C}_{n,m} \mathbf{C}_{m,m}^{-1} \mathbf{C}_{m,n}\right)
\end{equation}
Together, these properties along with the functional nature of GPs are often summarized using the following notation:
\begin{equation}
    y(t) \sim \mathcal{GP}(\mu(t), \mathcal{C}(t,t'))
\end{equation}
where the $\sim$ indicates ``is a realization of'' rather than the ``is of the same order-of-magnitude as'' definition usually used in the astrophysics literature.

Taken together, the above results imply that we can use GPs to quickly and easily generate realizations of our data $\mathbf{y}_n$ either completely from scratch or conditioning on some known values $\mathbf{y}_m$ based on some input mean $\mu(t)$ and covariance $\mathcal{C}(t,t')$ functions that we can easily replace with any of the ACFs derived in \S\ref{sec:model}. In particular, switching over to our variable of interest ($\ln {\rm SFR}$) and the model of interest (the Extended Regulator model discussed in \S\ref{subsec:regulator_ext_intro}), the model we explore in our paper takes the form
\begin{equation}
\boxed{
    \ln {\rm SFR}(t) \sim \mathcal{GP}(\ln {\rm SFR}_{\rm base}(t), \mathcal{C}_{\rm ExReg}(\tau))
    }
\end{equation}
where $\ln {\rm SFR}_{\rm base}(t)$ is some baseline SFH we are interested in studying and again $\tau = t - t'$. In other words, any given realization of the SFH
will depend on both the ``baseline'' (mean) SFH, $\ln {\rm SFR}_{\rm base}(t)$, as well as the particular ACF $\mathcal{C}_{\rm ExReg}(\tau)$ defined by the Extended Regulator model.

In practice, our GP is implemented following a similar procedure to \citet{2019ApJ...879..116I} using a multidimensional Gaussian prior initialized at every point of an input time array. This is done through an instance of the \verb|GP_SFH()| class, which is initialized with a user-determined kernel at a particular redshift, along with an \verb|astropy.cosmology()| object and a \verb|fsps.stellarpopulation()| object for generating spectra and other observables. Upon initialization, the instance computes the covariance matrix specified by the ACF at a range of time values ranging from 0 to the $t_{univ}$ at the specified redshift. Once this matrix is computed, realisations of SFHs can be generated simply by sampling a multivariate normal distribution at each time in the array with the covariance structure determined by the kernel. This can then be conditioned on observable constraints using eqn.\ref{eqn:gp_cov}.

\subsection{Implementation with Stellar Population Synthesis Models} \label{subsec:fsps}

To generate spectra corresponding to draws from the GP, we pass the star formation histories though the Flexible Stellar Population Synthesis (FSPS) code \citep{conroy2009propagation, conroy2010propagation}. To highlight the differences in galaxy spectra that arise solely from differences in the ACF, we choose a simple set of modeling assumptions (listed in Table \ref{tab:spec_generation}) when generating spectra while keeping everything else fixed to their default FSPS values. We also demonstrate the effects of varying some of these parameters on our observables of interest in Section 5.1.

In practice, changes to the stellar population parameters can be made simply by reassigning the input \verb|fsps.stellarpopulation| object linked to the \verb|GP-SFH| class instance. This modular implementation allows for a pre-computed covariance matrix to be rapidly associated with many different stellar population parameters while modeling and fitting SEDs, since that is the rate-limiting step to generating SFH realisations.

For spectral features, we choose to consider features sensitive to star formation on a range of timescales \citep[][]{kauffmann2003stellar}.
H$\alpha$ emission from O- and B-stars probes star formation on $4-10$ Myr timescales \citep{2021MNRAS.501.4812F, 2022MNRAS.513.2904T}.
H$\delta$ absorption traces star formation over the last $0.1-1$ Gyr \citep{1997ApJS..111..377W, 2020ApJ...892...87W}.
Finally, the 4000 \AA{} break strength (D$_n (4000)$) provides a reliable tracer of the median age of the stellar populations that make up a galaxy, as can be seen in Figure 2 of \citet{kauffmann2003stellar}, who point out that these indices are largely insensitive to dust attenuation effects, and the distribution of galaxies in this space is sensitive to stochasticity in star formation over the most recent $\sim 2$ Gyr in a galaxy's past.
This happens because galaxies that form stars at a steady rate tend to occupy a very narrow locus in H$\delta_{\rm EW}$-D$_n (4000)$ space. In addition to these, we also consider the equivalent width of the Ca H, K lines at $\lambda \simeq 3933.6, 3968.5 $ \AA. The Ca II K line traces older stellar populations, while the Ca H absorption line is blended with H$\epsilon$ and [Ne III] and effectively traces intermediate age populations \citep{2004ApJ...600..188M, 2013ApJ...773...16Z}. The resulting equivalent widths probe a range of timescales as seen in Figure \ref{fig:var_ExReg_params} and Appendix \ref{app:resp_curves}.

For our spectra, we get the H$\alpha$ line luminosity directly from the FSPS outputs and following a procedure similar to \citet{kauffmann2003stellar}, we use the $3850-3950$ and $4000-4100$ \AA{} continuum bands introduced by \citet{1999ApJ...527...54B} to compute the strength of the D$_n (4000)$ break, and compute the H$\delta_{\rm EW}$ using the bandpasses of $4083.50-4122.25$ \AA, $4041.60-4079.75$ \AA, and $4128.50- 4161.00$ \AA{} for the index and blue/red continuum bands respectively, defined in Table 1 of \citet{1997ApJS..111..377W}. For the Ca-K$_{\rm EW}$, we use $3929.51-3941.22$ \AA, $3907.01 - 3929.51$ \AA, and $3941.22 - 3961.02$ \AA{} and for Ca-H$_{\rm EW}$, we use $3961.02-3980.83$ \AA, $3941.22-3961.02$ \AA and $3980.82-3997.03$ \AA{} for the index, blue and red bandpasses respectively. For better comparison across the different stellar masses that could be produced due to bursts and troughs in individual realisations of SFHs, we normalize these quantities by the stellar masses of each realisation, effectively reporting e.g., the distribution of H$\alpha$ luminosity (in L$_\odot$) \textit{per solar mass} for the different cases discussed in Section \ref{sec:single}. The H$\delta_{\rm EW}$, D$_n (4000)$ and broad-band galaxy colors remain unaffected by this normalisation.

\begin{table}[]
    \centering
    \begin{tabular}{c|c}
        \hline
        Input/Parameter &  Fixed Option/Value \\
        \hline \hline
        Isochrones, Stellar tracks & MILES+MIST \\
        Redshift & 0.1 \\
        SFH$_{\rm base}(t)$ & Constant (1.0 M$_\odot /yr$) \\
        IMF & Chabrier \\
        Dust attenuation & Calzetti ($A_V = 0.2$) \\
        log Z/Z$_\odot$ & 0.0 (Solar) \\
        \hline
    \end{tabular}
    \caption{Modeling assumptions for generating spectra corresponding to different ACF cases.}
    \label{tab:spec_generation}
\end{table}

\section{Spectrophotometric Signals of Changing Stochastic Behavior} \label{sec:single}

Following the procedure highlighted in Figure \ref{fig:overview}, in this section we identify the particular spectrophotometric signals that can help to distinguish different types of stochastic behavior (i.e. varying correlation timescales $\tau_{\rm in}$, $\tau_{\rm eq}$, and $\tau_{\rm dyn}$ and fluctuation strengths $\sigma_{\rm gas}$ and $\sigma_{\rm dyn}$). We generate 10000 realizations of various SFHs ($\log {\rm SFR}(t)$) for each of the cases outlined in \S\ref{subsec:cases} with the scatter fixed (Figure \ref{fig:extended_regulator_kernel_cases1}) and the scatter matched to \citetalias{2020MNRAS.497..698T} (Figure \ref{fig:extended_regulator_kernel_cases2}). We then feed these into FSPS to generate a set of UV-to-IR galaxy SEDs as described in Section \ref{subsec:fsps}.

\begin{figure}
    \centering
    \includegraphics[width=\textwidth]{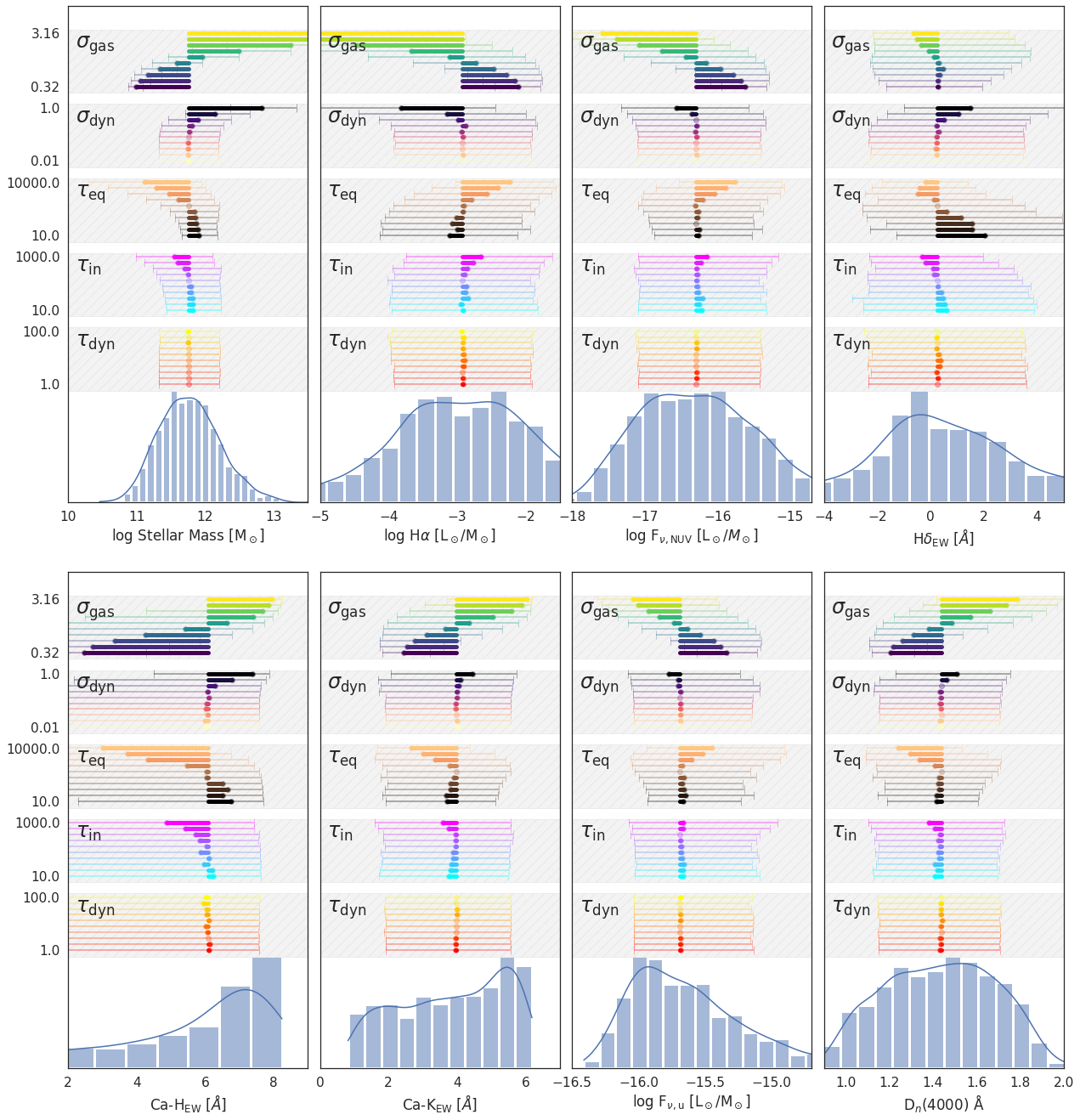}
    \caption{Effects of varying the Extended Regulator model parameters on the distributions of stellar mass and spectral features spanning a range of timescales: (in increasing order) H$\alpha$ \& NUV fluxes, equivalent widths of H$\delta$ \& Ca-H,K, u-band flux, and finally the strength of the D$_n$(4000)\AA{} break. Histograms at the bottom of each panel show the fiducial distribution, corresponding to $(\sigma_{gas}, \sigma_{dyn}, \tau_{in}, \tau_{eq}, \tau_{dyn}) \equiv (1.0,0.1,500,150,10)$, while the individual bars show the change in the median (thick solid line) and the width of the distribution, shown using the 16-84$^{th}$ percentiles (thin error bars) upon changing each of the ExReg model parameters.
    The non-degenerate changes in the observables upon perturbing the ExReg model allow us to use a combination of spectral features sensitive to a range of timescales to test and constrain the model parameters observationally.}
    \label{fig:var_ExReg_params}
\end{figure}

To highlight the behavior of our model, we first investigate overall effects that the parameters in the Extended Regulator model may have on a few key observables. Our results are highlighted in Figure \ref{fig:var_ExReg_params}, where we vary single parameters in the Extended Regulator model while holding the rest fixed at fiducial values ($\sigma_{\rm gas} = 1.0$, $\tau_{\rm eq} = 500$ Myr, $\tau_{\rm eq} = 150$ Myr, $\sigma_{\rm dyn} = 0.1$, $\tau_{\rm dyn} = 10$ Myr). We find that the exact mechanism for adjusting the ``burstiness'' of star formation -- whether through the overall level of variability in the gas inflows/cycling ($\sigma_{\rm gas}$) or GMC formation/disruption ($\sigma_{\rm dyn}$), or through the duration of the (gas equilibrium) correlation time $\tau_{\rm eq}$ -- leaves different imprints on various observables \textit{even at fixed SFR scatter}. In particular, while varying $\sigma_{\rm gas}$ affects both the stellar mass and (s)SFR distributions, varying $\tau_{\rm eq}$ \textit{only} affects the stellar mass distribution. This gives at least one way to distinguish populations with differing amounts of variability about the same base set of SFHs.

Since changing correlation timescales and the level of scatter can lead to large differences in the final stellar mass formed,
we choose to normalize all SEDs based on their final stellar mass before investigating possible (relative) differences. This helps to highlights trends as a function of specific star formation rate (sSFR) rather than just SFR, and helps to account for the increasing (expected) variance in the total stellar mass formed for more bursty SFHs.

\begin{figure}[t]
    \centering
    \includegraphics[width=\textwidth]{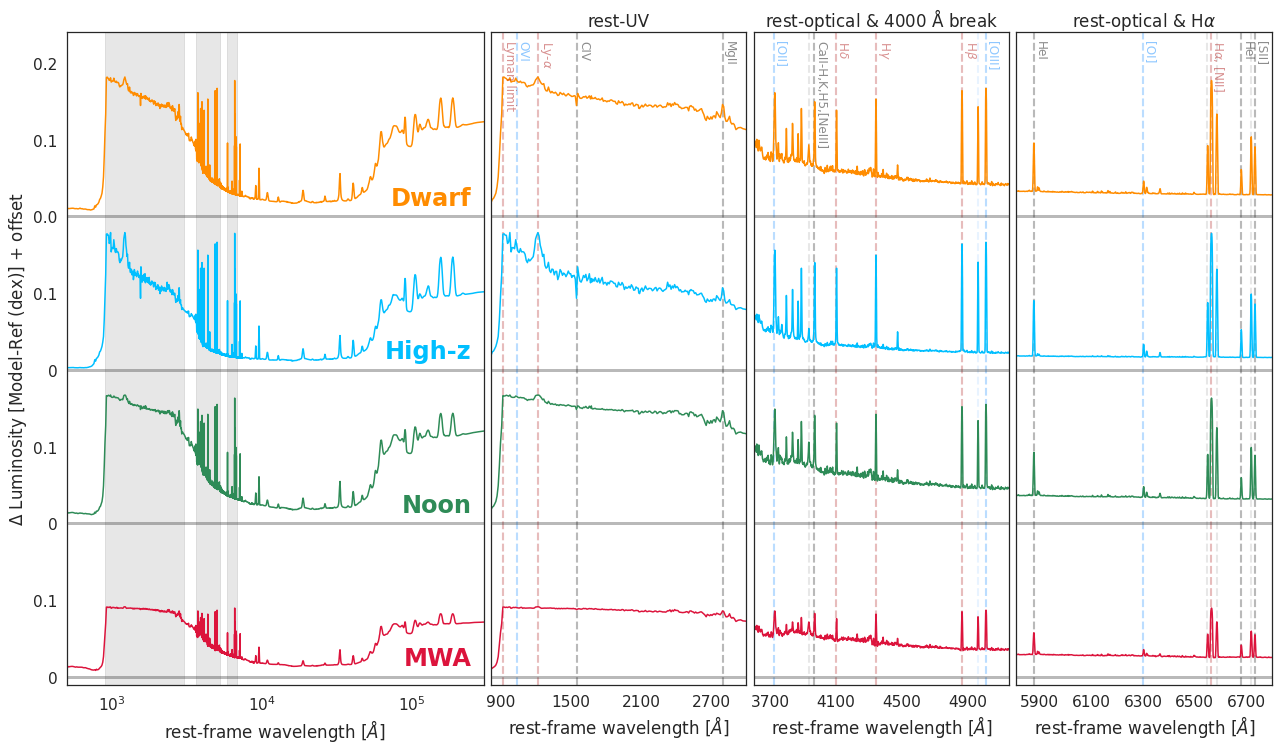}
    \caption{Observational signatures caused by \textit{only} varying correlation timescales $\tau_{\rm in}$, $\tau_{\rm eq}$, and $\tau_{\rm dyn}$ based on the four cases highlighted in Figure \ref{fig:extended_regulator_kernel_cases1} with \textit{fixed scatter}. Using 10000 SFH realizations for each case, we highlight differences in the median \textit{stellar mass-normalized} full UV-to-IR galaxy SEDs with respect to the reference SED of a galaxy with the base SFH (i.e., constant SFR of $1$M$_\odot$/yr).
    As in Figure \ref{fig:var_ExReg_params}, increasing the correlation timescales tends to decrease the final stellar mass and increase the variability due to longer bursts/quiescent periods above/below the base SFH. On a broad scale, this results in a decreased relative sSFR for models with more variability, leading to signatures in the rest-UV. More subtle signatures of the stochasticity are also visible in the relative strengths of the 4000\AA{} break and absorption lines.
    }
    \label{fig:extended_regulator_spec_cases1}
\end{figure}

\begin{figure}[t]
    \centering
    \includegraphics[width=\textwidth]{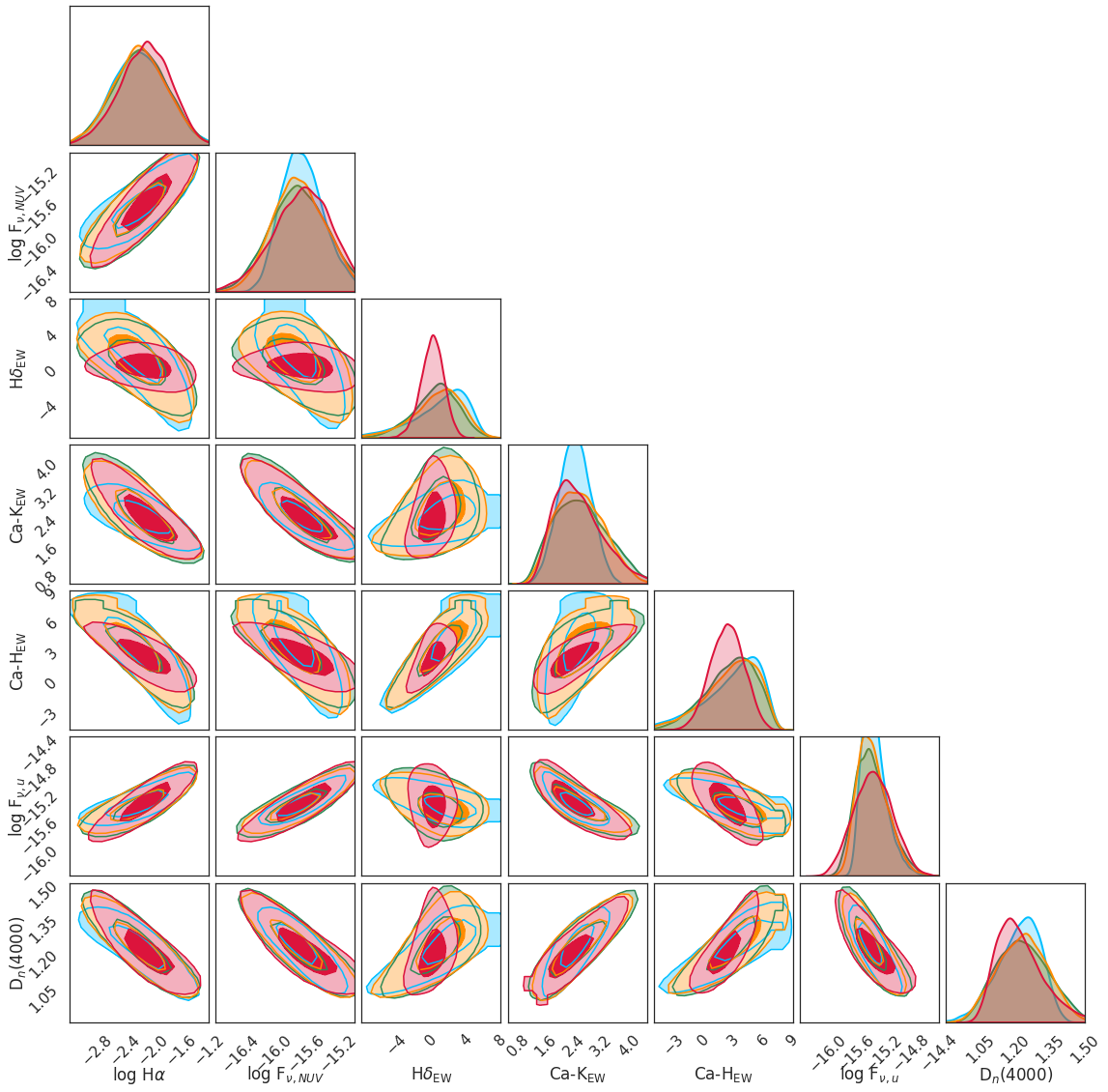}
    \caption{Observational signatures caused by \textit{only} varying correlation timescales $\tau_{\rm in}$, $\tau_{\rm eq}$, and $\tau_{\rm dyn}$ based on the four cases highlighted in Figure \ref{fig:extended_regulator_kernel_cases1} with \textit{fixed scatter}. Using 10000 SFH realizations for each case, we highlight differences in the distributions of spectral features sensitive to star formation across a range of timescales,  including H$\alpha$, H$\delta$, and $D_n(4000)$.
    While the differences in the median value of observables are subtle, the joint distributions do show some differences that arise from changes in the underlying sSFR distribution.}
    \label{fig:extended_regulator_spec_cases1_corner}
\end{figure}

\subsection{Fixed (0.4 dex) scatter}
\label{subsec:scatter_fixed}

The results of this procedure assuming a fixed (0.4 dex) scatter are the SFR is shown in Figures \ref{fig:extended_regulator_spec_cases1} and \ref{fig:extended_regulator_spec_cases1_corner} for all four of the models we consider. Figure \ref{fig:extended_regulator_spec_cases1} shows the median spectra for each model compared to  a reference spectrum corresponding to the base SFH, and Figure \ref{fig:extended_regulator_spec_cases1_corner} shows a corner plot comparing the full distributions of the spectral features described in Section \ref{subsec:fsps}. As expected based on our results in Figure \ref{fig:var_ExReg_params}, since the gas inflow/cycling physics dominates the main behavior of the model, decreasing the associated gas timescale (i.e. $\tau_{\rm gas} = \tau_{\rm in} + \tau_{\rm eq}$) to make SFHs more bursty leads to both larger stellar masses and a tighter overall distribution for a given scatter. While the mass-normalized spectra have very similar medians across the four archetypes, they diverge in the optical and ultraviolet.
This, combined with subtle differences in the H$\delta$ EW (rising with the decreasing $\tau_{\rm gas}$ as SFHs form more of their stars in recent bursts) and the slope of the H$\alpha$-$D_n(4000)$ correlation (becoming shallower and more dispersed as SFHs become less correlated) indicate that, even in the case where the \textit{only} thing varying are the timescales, constraining them should be possible with a large enough sample size.

\begin{figure}
    \centering
    \includegraphics[width=0.31\textwidth]{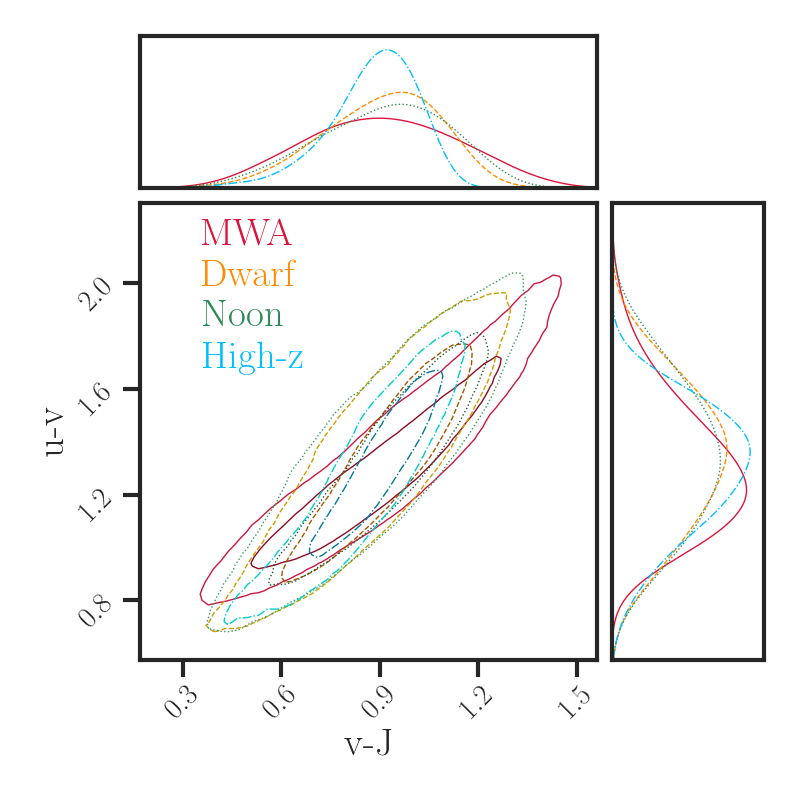}
    \includegraphics[width=0.31\textwidth]{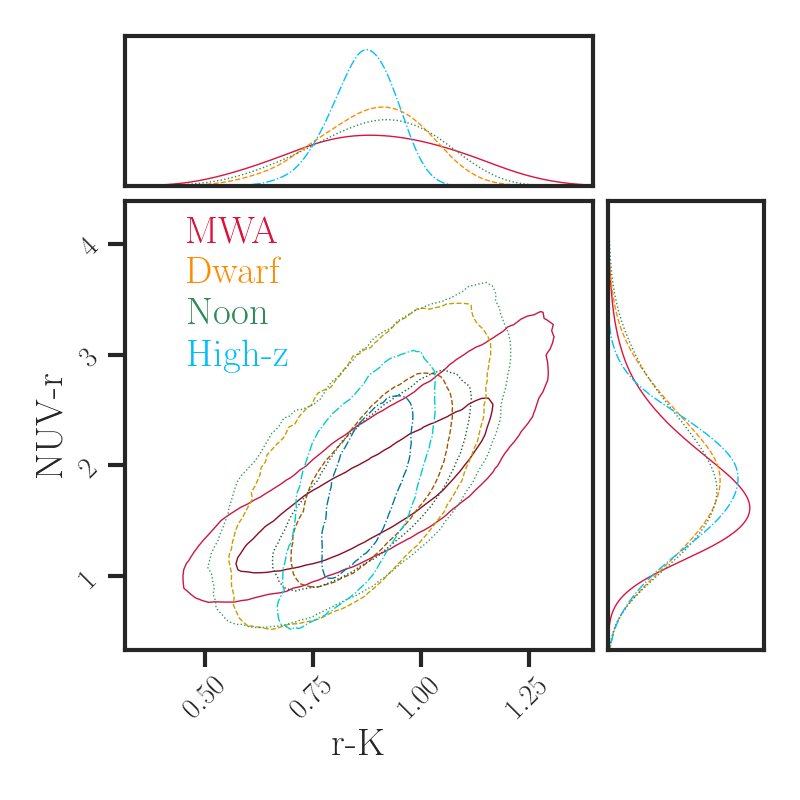}
    \includegraphics[width=0.31\textwidth]{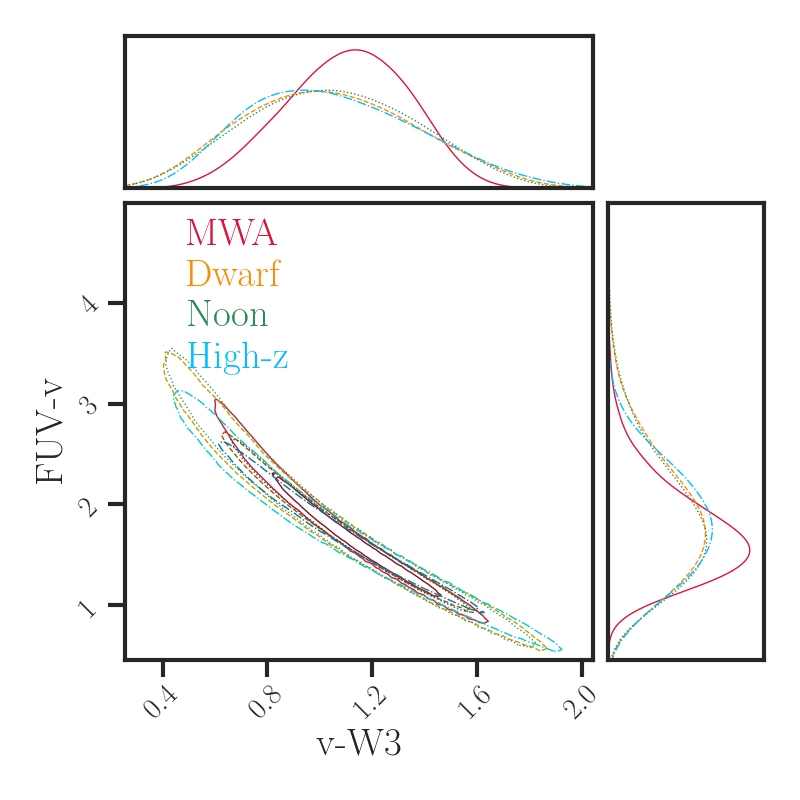}
    \caption{The distribution of galaxy colors corresponding to the populations described in Figure \ref{fig:extended_regulator_spec_cases1} in U-V-J (\textit{left}), NUV-r-K (\textit{middle}), and NUV-V-W3 (\textit{right}). While the broad distributions are similar among all populations due to their similar SFHs, we see meaningful shifts in both the wings of the distributions and the overall general centers.
    }
    \label{fig:traditional_colors_1dex}
\end{figure}

Since obtaining spectroscopic data for large ensembles of galaxies is expensive and may not always be feasible, we also consider distributions of broad-band colors corresponding to the different archetypes. We consider three different color-color spaces - (i) the commonly used UVJ diagram \citep{2007ApJ...655...51W,2009ApJ...691.1879W, 2013ApJ...777...18M}, where the different archetypes can be differentiated based on their sSFR distributions, (ii) the NUV-r-K diagram \citep{2013A&A...558A..67A,2016A&A...590A.103M}, which traces the differences in stellar mass distributions since the K$_S$ band probes the rest-frame 1.6$\mu$ feature, and the $NUV-r$ probes the effects of dust and SFR, and (iii) the recently introduced wide-baseline FUV-V-(Wise)W3 diagram following \citet{2019ApJ...880L...9L}, which is more sensitive to lower sSFRs than the UVJ diagram, where galaxies with low sSFRs tend to populate the top-left portion of the space. As shown in Figure \ref{fig:traditional_colors_1dex}, all three color-color spaces provide a means to differentiate between the different regimes of stochasticity typified by the four archetypes with sufficient sample sizes and SNR.
In addition to this, upcoming observations with JWST will help push rest-frame colors out to higher redshifts, and provide extremely high SNR probes of differences in the stochasticity across the four archetypes considered here.

\subsection{Variable (TCF20) scatter}
\label{subsec:scatter_tcf20}

\begin{figure}
    \centering
    \includegraphics[width=\textwidth]{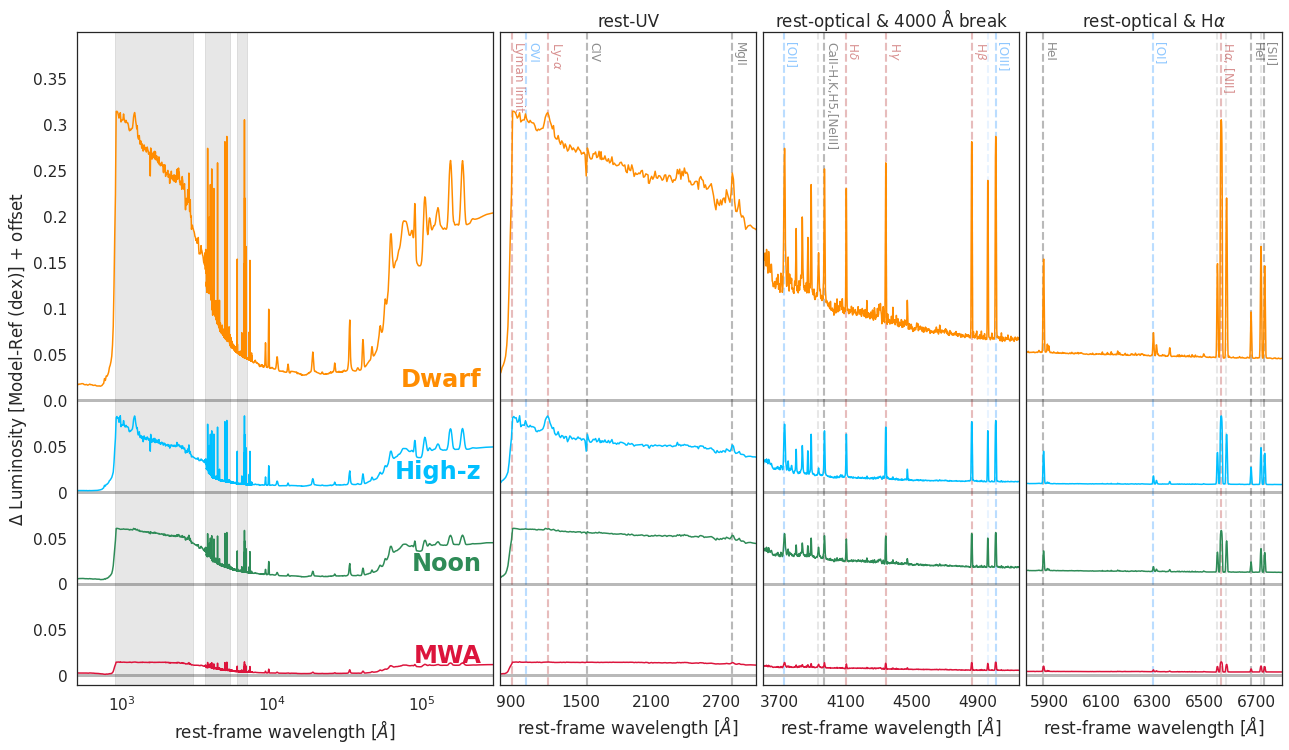}
    \caption{As in Figure \ref{fig:extended_regulator_spec_cases1}, but now for the four cases highlighted in Figure \ref{fig:extended_regulator_kernel_cases2} with \textit{variable (\citetalias{2020MNRAS.497..698T}) scatter}. The variation in scatter dramatically expands the set of sSFR distributions due to the changing distribution of total stellar mass formed after a fixed time given a constant SFH. These differences are most prominent when comparing the model with the smallest scatter and longest timescales (MW analogue; red) with the one with the largest scatter (dwarf; orange).}
    \label{fig:extended_regulator_spec_cases2}
\end{figure}

\begin{figure}
    \centering
    \includegraphics[width=\textwidth]{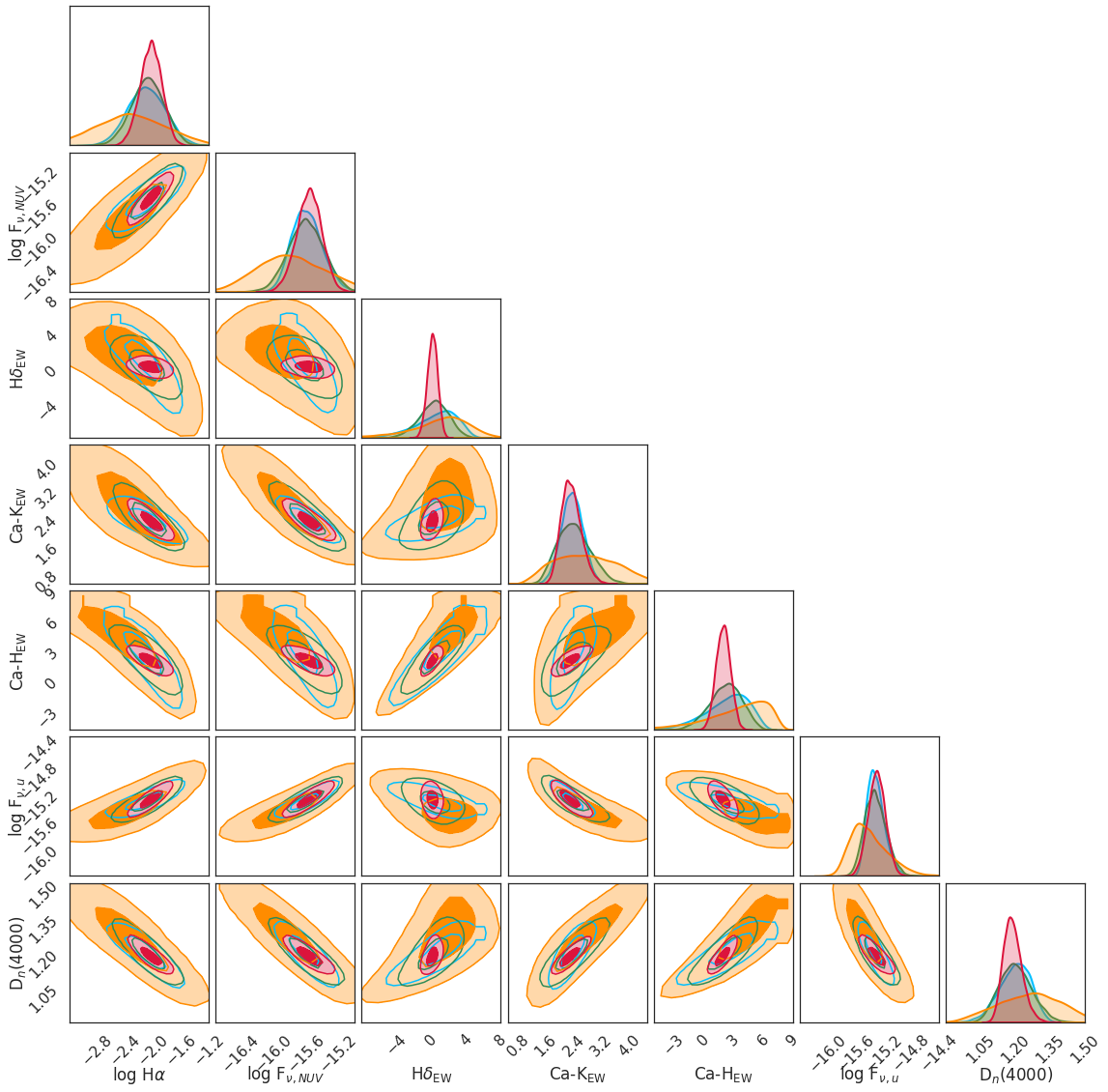}
    \caption{As in Figure \ref{fig:extended_regulator_spec_cases1_corner}, but now for the four cases highlighted in Figure \ref{fig:extended_regulator_kernel_cases2} with \textit{variable (\citetalias{2020MNRAS.497..698T}) scatter}. The variation in scatter dramatically expands the set of sSFR distributions due to the changing distribution of total stellar mass formed after a fixed time given a constant SFH. These differences lead to greater distinguishing power using the distributions of the spectral features we consider.}
    \label{fig:extended_regulator_spec_cases2_corner}
\end{figure}

In the more realistic case from \citetalias{2020MNRAS.497..698T} where the scatter among each archetype also varies, we see larger differences in the corresponding observables (Figures \ref{fig:extended_regulator_spec_cases2} and \ref{fig:extended_regulator_spec_cases2_corner}). To first order, this is associated with the large variations in $\sigma_{\rm gas}$, leading to shifts and/or broadening of the stellar mass and SFR distributions that propagate down to H$\alpha$, H$\delta_{\rm EW}$, and $D_n(4000)$. These can be seen most clear by looking at the distribution of values for the dwarf model (which has the largest $\sigma_{\rm gas}$) and the MW analogue (which has the longest $\tau_{\rm gas}$). Perhaps the biggest change in terms of the spectral features is the H$\delta_{\rm EW}$ distribution which, while still varying in width, is much more centered around a small mean amount of emission from recent SFR. This mainly happens due to the reduced scatter in all the models, which provides less opportunity for the bursty episodes of SF to form large amounts of stellar mass and thereby reduce the sSFRs on average. In terms of the broad-band colors, changing the scatter has the effect of scaling the corresponding distribution of colors by a proportional amount, while maintaining the differences in shape due to varying stochasticity.

\section{Discussion} \label{sec:disc}

Having established the GP-SFH formalism and demonstrated sensitivity to the extended regulator model parameters in Section \ref{sec:model}, we consider the implications and predictions we can make with JWST observations using this formalism in \ref{subsec:jwst}. We consider the effects of other stellar population variables such as dust and metallicity in Section \ref{subsec:dustmet}. Sec \ref{subsec:binnedsfh} demonstrates how the GP-SFH can be used as a stochasticity prior for binned SFHs, and \ref{subsec:psbpops} considers how populations of galaxies with specific spectral features can provide novel constraints on stochasticity. Sec \ref{subsec:varsfhs} considers relaxing the assumption of a fixed base SFH, and Sec \ref{subsec:nonstat} considers the assumption of stationarity and shows an example of relaxing it with a time-varying kernel and \ref{subsec:caveats} considers some caveats and challenges of the GP-SFH formalism.

\subsection{Implications for JWST}
\label{subsec:jwst}

The release of JWST data from the early release observations (ERO, aka ``Webb's First Deep Field"; \citealt{2022arXiv220713067P}), GLASS \citep{2022arXiv220711701M} and CEERS \citep{2022arXiv220712474F} have demonstrated the incredible potential for probing star formation in galaxies across an incredible range of redshifts, environments and stellar masses. This also enables studies of star formation stochasticity using the colors and spectral features in this work to constrain the ExReg model parameters - i.e. stochasticity amplitudes and timescales.

At redshifts around cosmic noon, we will see a major improvement in being able to directly measure rest-frame colors. In contrast to HST ACS+WFC3, which can measure rest-frame UVJ colors at $0.21\lesssim z\lesssim 0.29$, JWST's NIRCam filters will extend this to $1.51 \lesssim  z \lesssim  2.55$, and a combination of HST+JWST will span the entire $0.21\lesssim z\lesssim 2.55$ range. This is also similar to the redshift range that slitless spectroscopy using NIRISS will be able to measure the spectral features discussed in this work for large populations of galaxies \citep{2022PASP..134b5002W}.

At higher redshifts, as we attempt the challenging task of measuring the star formation rates and histories of these galaxies, care must be taken to account for the dependence of the results on the assumed priors \citep{2022ApJ...927..170T, 2022arXiv220605315W}, and use a combination of spectroscopic and photometric data where available \citep{2022arXiv220803281T}. The GP-SFH formalism described here can help motivate SFH priors for the next generation of SED fitting based studies based on estimates of the stochasticity from lower-redshift analogs or simulations.

Additionally, although we did not consider its effects in this work, the evolution of gas-phase and stellar metallicity in galaxies is also tied to their star formation, and further studies of their correlated properties \citep[as in][]{2022ApJ...933...44C, 2022MNRAS.513.5446Z} could provide further observable tests of the ExReg model timescales, and priors for SED fitting codes that explicitly allow for evolution in chemical enrichment over time \citep{2013ApJ...762L..15P, 2021MNRAS.505..540T}.

\subsection{Effects of varying other SED parameters}
\label{subsec:dustmet}

SED modeling depends on a host of assumptions about the stellar populations that make up a galaxy, in addition to dust attenuation and emission from dust heating, nebular regions and AGN. In our current analysis we have held most of these constant in order to isolate and study the effects of perturbing the SFH stochasticity model, but it is informative to consider the extent to which varying these additional parameters will broaden the distributions we expect to observe.

\begin{figure}
    \centering
    \includegraphics[width=\textwidth, page=4, trim=0.0cm 5.2cm 0.0cm 0.0cm, clip]{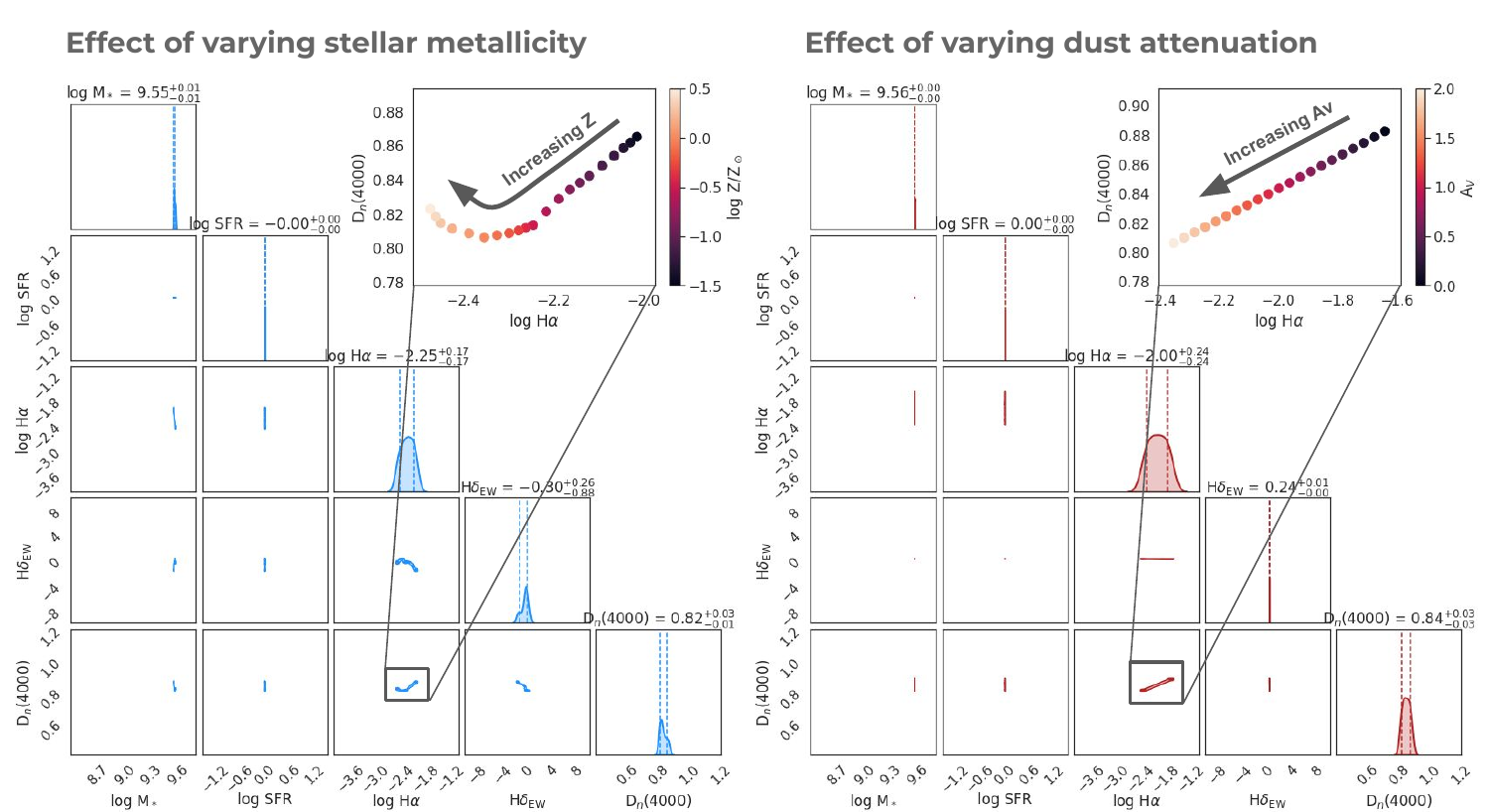}
    \caption{The impact that varying dust and metallicity has on the spectral features shown in Figure \ref{fig:extended_regulator_spec_cases1_corner} and \ref{fig:extended_regulator_spec_cases2_corner}. To highlight the size of the effect, we keep the axis ranges in the corner identical while highlighting the variability in the top inset panels. While we expect the distribution of these spectral features to be broadened due to variations in these quantities, the differences are still sufficient to tell the different toy models in Figure \ref{fig:extended_regulator_spec_cases2_corner} apart.}
    \label{fig:dust_met_effects}
\end{figure}

Figure \ref{fig:dust_met_effects} shows the effect of varying the stellar metallicity and the dust attenuation (assuming a Calzetti dust law; \citealt{2000ApJ...533..682C}) on the spectral indices we consider. This analysis is done for a galaxy with a fixed SFH of $1$M$_\odot/yr$ and other parameters corresponding to Table \ref{tab:spec_generation}. The effects of a distribution of values in either of these stellar population parameters would correspond to a broadening in the distribution of spectral indices by an amount proportional to the mean and width of the dust/metallicity distribution. For example, a distribution of $\sim 1$ dex in metallicity centered around solar metallicity would correspond to a spread of $\sim 0.04$ in $D_n(4000)$ and $\sim 0.3$ dex in log H$\alpha$ luminosity. While convolving the distributions in Figure \ref{fig:extended_regulator_spec_cases2} does make the different models harder to discriminate between, it is still distinct enough to be possible with a large enough sample size. This is additionally helped by the fact that the broadening of distributions in the spectral indices is not homogeneous, and in fact displays quite different signatures across the three indices for dust and metallicity - notably that dust attenuation does not affect the H$\delta_{\rm EW}$. We have not shown the effects of varying SPS models or the IMF, since that would correspond to an overall shift in the indices rather than a broadening of the distribution. In the rest-optical part of the SED that we study in this work, we are also not significantly affected by AGN, dust re-emission and other factors that manifest in the mid-to-far IR portions of the SED. These effects have also been studied in relation to SFR stochasticity in the literature \citep[][]{2019ApJ...873...74B, 2019ApJ...881...71E, 2020ApJ...892...87W}

One additional factor to note is that the broadening predicted by Figure \ref{fig:dust_met_effects} assumes that variations in dust and metallicity are independent of SFR stochasticity and star formation history. However, given that stochasticity determines the frequency of sharp bursts of star formation, it is likely that it will be correlated with the chemical enrichment of the galaxy. Although this is outside the scope of this work, cosmological simulations of galaxy evolution could shed light on the link between these parameters and help develop correlated priors for use with future observations.

\subsection{A prior for binned SFHs}
\label{subsec:binnedsfh}

\begin{figure}
    \centering
    \includegraphics[width=\textwidth]{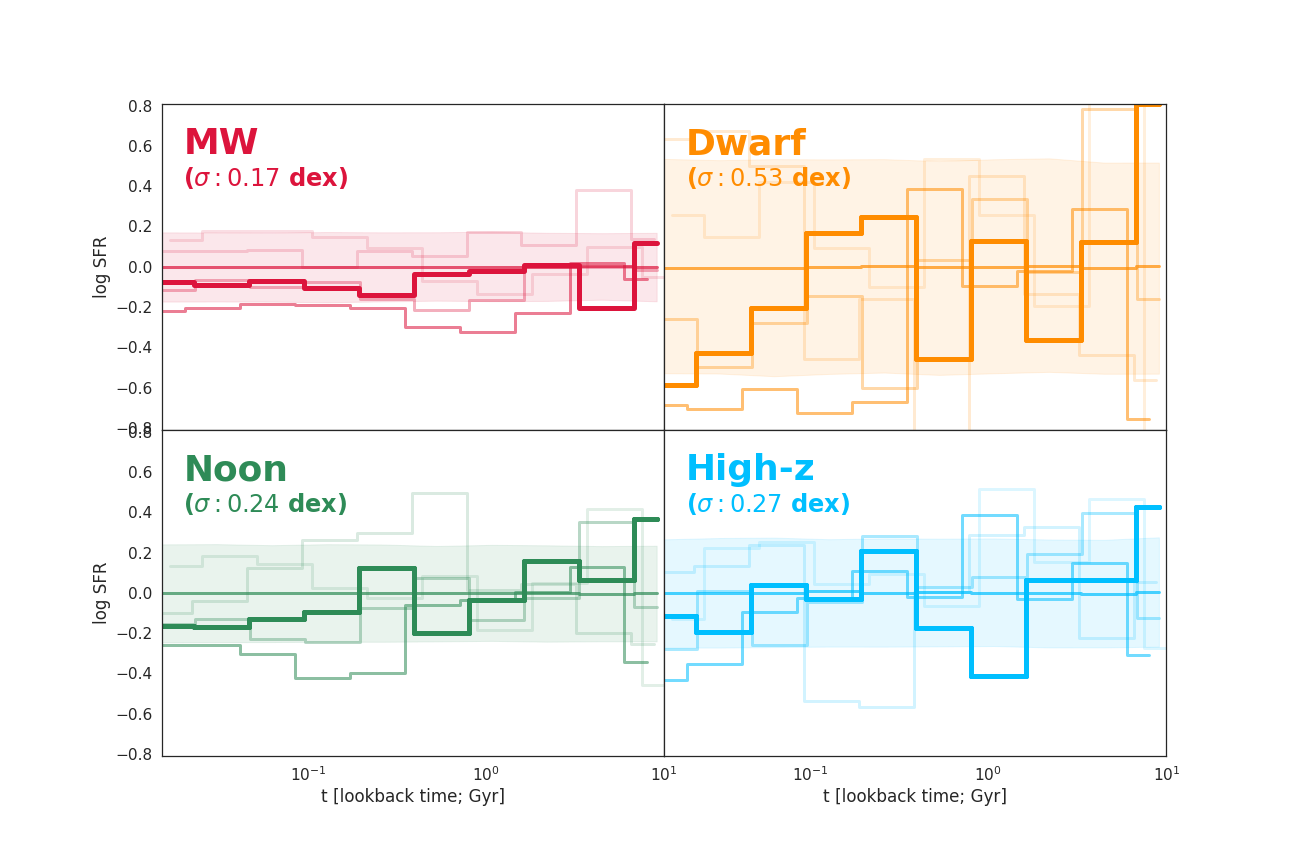}
    \caption{Log-binned SFHs sampled from Gaussian processes using the Extended Regulator model kernel, corresponding to the four stochasticity regimes described in Section \ref{sec:model}.}
    \label{fig:binnedsfhsamples}
\end{figure}

The Gaussian process implementation described in this paper can also be adapted as a prior for binned SFHs that are used by spectrophotometric fitting codes like Prospector \citep{leja2017deriving, ben_johnson_2021_4737461} or CMD based methods like Match \citep{2002MNRAS.332...91D, 2014ApJ...789..147W}. Prospector in particular uses either a Dirichlet prior or a continuity prior which parametrizes the log SFR ratios between bins using a Student's-t distribution. To incorporate the covariance models described in this work, it suffices to replace these priors with the covariance values $\mathcal{C}_{ExReg}(t_i, t_j)$ where $t_i, t_j$ correspond to the centers of the individual bins. In Appendix \ref{app:binnedsfh_validation}, we verify that this procedure yields SFHs identical to sampling from the high-resolution GP-SFH and degrading the resolution to match the logarithmically spaced time bins used in these codes. While we do not reproduce the distribution of observational metrics in Figures \ref{fig:extended_regulator_spec_cases1}-\ref{fig:extended_regulator_spec_cases2} using this formalism, we expect them to be very similar due to the lack of information encoded in galaxy SEDs about short-term variability at large lookback times. Figure \ref{fig:binnedsfhsamples} shows samples of binned SFHs corresponding to the four stochasticity regimes used in this work.

\subsection{How do we choose a `population of galaxies' to study observationally?}
\label{subsec:psbpops}

An underlying assumption in comparing distributions of spectral features or colors is that galaxies in the sample belong to the same underlying population and thus can be described using the same model for star formation stochasticity. Caution should therefore be exercised when creating a galaxy sample to minimize contamination by other populations. Methods like clustering in colors, physical properties, or SFHs can be used to select galaxies that are likely to belong to the same population. Additional methods like selecting specific galaxy sub-populations using self-organising maps \citep{2022AJ....163...71T, 2022MNRAS.513.1972H, 2022arXiv220606373D} or in the latent space of variational auto-encoders \citep{2020AJ....160...45P} can also be used for this purpose.

It is currently not well understood whether properties of SF stochasticity (i.e. properties like $\sigma_{gas}$ or $\tau_{eq}$ in the Extended Regulator model) correlate strongly with other physical properties like galaxy size, environment, morphology or dynamics. Since the variability in these quantities can span a wide range of timescales, that may or may not relate to the timescales on which SFR fluctuations are correlated. It would be interesting to study this further, since the stochasticity model can also influence the presence of certain galaxy populations.

\begin{figure}
    \centering
    \includegraphics[width=0.36\textwidth]{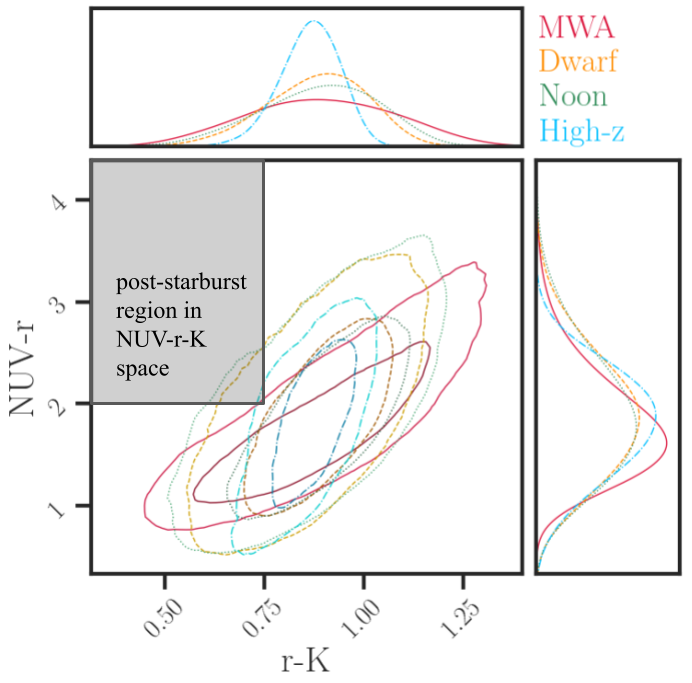}
    \includegraphics[width=0.6\textwidth]{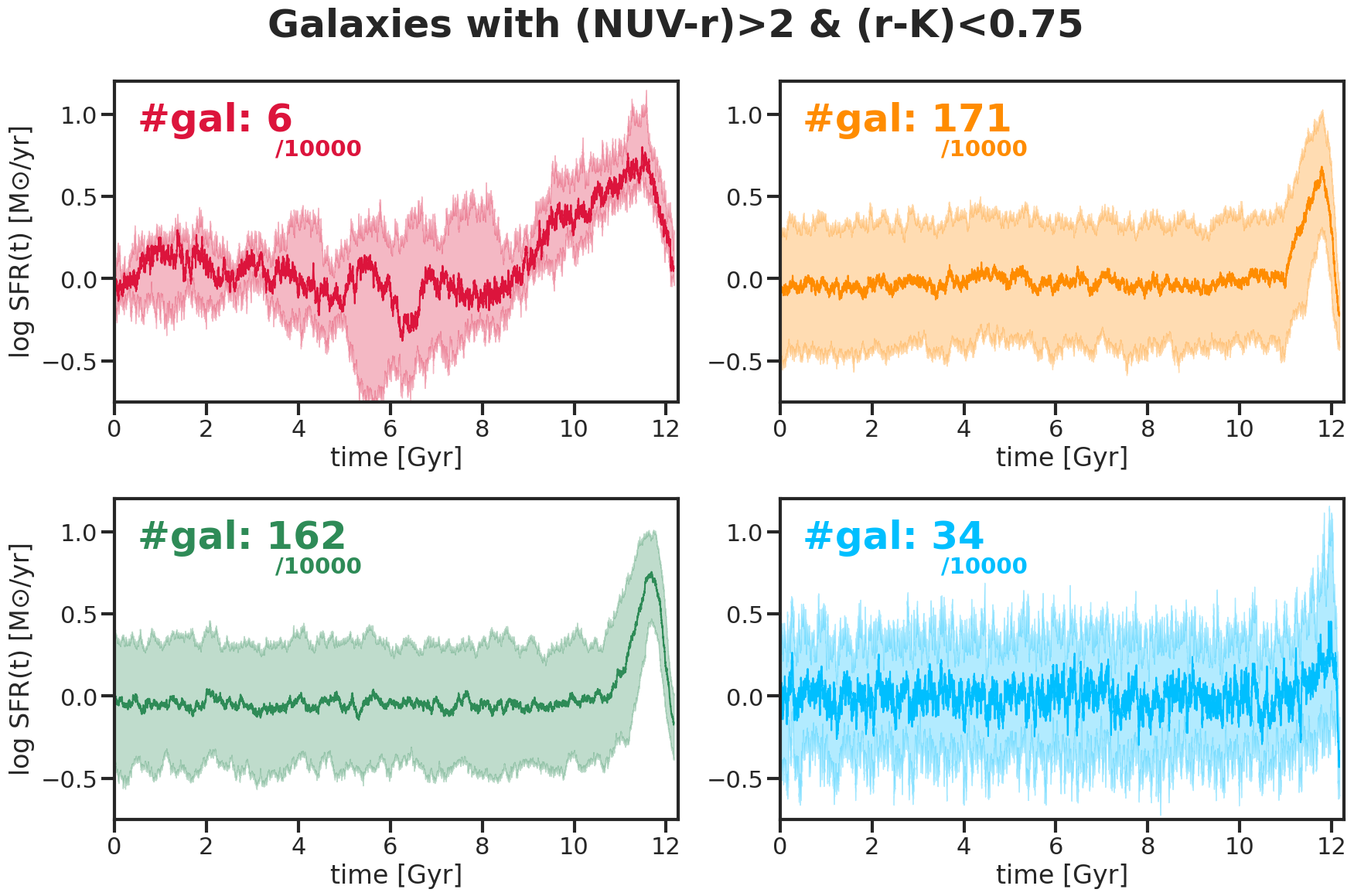}
    \caption{Post starburst galaxy SFHs are naturally produced by the different models, although the fraction of these galaxies with a recent burst on a $\sim 100$ Myr timescale in any given sample is higher for the dwarf and cosmic noon stochasticity regimes. This is picked up by a color-color selection shown in the bottom panels, similar to Figure \ref{fig:traditional_colors_1dex} where we select galaxies with $R-K_s<0.75$ and $NUV-R>2$. Median SFHs along with the $68\%$ variance for these galaxies is shown in the top panel, where the dwarf and noon samples show a clear post-starburst feature in the last $\sim 1$ Gyr. This is unlikely in the MWA case due to the much longer correlation timescales of SFHs, and in the High-z case because of the much shorter effective correlation timescales. Statistical samples of galaxies can therefore allow us to use the relative abundance of post-starburst galaxies to constrain the overall sample stochasticity.
    }
    \label{fig:colors_of_PSB_galaxies}
\end{figure}

For example, in Figure \ref{fig:traditional_colors_1dex}, we notice that the Dwarf and Cosmic noon populations show a slight excess of galaxies with $(r-K < 0.75)$ and $(NUV-r \sim 3)$. This region is highlighted in the left panel of Figure \ref{fig:colors_of_PSB_galaxies}.
Since spectral sensitivity falls off as a function of age at different rates depending on the wavelengths under consideration, an assumed model for SFH stochasticity can produce unique spectral signatures depending on a combination of broadband filters.
In the NUV-r-K color-color space for example, the r-K color has a mild linear dependence with age, except for a short period between $\sim 5-60$Myr during which the color sharply decreases. In complement to this, the NUV-r color is relatively flat until $\sim 20$Myr, after which it shows a linear dependence with age. Because of this combination of sensitivities, a portion of NUV-r-K color space ($\approx (NUV-r >2)$ \& $(r-K < 0.5)$) is uniquely sensitive to galaxies that recently experienced a sudden recent rise and fall in their star formation histories\footnote{A similar region exists in the UVJ diagram as well \citep{2020MNRAS.494..529W, 2020ApJ...899L..26S, 2022ApJ...929...94A}.}.

Since the probability of such an event is directly proportional to the amount of burstiness and effective timescales over which SFR is correlated, the four stochasticity models considered above make differing predictions for the probability that a galaxy can have such an event (and therefore on the number of galaxies in a given sample).
We examine this portion of the NUV-r-K color space better in Figure \ref{fig:colors_of_PSB_galaxies}, finding that the SFHs of galaxies with these colors tend to show a strong post-starburst feature in their SFHs, with the Dwarf and Cosmic noon ACFs resulting in a higher number of these galaxies compared to the MW or high-z populations.
Additional quantities like the fraction of PSB galaxies and the timescales of the recent burst could therefore be useful tracers of SF stochasticity in future studies.

\subsection{Varying the base SFHs}
\label{subsec:varsfhs}

The analysis in Sections \ref{subsec:scatter_fixed} and \ref{subsec:scatter_tcf20} assumes that the individual SFHs can be described as perturbations around a base SFH, which we assume to a constant SFH with $SFR=1$M$_\odot$/yr. While it is possible to relax this assumption in our current framework, caution needs to be exercised when the variability of the base SFHs is comparable to that of the timescales in the extended regulator kernel, since this will modify the ACF by adding power on the longer timescales when the autocorrelation time of the base SFH is of the same order as that for the extended regulator model $\tau_{\rm A, baseSFH} \simeq \tau_{\rm A,ExReg}$.
Appendix \ref{app:varbase} presents a detailed discussion of the effects on spectral features with an implementation where the base SFHs themselves vary across the sample and are drawn from a distribution. While this leads to a broadening of the distributions of individual spectral features, we find that it is still possible to differentiate between the models by comparing distributions of spectral features.

\subsection{The assumption of stationarity and ergodicity}
\label{subsec:nonstat}

For simplicity, the derivation in Section \ref{sec:model} assumes that the star formation histories of a population of galaxies are stationary and ergodic. The assumption of stationarity requires that the PSD or ACF of a galaxy SFH does not have an explicit time dependence. However, it is not necessary that SFHs in the real universe follow this, with either the stochasticity or timescales of the PSD model evolving with time. However, (i) for most science cases that discriminate between different models of stochasticity, the evolution is slow enough that this assumption is expected to hold \citep[see the discussion in \S 3.2 of ][]{2020ApJ...895...25W}, and (ii) if/when we decide to relax this assumption of stationarity, the kernel in our Gaussian process formalism can be updated to account for that. Indeed, non-stationary kernels are an open topic of research in Gaussian processes \citep{rhode2020non} and models for the time-evolution of the ACF are an important part of the future work enabled by this formalism.

\begin{figure}
    \centering
    \includegraphics[width=0.39\textwidth]{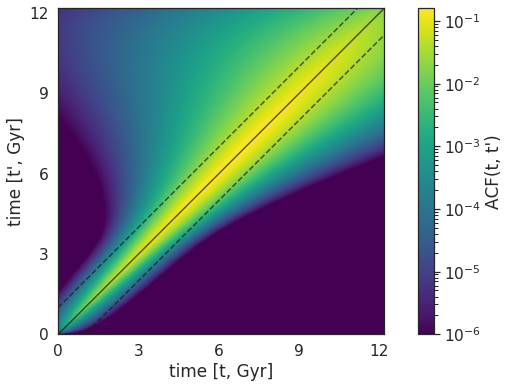}
    \includegraphics[width=0.59\textwidth]{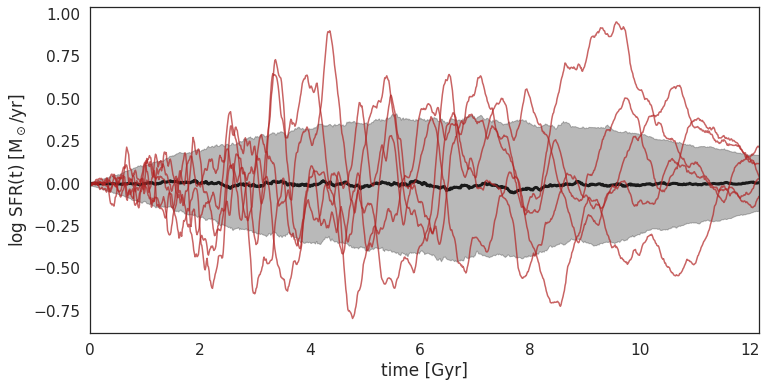}
    \caption{Implementing a non-stationary kernel in the GP-SFH formalism. While following most of the behaviour of the MW analogs, this kernel has a time-varying component to (i) $\sigma_{\rm gas}$, which rises to its maximum value at $z\sim 1$, after which it falls again, and (ii) $\tau_{\rm eq}$, which keeps rising with time, leading to smoother, less `bursty' SFHs at lower redshifts. (Left) The time-dependent `kernel' for the GP with the dotted lines showing a deviation of 1 Gyr to the future and past, and (right) the median behaviour of the galaxy sample (black line and shaded region) and individual realisations (red lines) of galaxy SFHs. }
    \label{fig:nonstationary_kernel}
\end{figure}

As observational data with future telescopes unlocks new timescales and large populations of galaxies across different cosmic epochs, our models can be updated to include variations in the extended regulator model parameters as a function of time. In this case, it is possible to relax the assumption of stationarity (i.e., Eqn. 2) and implement a more general Gaussian process for galaxy (log) SFRs, as shown in Figure \ref{fig:nonstationary_kernel}. A simple extension in this direction would be to make one or more of the parameters in the extended regulator model time-dependent. As an example of this case, we consider a simple time-dependence to both $\sigma_{\mathrm gas}$ and $\tau_{\mathrm eq}$ given by
\begin{equation}
    \sigma_{\mathrm gas} = 0.3 (-0.03t_{\mathrm univ}^2 + 0.4 t_{\mathrm univ})~~~\&~~~\tau_{\mathrm eq} = 0.01(t_{\mathrm univ}^2+1)
\end{equation}
This form is chosen for the $\sigma$ to allow for increased variability at higher redshifts, while the increasing $\tau_{\mathrm eq}$ is linked to the increasing dynamical times of galaxies with decreasing redshifts.

Ergodicity is a more subtle issue, and concerns the fact that we perform our analysis using populations of galaxies. For a dynamical system, ergodicity implies that the variability of a single galaxy's SFH over time is equivalent to the variability of an ensemble of SFHs observed at a given epoch, after accounting for selection functions and completeness. This is considered in Sec 3.2 of  \citet{2020ApJ...895...25W}, which also considers the need for ergodicity while working with galaxy PSDs. While it is unlikely that galaxy SFHs are fully ergodic, due to the changing conditions in which galaxies form stars at different epochs, the extent to which this is violated is expected to be minor and can be tested in the future using cosmological simulations.

\subsection{Challenges and caveats}
\label{subsec:caveats}

Techniques studying the stochasticity of star formation across timescales rely on a host of modeling assumptions, and these should be kept in mind while using inference to determine the SFH kernel parameters. This paper aims to provide a framework for forward modeling observables that can be combined with realistic distributions in other properties like the mean SFH model, stellar and gas-phase metallicity and dust attenuation, along with assumptions of stellar tracks, isochrones, IMF, dust law, binary fraction, TP-AGB contributions, and other factors that could systematically bias results in any SED-based analysis.

This is mitigated somewhat by the fact that the code developed here is modular and runs off FSPS. As a result, the SFH stochasticity model can be folded into existing SED fitting codes like dense\_basis \citep{2019ApJ...879..116I} and Prospector \citep{leja2017deriving, 2021ApJS..254...22J}, or combined with realistic models for physical properties like dust \citep{2022ApJ...926..122H, 2022ApJ...932...54N} in a population-level forward modeling setup.

Another concern is that the assumed form for the GP covariance does not take into account all the possible physical processes that induce variability in galaxy SFRs. While this is undoubtedly true, the processes modeled here are `effective' timescales as a result of a wide range of physical processes - $\tau_{\rm in}$ can include changes in the gas reservoir due to pristine inflows, outflows, mergers, and the re-integration of gas blown out in previous cycles; $\tau_{\rm eq}$ accounts for the fact that multiphase gas undergoes complex transformations across a galaxy as a result of stellar and AGN feedback, and $\tau_{\rm dyn}$ accounts for dynamical processes that create and destroy GMCs, which can be due to stellar feedback but also gas compression by spiral arms or compaction-induced starbursts. The next generation of large-volume, high-resolution simulations \citep{2019MNRAS.490.3196P, 2019MNRAS.490.3234N,2022arXiv220515325F} and studies that systematically vary feedback prescriptions \citep{2021ApJ...915...71V,2022arXiv220101300V} are needed to robustly understand these effective timescales, and further refine the GP kernel as needed.

A minor concern is the scalability of the GP-formalism in the limit of high-resolution SFHs and/or computationally intensive forward-modeling comparisons to large datasets of observations. However, fast GP-based methods that include automatic differentiation \citep[e.g. tinyGP;][]{dan_foreman_mackey_2022_6473662} can provide workarounds to these issues as they arise.

\section{Conclusion} \label{sec:conc}

Modeling the stochasticity of star formation over a range of timescales provides a way to connect observations of galaxy star formation histories to the underlying physical processes driving galaxy growth and quenching.

In this paper, we propose a fast, modular, Gaussian process-based (GP) formalism implementing the extended regulator model based on \citetalias{2020MNRAS.497..698T} as an autocovariance function (ACF) or kernel. This model provides a parametric form for SFR stochasticity as a combination of different physical processes, and is completely characterised by three effective timescales corresponding to stochastic gas inflows, equilibrium and dynamical processes influencing GMC creation and destruction.

Implementing a GP with this kernel allows us to make extremely fast draws of galaxy SFHs with a particular SFH autocovariance structure and to use them to forward-model galaxy spectra and dependent observables. Studying these observables as a function of the kernel parameters allows us to quantify differences as a function of the extended regulator model's timescales and thus differentiate between different regimes of stochasticity. This is illustrated by considering four toy-models for galaxy populations - Milky Way analogs, dwarf galaxies, massive galaxies at cosmic noon, and galaxies at high redshifts. We model the spectra of these galaxies using FSPS and study distributions of spectral features including H$\alpha$ and UV-based SFR indicators, H$\delta$ and Ca-H,K absorption line equivalent widths, the D$_n$(4000) spectral break and broadband galaxy colors, finding that these distributions are sensitive to the Extended Regulator model parameters, and their distributions and covariances can be used to discriminate between the models.

Since increasing the amount of stochasticity leads to greater stellar masses formed in intense bursts of star formation, the sSFR distribution, and therefore the flux in SFR tracers like H$\alpha$ and rest-$UV$, are sensitive to the overall level of stochasticity. Complementary to this, the H$\delta_{\rm EW}$ vs D$_n$(4000) space traces star formation over longer timescales, and is extremely sensitive to both the overall level of variability as well as the timescales on which SFR is correlated. We also find that the rest-frame broadband colors reveal populations of galaxies such as post-starbursts that are preferentially found in models that allow SFR correlations on the timescales that the colors are sensitive to, and can thus be used as additional constraints on the ExReg model parameters.

The GP-SFH formalism can also be easily incorporated into existing SED fitting codes to provide realistic priors for SFH covariance or infer them from future high S/N spectrophotometric observations using JWST, used to study the effective timescales in cosmological simulations, and further expanded to include factors like non-stationarity. Code to reproduce our results can be found at \faGithub\href{https://github.com/kartheikiyer/GP-SFH}{kartheikiyer/GP-SFH}.

\begin{acknowledgements}
JSS \& KGI jointly led the work and drafted the majority of the text. NC to ST contributed ideas and helped to draft and edit the text.
The authors are grateful to Harry Ferguson, Camilla Pacifici, and Vince Estrada-Carpenter for their comments and suggestions.
JSS is grateful to Rebecca Bleich for continuing to tolerate his excessive patronage of McDonald's.
KGI is grateful to the nonzero intersection in the Venn diagram of Noopur Gosavi, 5e, sci-fi and potatoes.
The Dunlap Institute is funded through an endowment established by the David Dunlap family and the University of Toronto.
EG acknowledges support from NASA ADAP grant 80NSSC22K0487.

\textbf{Code acknowledgements:}
Astropy \citep{astropy:2013, price2018astropy},
matplotlib \citep{caswell2019matplotlib},
scipy \citep{virtanen2020scipy},
numpy \citep{walt2011numpy},
corner \citep{Foreman-Mackey2016},
FSPS \citep{ben_johnson_2021_4737461},
dense basis \citep{2019ApJ...879..116I} and
hickle \citep{Price2018}.

\end{acknowledgements}

\bibliography{GPSFH}{}

\begin{thebibliography}{}
\expandafter\ifx\csname natexlab\endcsname\relax\def\natexlab#1{#1}\fi
\providecommand{\url}[1]{\href{#1}{#1}}
\providecommand{\dodoi}[1]{doi:~\href{http://doi.org/#1}{\nolinkurl{#1}}}
\providecommand{\doeprint}[1]{\href{http://ascl.net/#1}{\nolinkurl{http://ascl.net/#1}}}
\providecommand{\doarXiv}[1]{\href{https://arxiv.org/abs/#1}{\nolinkurl{https://arxiv.org/abs/#1}}}

\bibitem[{{Abazajian} {et~al.}(2009){Abazajian}, {Adelman-McCarthy},
  {Ag{\"u}eros}, {Allam}, {Allende Prieto}, {An}, {Anderson}, {Anderson},
  {Annis}, {Bahcall}, {Bailer-Jones}, {Barentine}, {Bassett}, {Becker},
  {Beers}, {Bell}, {Belokurov}, {Berlind}, {Berman}, {Bernardi}, {Bickerton},
  {Bizyaev}, {Blakeslee}, {Blanton}, {Bochanski}, {Boroski}, {Brewington},
  {Brinchmann}, {Brinkmann}, {Brunner}, {Budav{\'a}ri}, {Carey}, {Carliles},
  {Carr}, {Castander}, {Cinabro}, {Connolly}, {Csabai}, {Cunha}, {Czarapata},
  {Davenport}, {de Haas}, {Dilday}, {Doi}, {Eisenstein}, {Evans}, {Evans},
  {Fan}, {Friedman}, {Frieman}, {Fukugita}, {G{\"a}nsicke}, {Gates},
  {Gillespie}, {Gilmore}, {Gonzalez}, {Gonzalez}, {Grebel}, {Gunn},
  {Gy{\"o}ry}, {Hall}, {Harding}, {Harris}, {Harvanek}, {Hawley}, {Hayes},
  {Heckman}, {Hendry}, {Hennessy}, {Hindsley}, {Hoblitt}, {Hogan}, {Hogg},
  {Holtzman}, {Hyde}, {Ichikawa}, {Ichikawa}, {Im}, {Ivezi{\'c}}, {Jester},
  {Jiang}, {Johnson}, {Jorgensen}, {Juri{\'c}}, {Kent}, {Kessler}, {Kleinman},
  {Knapp}, {Konishi}, {Kron}, {Krzesinski}, {Kuropatkin}, {Lampeitl},
  {Lebedeva}, {Lee}, {Lee}, {French Leger}, {L{\'e}pine}, {Li}, {Lima}, {Lin},
  {Long}, {Loomis}, {Loveday}, {Lupton}, {Magnier}, {Malanushenko},
  {Malanushenko}, {Mandelbaum}, {Margon}, {Marriner}, {Mart{\'\i}nez-Delgado},
  {Matsubara}, {McGehee}, {McKay}, {Meiksin}, {Morrison}, {Mullally}, {Munn},
  {Murphy}, {Nash}, {Nebot}, {Neilsen}, {Newberg}, {Newman}, {Nichol},
  {Nicinski}, {Nieto-Santisteban}, {Nitta}, {Okamura}, {Oravetz}, {Ostriker},
  {Owen}, {Padmanabhan}, {Pan}, {Park}, {Pauls}, {Peoples}, {Percival}, {Pier},
  {Pope}, {Pourbaix}, {Price}, {Purger}, {Quinn}, {Raddick}, {Re Fiorentin},
  {Richards}, {Richmond}, {Riess}, {Rix}, {Rockosi}, {Sako}, {Schlegel},
  {Schneider}, {Scholz}, {Schreiber}, {Schwope}, {Seljak}, {Sesar}, {Sheldon},
  {Shimasaku}, {Sibley}, {Simmons}, {Sivarani}, {Allyn Smith}, {Smith},
  {Smol{\v{c}}i{\'c}}, {Snedden}, {Stebbins}, {Steinmetz}, {Stoughton},
  {Strauss}, {SubbaRao}, {Suto}, {Szalay}, {Szapudi}, {Szkody}, {Tanaka},
  {Tegmark}, {Teodoro}, {Thakar}, {Tremonti}, {Tucker}, {Uomoto}, {Vanden
  Berk}, {Vandenberg}, {Vidrih}, {Vogeley}, {Voges}, {Vogt}, {Wadadekar},
  {Watters}, {Weinberg}, {West}, {White}, {Wilhite}, {Wonders}, {Yanny},
  {Yocum}, {York}, {Zehavi}, {Zibetti}, \& {Zucker}}]{2009ApJS..182..543A}
{Abazajian}, K.~N., {Adelman-McCarthy}, J.~K., {Ag{\"u}eros}, M.~A., {et~al.}
  2009, \apjs, 182, 543, \dodoi{10.1088/0067-0049/182/2/543}

\bibitem[{{Abramson} {et~al.}(2016){Abramson}, {Gladders}, {Dressler},
  {Oemler}, {Poggianti}, \& {Vulcani}}]{2016ApJ...832....7A}
{Abramson}, L.~E., {Gladders}, M.~D., {Dressler}, A., {et~al.} 2016, \apj, 832,
  7, \dodoi{10.3847/0004-637X/832/1/7}

\bibitem[{{Akins} {et~al.}(2022){Akins}, {Narayanan}, {Whitaker}, {Dav{\'e}},
  {Lower}, {Bezanson}, {Feldmann}, \& {Kriek}}]{2022ApJ...929...94A}
{Akins}, H.~B., {Narayanan}, D., {Whitaker}, K.~E., {et~al.} 2022, \apj, 929,
  94, \dodoi{10.3847/1538-4357/ac5d3a}

\bibitem[{{Alarcon} {et~al.}(2022){Alarcon}, {Hearin}, {Becker}, \&
  {Chaves-Montero}}]{2022arXiv220504273A}
{Alarcon}, A., {Hearin}, A.~P., {Becker}, M.~R., \& {Chaves-Montero}, J. 2022,
  arXiv e-prints, arXiv:2205.04273.
\newblock \doarXiv{2205.04273}

\bibitem[{{Arnouts} {et~al.}(2013){Arnouts}, {Le Floc'h}, {Chevallard},
  {Johnson}, {Ilbert}, {Treyer}, {Aussel}, {Capak}, {Sanders}, {Scoville},
  {McCracken}, {Milliard}, {Pozzetti}, \& {Salvato}}]{2013A&A...558A..67A}
{Arnouts}, S., {Le Floc'h}, E., {Chevallard}, J., {et~al.} 2013, \aap, 558,
  A67, \dodoi{10.1051/0004-6361/201321768}

\bibitem[{{Astropy Collaboration} {et~al.}(2013){Astropy Collaboration},
  {Robitaille}, {Tollerud}, {Greenfield}, {Droettboom}, {Bray}, {Aldcroft},
  {Davis}, {Ginsburg}, {Price-Whelan}, {Kerzendorf}, {Conley}, {Crighton},
  {Barbary}, {Muna}, {Ferguson}, {Grollier}, {Parikh}, {Nair}, {Unther},
  {Deil}, {Woillez}, {Conseil}, {Kramer}, {Turner}, {Singer}, {Fox}, {Weaver},
  {Zabalza}, {Edwards}, {Azalee Bostroem}, {Burke}, {Casey}, {Crawford},
  {Dencheva}, {Ely}, {Jenness}, {Labrie}, {Lim}, {Pierfederici}, {Pontzen},
  {Ptak}, {Refsdal}, {Servillat}, \& {Streicher}}]{astropy:2013}
{Astropy Collaboration}, {Robitaille}, T.~P., {Tollerud}, E.~J., {et~al.} 2013,
  \aap, 558, A33, \dodoi{10.1051/0004-6361/201322068}

\bibitem[{{Balogh} {et~al.}(1999){Balogh}, {Morris}, {Yee}, {Carlberg}, \&
  {Ellingson}}]{1999ApJ...527...54B}
{Balogh}, M.~L., {Morris}, S.~L., {Yee}, H.~K.~C., {Carlberg}, R.~G., \&
  {Ellingson}, E. 1999, \apj, 527, 54, \dodoi{10.1086/308056}

\bibitem[{{Broussard} {et~al.}(2019{\natexlab{a}}){Broussard}, {Gawiser},
  {Iyer}, {Kurczynski}, {Somerville}, {Dav{\'e}}, {Finkelstein}, {Jung}, \&
  {Pacifici}}]{broussard2019stochastic}
{Broussard}, A., {Gawiser}, E., {Iyer}, K., {et~al.} 2019{\natexlab{a}}, \apj,
  873, 74, \dodoi{10.3847/1538-4357/ab04ad}

\bibitem[{{Broussard} {et~al.}(2019{\natexlab{b}}){Broussard}, {Gawiser},
  {Iyer}, {Kurczynski}, {Somerville}, {Dav{\'e}}, {Finkelstein}, {Jung}, \&
  {Pacifici}}]{2019ApJ...873...74B}
---. 2019{\natexlab{b}}, \apj, 873, 74, \dodoi{10.3847/1538-4357/ab04ad}

\bibitem[{{Calzetti} {et~al.}(2000){Calzetti}, {Armus}, {Bohlin}, {Kinney},
  {Koornneef}, \& {Storchi-Bergmann}}]{2000ApJ...533..682C}
{Calzetti}, D., {Armus}, L., {Bohlin}, R.~C., {et~al.} 2000, \apj, 533, 682,
  \dodoi{10.1086/308692}

\bibitem[{{Camps-Fari{\~n}a} {et~al.}(2022){Camps-Fari{\~n}a}, {S{\'a}nchez},
  {Mej{\'\i}a-Narv{\'a}ez}, {Lacerda}, {Carigi}, {Bruzual}, {Alvarez-Hurtado},
  {Drory}, {Lane}, {Boardman}, \& {Blanc}}]{2022ApJ...933...44C}
{Camps-Fari{\~n}a}, A., {S{\'a}nchez}, S.~F., {Mej{\'\i}a-Narv{\'a}ez}, A.,
  {et~al.} 2022, \apj, 933, 44, \dodoi{10.3847/1538-4357/ac6cea}

\bibitem[{{Caplar} \& {Tacchella}(2019)}]{2019MNRAS.487.3845C}
{Caplar}, N., \& {Tacchella}, S. 2019, \mnras, 487, 3845,
  \dodoi{10.1093/mnras/stz1449}

\bibitem[{Carnall {et~al.}(2018)Carnall, McLure, Dunlop, \&
  Dav{\'e}}]{carnall2018inferring}
Carnall, A., McLure, R., Dunlop, J., \& Dav{\'e}, R. 2018, Monthly Notices of
  the Royal Astronomical Society, 480, 4379

\bibitem[{Caswell {et~al.}(2019)Caswell, Droettboom, Hunter, Firing, Lee,
  Klymak, Stansby, de~Andrade, Nielsen, Varoquaux,
  {et~al.}}]{caswell2019matplotlib}
Caswell, T., Droettboom, M., Hunter, J., {et~al.} 2019, matplotlib/matplotlib
  v3. 1.0,  May

\bibitem[{Conroy \& Gunn(2010)}]{conroy2010propagation}
Conroy, C., \& Gunn, J.~E. 2010, The Astrophysical Journal, 712, 833

\bibitem[{Conroy {et~al.}(2009)Conroy, Gunn, \& White}]{conroy2009propagation}
Conroy, C., Gunn, J.~E., \& White, M. 2009, The Astrophysical Journal, 699, 486

\bibitem[{Daddi {et~al.}(2007)Daddi, Dickinson, Morrison, Chary, Cimatti,
  Elbaz, Frayer, Renzini, Pope, Alexander, {et~al.}}]{daddi2007multiwavelength}
Daddi, E., Dickinson, M., Morrison, G., {et~al.} 2007, The Astrophysical
  Journal, 670, 156

\bibitem[{{Dav{\'e}} {et~al.}(2012){Dav{\'e}}, {Finlator}, \&
  {Oppenheimer}}]{2012MNRAS.421...98D}
{Dav{\'e}}, R., {Finlator}, K., \& {Oppenheimer}, B.~D. 2012, \mnras, 421, 98,
  \dodoi{10.1111/j.1365-2966.2011.20148.x}

\bibitem[{{Davidzon} {et~al.}(2022){Davidzon}, {Jegatheesan}, {Ilbert}, {de la
  Torre}, {Leslie}, {Laigle}, {Hemmati}, {Masters}, {Blanquez-Sese},
  {Kauffmann}, {Magdis}, {Ma{\l}ek}, {McCracken}, {Mobasher}, {Moneti},
  {Sanders}, {Shuntov}, {Toft}, \& {Weaver}}]{2022arXiv220606373D}
{Davidzon}, I., {Jegatheesan}, K., {Ilbert}, O., {et~al.} 2022, arXiv e-prints,
  arXiv:2206.06373.
\newblock \doarXiv{2206.06373}

\bibitem[{{Dolphin}(2002)}]{2002MNRAS.332...91D}
{Dolphin}, A.~E. 2002, \mnras, 332, 91,
  \dodoi{10.1046/j.1365-8711.2002.05271.x}

\bibitem[{{Driver} {et~al.}(2009){Driver}, {Norberg}, {Baldry}, {Bamford},
  {Hopkins}, {Liske}, {Loveday}, {Peacock}, {Hill}, {Kelvin}, {Robotham},
  {Cross}, {Parkinson}, {Prescott}, {Conselice}, {Dunne}, {Brough}, {Jones},
  {Sharp}, {van Kampen}, {Oliver}, {Roseboom}, {Bland-Hawthorn}, {Croom},
  {Ellis}, {Cameron}, {Cole}, {Frenk}, {Couch}, {Graham}, {Proctor}, {De
  Propris}, {Doyle}, {Edmondson}, {Nichol}, {Thomas}, {Eales}, {Jarvis},
  {Kuijken}, {Lahav}, {Madore}, {Seibert}, {Meyer}, {Staveley-Smith},
  {Phillipps}, {Popescu}, {Sansom}, {Sutherland}, {Tuffs}, \&
  {Warren}}]{2009A&G....50e..12D}
{Driver}, S.~P., {Norberg}, P., {Baldry}, I.~K., {et~al.} 2009, Astronomy and
  Geophysics, 50, 5.12, \dodoi{10.1111/j.1468-4004.2009.50512.x}

\bibitem[{Dye(2008)}]{dye}
Dye, S. 2008, Monthly Notices of the Royal Astronomical Society, 389, 1293

\bibitem[{Elbaz {et~al.}(2007)Elbaz, Daddi, Le~Borgne, Dickinson, Alexander,
  Chary, Starck, Brandt, Kitzbichler, MacDonald, {et~al.}}]{elbaz2007reversal}
Elbaz, D., Daddi, E., Le~Borgne, D., {et~al.} 2007, Astronomy \& Astrophysics,
  468, 33

\bibitem[{{Emami} {et~al.}(2019{\natexlab{a}}){Emami}, {Siana}, {Weisz},
  {Johnson}, {Ma}, \& {El-Badry}}]{emami2018closer}
{Emami}, N., {Siana}, B., {Weisz}, D.~R., {et~al.} 2019{\natexlab{a}}, \apj,
  881, 71, \dodoi{10.3847/1538-4357/ab211a}

\bibitem[{{Emami} {et~al.}(2019{\natexlab{b}}){Emami}, {Siana}, {Weisz},
  {Johnson}, {Ma}, \& {El-Badry}}]{2019ApJ...881...71E}
---. 2019{\natexlab{b}}, \apj, 881, 71, \dodoi{10.3847/1538-4357/ab211a}

\bibitem[{{Faisst} {et~al.}(2019){Faisst}, {Capak}, {Emami}, {Tacchella}, \&
  {Larson}}]{2019ApJ...884..133F}
{Faisst}, A.~L., {Capak}, P.~L., {Emami}, N., {Tacchella}, S., \& {Larson},
  K.~L. 2019, \apj, 884, 133, \dodoi{10.3847/1538-4357/ab425b}

\bibitem[{{Feldmann} {et~al.}(2022){Feldmann}, {Quataert},
  {Faucher-Gigu{\`e}re}, {Hopkins}, {{\c{C}}atmabacak}, {Kere{\v{s}}},
  {Bassini}, {Bernardini}, {Bullock}, {Cenci}, {Gensior}, {Liang}, {Moreno}, \&
  {Wetzel}}]{2022arXiv220515325F}
{Feldmann}, R., {Quataert}, E., {Faucher-Gigu{\`e}re}, C.-A., {et~al.} 2022,
  arXiv e-prints, arXiv:2205.15325.
\newblock \doarXiv{2205.15325}

\bibitem[{{Finkelstein} {et~al.}(2022){Finkelstein}, {Bagley}, {Arrabal Haro},
  {Dickinson}, {Ferguson}, {Kartaltepe}, {Papovich}, {Burgarella}, {Kocevski},
  {Huertas-Company}, {Iyer}, {Larson}, {P{\'e}rez-Gonz{\'a}lez}, {Rose},
  {Tacchella}, {Wilkins}, {Chworowsky}, {Medrano}, {Morales}, {Somerville},
  {Yung}, {Fontana}, {Giavalisco}, {Grazian}, {Grogin}, {Kewley}, {Koekemoer},
  {Kirkpatrick}, {Kurczynski}, {Lotz}, {Pentericci}, {Pirzkal}, {Ravindranath},
  {Ryan}, {Trump}, {Yang}, {Almaini}, {Amor{\'\i}n}, {Annunziatella},
  {Backhaus}, {Barro}, {Behroozi}, {Bell}, {Bhatawdekar}, {Bisigello}, {Bromm},
  {Buat}, {Buitrago}, {Calabr{\'o}}, {Casey}, {Castellano}, {Ch{\'a}vez Ortiz},
  {Ciesla}, {Cleri}, {Cohen}, {Cole}, {Cooke}, {Cooper}, {Cooray}, {Costantin},
  {Cox}, {Croton}, {Daddi}, {Dav{\'e}}, {de la Vega}, {Dekel}, {Elbaz},
  {Estrada-Carpenter}, {Faber}, {Fern{\'a}ndez}, {Finkelstein}, {Freundlich},
  {Fujimoto}, {Garc{\'\i}a-Argum{\'a}nez}, {Gardner}, {Gawiser},
  {G{\'o}mez-Guijarro}, {Guo}, {Hamilton}, {Hathi}, {Holwerda}, {Hirschmann},
  {Hutchison}, {Jha}, {Jogee}, {Juneau}, {Jung}, {Kassin}, {Le Bail}, {Leung},
  {Lucas}, {Magnelli}, {Mantha}, {Matharu}, {McGrath}, {McIntosh}, {Merlin},
  {Mobasher}, {Newman}, {Nicholls}, {Pandya}, {Rafelski}, {Ronayne}, {Santini},
  {Seill{\'e}}, {Shah}, {Shen}, {Simons}, {Snyder}, {Stanway}, {Straughn},
  {Teplitz}, {Vanderhoof}, {Vega-Ferrero}, {Wang}, {Weiner}, {Willmer},
  {Wuyts}, \& {Zavala}}]{2022arXiv220712474F}
{Finkelstein}, S.~L., {Bagley}, M.~B., {Arrabal Haro}, P., {et~al.} 2022, arXiv
  e-prints, arXiv:2207.12474.
\newblock \doarXiv{2207.12474}

\bibitem[{{Flores Vel{\'a}zquez} {et~al.}(2021){Flores Vel{\'a}zquez},
  {Gurvich}, {Faucher-Gigu{\`e}re}, {Bullock}, {Starkenburg}, {Moreno},
  {Lazar}, {Mercado}, {Stern}, {Sparre}, {Hayward}, {Wetzel}, \&
  {El-Badry}}]{2021MNRAS.501.4812F}
{Flores Vel{\'a}zquez}, J.~A., {Gurvich}, A.~B., {Faucher-Gigu{\`e}re}, C.-A.,
  {et~al.} 2021, \mnras, 501, 4812, \dodoi{10.1093/mnras/staa3893}

\bibitem[{{Forbes} {et~al.}(2014){Forbes}, {Krumholz}, {Burkert}, \&
  {Dekel}}]{2014MNRAS.443..168F}
{Forbes}, J.~C., {Krumholz}, M.~R., {Burkert}, A., \& {Dekel}, A. 2014, \mnras,
  443, 168, \dodoi{10.1093/mnras/stu1142}

\bibitem[{{Forbes} {et~al.}(2019){Forbes}, {Krumholz}, \&
  {Speagle}}]{2019MNRAS.487.3581F}
{Forbes}, J.~C., {Krumholz}, M.~R., \& {Speagle}, J.~S. 2019, \mnras, 487,
  3581, \dodoi{10.1093/mnras/stz1473}

\bibitem[{Foreman-Mackey(2016)}]{Foreman-Mackey2016}
Foreman-Mackey, D. 2016, Journal of Open Source Software, 1, 24,
  \dodoi{10.21105/joss.00024}

\bibitem[{Foreman-Mackey {et~al.}(2022)Foreman-Mackey, Yadav, Tronsgaard,
  Schmerler, \& theorashid}]{dan_foreman_mackey_2022_6473662}
Foreman-Mackey, D., Yadav, S., Tronsgaard, R., Schmerler, S., \& theorashid.
  2022, dfm/tinygp: tinygp v0.2.2, v0.2.2,  Zenodo,
  \dodoi{10.5281/zenodo.6473662}

\bibitem[{{Gon{\c{c}}alves} {et~al.}(2012){Gon{\c{c}}alves}, {Martin},
  {Men{\'e}ndez-Delmestre}, {Wyder}, \& {Koekemoer}}]{2012ApJ...759...67G}
{Gon{\c{c}}alves}, T.~S., {Martin}, D.~C., {Men{\'e}ndez-Delmestre}, K.,
  {Wyder}, T.~K., \& {Koekemoer}, A. 2012, \apj, 759, 67,
  \dodoi{10.1088/0004-637X/759/1/67}

\bibitem[{{Guo} {et~al.}(2014){Guo}, {Koo}, {Primack}, \& {CANDELS
  Collaboration}}]{2014AAS...22314511G}
{Guo}, Y., {Koo}, D.~C., {Primack}, J.~R., \& {CANDELS Collaboration}. 2014, in
  American Astronomical Society Meeting Abstracts, Vol. 223, American
  Astronomical Society Meeting Abstracts \#223, 145.11

\bibitem[{{Guo} {et~al.}(2016){Guo}, {Rafelski}, {Faber}, {Koo}, {Krumholz},
  {Trump}, {Willner}, {Amor{\'\i}n}, {Barro}, {Bell}, {Gardner}, {Gawiser},
  {Hathi}, {Koekemoer}, {Pacifici}, {P{\'e}rez-Gonz{\'a}lez}, {Ravindranath},
  {Reddy}, {Teplitz}, \& {Yesuf}}]{guo2016bursty}
{Guo}, Y., {Rafelski}, M., {Faber}, S.~M., {et~al.} 2016, \apj, 833, 37,
  \dodoi{10.3847/1538-4357/833/1/37}

\bibitem[{Guszejnov {et~al.}(2018)Guszejnov, Hopkins, \&
  Grudi{\'c}}]{guszejnov2018universal}
Guszejnov, D., Hopkins, P.~F., \& Grudi{\'c}, M.~Y. 2018, Monthly Notices of
  the Royal Astronomical Society, 477, 5139

\bibitem[{{Hahn} {et~al.}(2022){Hahn}, {Starkenburg}, {Angl{\'e}s-Alc{\'a}zar},
  {Choi}, {Dav{\'e}}, {Dickey}, {Iyer}, {Maller}, {Somerville}, {Tinker}, \&
  {Yung}}]{2022ApJ...926..122H}
{Hahn}, C., {Starkenburg}, T.~K., {Angl{\'e}s-Alc{\'a}zar}, D., {et~al.} 2022,
  \apj, 926, 122, \dodoi{10.3847/1538-4357/ac4253}

\bibitem[{Heavens {et~al.}(2000)Heavens, Jimenez, \& Lahav}]{moped}
Heavens, A.~F., Jimenez, R., \& Lahav, O. 2000, Monthly Notices of the Royal
  Astronomical Society, 317, 965

\bibitem[{{Hirashita} \& {Kamaya}(2000)}]{2000AJ....120..728H}
{Hirashita}, H., \& {Kamaya}, H. 2000, \aj, 120, 728, \dodoi{10.1086/301497}

\bibitem[{{Holwerda} {et~al.}(2022){Holwerda}, {Smith}, {Porter}, {Henry},
  {Porter-Temple}, {Cook}, {Pimbblet}, {Hopkins}, {Bilicki}, {Turner},
  {Acquaviva}, {Wang}, {Wright}, {Kelvin}, \& {Grootes}}]{2022MNRAS.513.1972H}
{Holwerda}, B.~W., {Smith}, D., {Porter}, L., {et~al.} 2022, \mnras, 513, 1972,
  \dodoi{10.1093/mnras/stac889}

\bibitem[{{Huertas-Company} {et~al.}(2020){Huertas-Company}, {Guo}, {Ginzburg},
  {Lee}, {Mandelker}, {Metter}, {Primack}, {Dekel}, {Ceverino}, {Faber}, {Koo},
  {Koekemoer}, {Snyder}, {Giavalisco}, \& {Zhang}}]{2020MNRAS.499..814H}
{Huertas-Company}, M., {Guo}, Y., {Ginzburg}, O., {et~al.} 2020, \mnras, 499,
  814, \dodoi{10.1093/mnras/staa2777}

\bibitem[{{Iyer} \& {Gawiser}(2017)}]{2017ApJ...838..127I}
{Iyer}, K., \& {Gawiser}, E. 2017, \apj, 838, 127,
  \dodoi{10.3847/1538-4357/aa63f0}

\bibitem[{{Iyer} {et~al.}(2018){Iyer}, {Gawiser}, {Dav{\'e}}, {Davis},
  {Finkelstein}, {Kodra}, {Koekemoer}, {Kurczynski}, {Newman}, {Pacifici}, \&
  {Somerville}}]{2018ApJ...866..120I}
{Iyer}, K., {Gawiser}, E., {Dav{\'e}}, R., {et~al.} 2018, \apj, 866, 120,
  \dodoi{10.3847/1538-4357/aae0fa}

\bibitem[{{Iyer} {et~al.}(2019){Iyer}, {Gawiser}, {Faber}, {Ferguson},
  {Kartaltepe}, {Koekemoer}, {Pacifici}, \& {Somerville}}]{2019ApJ...879..116I}
{Iyer}, K.~G., {Gawiser}, E., {Faber}, S.~M., {et~al.} 2019, \apj, 879, 116,
  \dodoi{10.3847/1538-4357/ab2052}

\bibitem[{{Iyer} {et~al.}(2020){Iyer}, {Tacchella}, {Genel}, {Hayward},
  {Hernquist}, {Brooks}, {Caplar}, {Dav{\'e}}, {Diemer}, {Forbes}, {Gawiser},
  {Somerville}, \& {Starkenburg}}]{2020MNRAS.498..430I}
{Iyer}, K.~G., {Tacchella}, S., {Genel}, S., {et~al.} 2020, \mnras, 498, 430,
  \dodoi{10.1093/mnras/staa2150}

\bibitem[{Johnson {et~al.}(2021)Johnson, Foreman-Mackey, Sick, Leja, Byler,
  Walmsley, Tollerud, Leung, \& Scott}]{ben_johnson_2021_4737461}
Johnson, B., Foreman-Mackey, D., Sick, J., {et~al.} 2021, dfm/python-fsps:
  python-fsps v0.4.1rc1, v0.4.1rc1,  Zenodo, \dodoi{10.5281/zenodo.4737461}

\bibitem[{{Johnson} {et~al.}(2021){Johnson}, {Leja}, {Conroy}, \&
  {Speagle}}]{2021ApJS..254...22J}
{Johnson}, B.~D., {Leja}, J., {Conroy}, C., \& {Speagle}, J.~S. 2021, \apjs,
  254, 22, \dodoi{10.3847/1538-4365/abef67}

\bibitem[{{Kauffmann} {et~al.}(2006){Kauffmann}, {Heckman}, {De Lucia},
  {Brinchmann}, {Charlot}, {Tremonti}, {White}, \&
  {Brinkmann}}]{kauffmann2006gas}
{Kauffmann}, G., {Heckman}, T.~M., {De Lucia}, G., {et~al.} 2006, \mnras, 367,
  1394, \dodoi{10.1111/j.1365-2966.2006.10061.x}

\bibitem[{Kauffmann {et~al.}(2003)Kauffmann, Heckman, White, Charlot, Tremonti,
  Brinchmann, Bruzual, Peng, Seibert, Bernardi,
  {et~al.}}]{kauffmann2003stellar}
Kauffmann, G., Heckman, T.~M., White, S.~D., {et~al.} 2003, Monthly Notices of
  the Royal Astronomical Society, 341, 33

\bibitem[{{Kelson}(2014)}]{2014arXiv1406.5191K}
{Kelson}, D.~D. 2014, arXiv e-prints, arXiv:1406.5191.
\newblock \doarXiv{1406.5191}

\bibitem[{{Kelson} {et~al.}(2020){Kelson}, {Abramson}, {Benson}, {Patel},
  {Shectman}, {Dressler}, {McCarthy}, {Mulchaey}, \&
  {Williams}}]{kelson2020gravity}
{Kelson}, D.~D., {Abramson}, L.~E., {Benson}, A.~J., {et~al.} 2020, \mnras,
  494, 2628, \dodoi{10.1093/mnras/staa100}

\bibitem[{Khinchin(1938)}]{khinchin1938theory}
Khinchin, A.~Y. 1938, Uspekhi matematicheskikh nauk, 42

\bibitem[{{Krumholz} \& {Kruijssen}(2015)}]{2015MNRAS.453..739K}
{Krumholz}, M.~R., \& {Kruijssen}, J.~M.~D. 2015, \mnras, 453, 739,
  \dodoi{10.1093/mnras/stv1670}

\bibitem[{{Kurczynski} {et~al.}(2016){Kurczynski}, {Gawiser}, {Acquaviva},
  {Bell}, {Dekel}, {de Mello}, {Ferguson}, {Gardner}, {Grogin}, {Guo},
  {Hopkins}, {Koekemoer}, {Koo}, {Lee}, {Mobasher}, {Primack}, {Rafelski},
  {Soto}, \& {Teplitz}}]{2016ApJ...820L...1K}
{Kurczynski}, P., {Gawiser}, E., {Acquaviva}, V., {et~al.} 2016, \apjl, 820,
  L1, \dodoi{10.3847/2041-8205/820/1/L1}

\bibitem[{{Leja} {et~al.}(2019{\natexlab{a}}){Leja}, {Carnall}, {Johnson},
  {Conroy}, \& {Speagle}}]{leja2018measure}
{Leja}, J., {Carnall}, A.~C., {Johnson}, B.~D., {Conroy}, C., \& {Speagle},
  J.~S. 2019{\natexlab{a}}, \apj, 876, 3, \dodoi{10.3847/1538-4357/ab133c}

\bibitem[{Leja {et~al.}(2017)Leja, Johnson, Conroy, van Dokkum, \&
  Byler}]{leja2017deriving}
Leja, J., Johnson, B.~D., Conroy, C., van Dokkum, P.~G., \& Byler, N. 2017, The
  Astrophysical Journal, 837, 170

\bibitem[{{Leja} {et~al.}(2019{\natexlab{b}}){Leja}, {Tacchella}, \&
  {Conroy}}]{2019ApJ...880L...9L}
{Leja}, J., {Tacchella}, S., \& {Conroy}, C. 2019{\natexlab{b}}, \apjl, 880,
  L9, \dodoi{10.3847/2041-8213/ab2f8c}

\bibitem[{{Leja} {et~al.}(2021){Leja}, {Speagle}, {Ting}, {Johnson}, {Conroy},
  {Whitaker}, {Nelson}, {van Dokkum}, \& {Franx}}]{2021arXiv211004314L}
{Leja}, J., {Speagle}, J.~S., {Ting}, Y.-S., {et~al.} 2021, arXiv e-prints,
  arXiv:2110.04314.
\newblock \doarXiv{2110.04314}

\bibitem[{{Lilly} {et~al.}(2013){Lilly}, {Carollo}, {Pipino}, {Renzini}, \&
  {Peng}}]{2013ApJ...772..119L}
{Lilly}, S.~J., {Carollo}, C.~M., {Pipino}, A., {Renzini}, A., \& {Peng}, Y.
  2013, \apj, 772, 119, \dodoi{10.1088/0004-637X/772/2/119}

\bibitem[{{Lilly} {et~al.}(2007){Lilly}, {Le F{\`e}vre}, {Renzini}, {Zamorani},
  {Scodeggio}, {Contini}, {Carollo}, {Hasinger}, {Kneib}, {Iovino}, {Le Brun},
  {Maier}, {Mainieri}, {Mignoli}, {Silverman}, {Tasca}, {Bolzonella},
  {Bongiorno}, {Bottini}, {Capak}, {Caputi}, {Cimatti}, {Cucciati}, {Daddi},
  {Feldmann}, {Franzetti}, {Garilli}, {Guzzo}, {Ilbert}, {Kampczyk}, {Kovac},
  {Lamareille}, {Leauthaud}, {Le Borgne}, {McCracken}, {Marinoni}, {Pello},
  {Ricciardelli}, {Scarlata}, {Vergani}, {Sanders}, {Schinnerer}, {Scoville},
  {Taniguchi}, {Arnouts}, {Aussel}, {Bardelli}, {Brusa}, {Cappi}, {Ciliegi},
  {Finoguenov}, {Foucaud}, {Franceschini}, {Halliday}, {Impey}, {Knobel},
  {Koekemoer}, {Kurk}, {Maccagni}, {Maddox}, {Marano}, {Marconi}, {Meneux},
  {Mobasher}, {Moreau}, {Peacock}, {Porciani}, {Pozzetti}, {Scaramella},
  {Schiminovich}, {Shopbell}, {Smail}, {Thompson}, {Tresse}, {Vettolani},
  {Zanichelli}, \& {Zucca}}]{2007ApJS..172...70L}
{Lilly}, S.~J., {Le F{\`e}vre}, O., {Renzini}, A., {et~al.} 2007, \apjs, 172,
  70, \dodoi{10.1086/516589}

\bibitem[{{Madau} \& {Dickinson}(2014)}]{2014ARA&A..52..415M}
{Madau}, P., \& {Dickinson}, M. 2014, \araa, 52, 415,
  \dodoi{10.1146/annurev-astro-081811-125615}

\bibitem[{{Matthee} \& {Schaye}(2019)}]{2019MNRAS.484..915M}
{Matthee}, J., \& {Schaye}, J. 2019, \mnras, 484, 915,
  \dodoi{10.1093/mnras/stz030}

\bibitem[{{Mayya} {et~al.}(2004){Mayya}, {Bressan}, {Rodr{\'\i}guez}, {Valdes},
  \& {Chavez}}]{2004ApJ...600..188M}
{Mayya}, Y.~D., {Bressan}, A., {Rodr{\'\i}guez}, M., {Valdes}, J.~R., \&
  {Chavez}, M. 2004, \apj, 600, 188, \dodoi{10.1086/379707}

\bibitem[{{Merlin} {et~al.}(2022){Merlin}, {Bonchi}, {Paris}, {Belfiori},
  {Fontana}, {Castellano}, {Nonino}, {Polenta}, {Santini}, {Yang},
  {Glazebrook}, {Treu}, {Roberts-Borsani}, {Trenti}, {Birrer}, {Brammer},
  {Grillo}, {Calabr{\`o}}, {Marchesini}, {Mason}, {Mercurio}, {Morishita},
  {Strait}, {Boyett}, {Leethochawalit}, {Nanayakkara}, {Vulcani}, {Bradac}, \&
  {Wang}}]{2022arXiv220711701M}
{Merlin}, E., {Bonchi}, A., {Paris}, D., {et~al.} 2022, arXiv e-prints,
  arXiv:2207.11701.
\newblock \doarXiv{2207.11701}

\bibitem[{{Moutard} {et~al.}(2016){Moutard}, {Arnouts}, {Ilbert}, {Coupon},
  {Davidzon}, {Guzzo}, {Hudelot}, {McCracken}, {Van Waerbeke}, {Morrison}, {Le
  F{\`e}vre}, {Comte}, {Bolzonella}, {Fritz}, {Garilli}, \&
  {Scodeggio}}]{2016A&A...590A.103M}
{Moutard}, T., {Arnouts}, S., {Ilbert}, O., {et~al.} 2016, \aap, 590, A103,
  \dodoi{10.1051/0004-6361/201527294}

\bibitem[{{Muzzin} {et~al.}(2013){Muzzin}, {Marchesini}, {Stefanon}, {Franx},
  {McCracken}, {Milvang-Jensen}, {Dunlop}, {Fynbo}, {Brammer}, {Labb{\'e}}, \&
  {van Dokkum}}]{2013ApJ...777...18M}
{Muzzin}, A., {Marchesini}, D., {Stefanon}, M., {et~al.} 2013, \apj, 777, 18,
  \dodoi{10.1088/0004-637X/777/1/18}

\bibitem[{{Nagaraj} {et~al.}(2022){Nagaraj}, {Forbes}, {Leja},
  {Foreman-Mackey}, \& {Hayward}}]{2022ApJ...932...54N}
{Nagaraj}, G., {Forbes}, J.~C., {Leja}, J., {Foreman-Mackey}, D., \& {Hayward},
  C.~C. 2022, \apj, 932, 54, \dodoi{10.3847/1538-4357/ac6c80}

\bibitem[{{Nelson} {et~al.}(2019){Nelson}, {Pillepich}, {Springel}, {Pakmor},
  {Weinberger}, {Genel}, {Torrey}, {Vogelsberger}, {Marinacci}, \&
  {Hernquist}}]{2019MNRAS.490.3234N}
{Nelson}, D., {Pillepich}, A., {Springel}, V., {et~al.} 2019, \mnras, 490,
  3234, \dodoi{10.1093/mnras/stz2306}

\bibitem[{Noeske {et~al.}(2007)Noeske, Weiner, Faber, Papovich, Koo,
  Somerville, Bundy, Conselice, Newman, Schiminovich,
  {et~al.}}]{noeske2007star}
Noeske, K., Weiner, B., Faber, S., {et~al.} 2007, The Astrophysical Journal
  Letters, 660, L43

\bibitem[{Pacifici {et~al.}(2012)Pacifici, Kassin, Weiner, Charlot, \&
  Gardner}]{pacifici}
Pacifici, C., Kassin, S.~A., Weiner, B., Charlot, S., \& Gardner, J.~P. 2012,
  The Astrophysical Journal Letters, 762, L15

\bibitem[{{Pacifici} {et~al.}(2013){Pacifici}, {Kassin}, {Weiner}, {Charlot},
  \& {Gardner}}]{2013ApJ...762L..15P}
{Pacifici}, C., {Kassin}, S.~A., {Weiner}, B., {Charlot}, S., \& {Gardner},
  J.~P. 2013, \apjl, 762, L15, \dodoi{10.1088/2041-8205/762/1/L15}

\bibitem[{{Pacifici} {et~al.}(2016){Pacifici}, {Oh}, {Oh}, {Lee}, \&
  {Yi}}]{pacifici2016timing}
{Pacifici}, C., {Oh}, S., {Oh}, K., {Lee}, J., \& {Yi}, S.~K. 2016, \apj, 824,
  45, \dodoi{10.3847/0004-637X/824/1/45}

\bibitem[{{Parul} \& {Bailin}(2021)}]{2021AAS...23734208P}
{Parul}, H., \& {Bailin}, J. 2021, in American Astronomical Society Meeting
  Abstracts, Vol.~53, American Astronomical Society Meeting Abstracts, 342.08

\bibitem[{{Pillepich} {et~al.}(2019){Pillepich}, {Nelson}, {Springel},
  {Pakmor}, {Torrey}, {Weinberger}, {Vogelsberger}, {Marinacci}, {Genel}, {van
  der Wel}, \& {Hernquist}}]{2019MNRAS.490.3196P}
{Pillepich}, A., {Nelson}, D., {Springel}, V., {et~al.} 2019, \mnras, 490,
  3196, \dodoi{10.1093/mnras/stz2338}

\bibitem[{{Pontoppidan} {et~al.}(2022){Pontoppidan}, {Blome}, {Braun}, {Brown},
  {Carruthers}, {Coe}, {DePasquale}, {Espinoza}, {Garcia Marin}, {Gordon},
  {Henry}, {Hustak}, {James}, {Koekemoer}, {LaMassa}, {Law}, {Lockwood},
  {Moro-Martin}, {Mullally}, {Pagan}, {Player}, {Proffitt}, {Pulliam},
  {Ramsay}, {Ravindranath}, {Reid}, {Robberto}, {Sabbi}, \&
  {Ubeda}}]{2022arXiv220713067P}
{Pontoppidan}, K., {Blome}, C., {Braun}, H., {et~al.} 2022, arXiv e-prints,
  arXiv:2207.13067.
\newblock \doarXiv{2207.13067}

\bibitem[{{Portillo} {et~al.}(2020){Portillo}, {Parejko}, {Vergara}, \&
  {Connolly}}]{2020AJ....160...45P}
{Portillo}, S. K.~N., {Parejko}, J.~K., {Vergara}, J.~R., \& {Connolly}, A.~J.
  2020, \aj, 160, 45, \dodoi{10.3847/1538-3881/ab9644}

\bibitem[{Price {et~al.}(2018)Price, van~der Velden, Celles, Eendebak, McKerns,
  Olson, Raffel, Yi, \& Ash}]{Price2018}
Price, D.~C., van~der Velden, E., Celles, S., {et~al.} 2018, Journal of Open
  Source Software, 3, 1115, \dodoi{10.21105/joss.01115}

\bibitem[{Price-Whelan {et~al.}(2018)Price-Whelan, Sip{\H{o}}cz, G{\"u}nther,
  Lim, Crawford, Conseil, Shupe, Craig, Dencheva, Ginsburg,
  {et~al.}}]{price2018astropy}
Price-Whelan, A.~M., Sip{\H{o}}cz, B., G{\"u}nther, H., {et~al.} 2018, The
  Astronomical Journal, 156, 123

\bibitem[{Rasmussen \& Williams(2006)}]{gp_book}
Rasmussen, C.~E., \& Williams, C.~K. 2006, The MIT Press, Cambridge, MA, USA,
  38, 715

\bibitem[{Rhode(2020)}]{rhode2020non}
Rhode, S. 2020, Engineering Applications of Artificial Intelligence, 93, 103716

\bibitem[{{Scoville} {et~al.}(2007){Scoville}, {Aussel}, {Brusa}, {Capak},
  {Carollo}, {Elvis}, {Giavalisco}, {Guzzo}, {Hasinger}, {Impey}, {Kneib},
  {LeFevre}, {Lilly}, {Mobasher}, {Renzini}, {Rich}, {Sanders}, {Schinnerer},
  {Schminovich}, {Shopbell}, {Taniguchi}, \& {Tyson}}]{2007ApJS..172....1S}
{Scoville}, N., {Aussel}, H., {Brusa}, M., {et~al.} 2007, \apjs, 172, 1,
  \dodoi{10.1086/516585}

\bibitem[{{Semenov} {et~al.}(2018){Semenov}, {Kravtsov}, \&
  {Gnedin}}]{2018ApJ...861....4S}
{Semenov}, V.~A., {Kravtsov}, A.~V., \& {Gnedin}, N.~Y. 2018, \apj, 861, 4,
  \dodoi{10.3847/1538-4357/aac6eb}

\bibitem[{{Semenov} {et~al.}(2021){Semenov}, {Kravtsov}, \&
  {Gnedin}}]{2021ApJ...918...13S}
---. 2021, \apj, 918, 13, \dodoi{10.3847/1538-4357/ac0a77}

\bibitem[{Smith \& Hayward(2015)}]{smith2015deriving}
Smith, D.~J., \& Hayward, C.~C. 2015, Monthly Notices of the Royal Astronomical
  Society, 453, 1597

\bibitem[{{Somerville} \& {Dav{\'e}}(2015)}]{2015ARA&A..53...51S}
{Somerville}, R.~S., \& {Dav{\'e}}, R. 2015, \araa, 53, 51,
  \dodoi{10.1146/annurev-astro-082812-140951}

\bibitem[{{Strauss} {et~al.}(2002){Strauss}, {Weinberg}, {Lupton}, {Narayanan},
  {Annis}, {Bernardi}, {Blanton}, {Burles}, {Connolly}, {Dalcanton}, {Doi},
  {Eisenstein}, {Frieman}, {Fukugita}, {Gunn}, {Ivezi{\'c}}, {Kent}, {Kim},
  {Knapp}, {Kron}, {Munn}, {Newberg}, {Nichol}, {Okamura}, {Quinn}, {Richmond},
  {Schlegel}, {Shimasaku}, {SubbaRao}, {Szalay}, {Vanden Berk}, {Vogeley},
  {Yanny}, {Yasuda}, {York}, \& {Zehavi}}]{2002AJ....124.1810S}
{Strauss}, M.~A., {Weinberg}, D.~H., {Lupton}, R.~H., {et~al.} 2002, \aj, 124,
  1810, \dodoi{10.1086/342343}

\bibitem[{{Suess} {et~al.}(2020){Suess}, {Kriek}, {Price}, \&
  {Barro}}]{2020ApJ...899L..26S}
{Suess}, K.~A., {Kriek}, M., {Price}, S.~H., \& {Barro}, G. 2020, \apjl, 899,
  L26, \dodoi{10.3847/2041-8213/abacc9}

\bibitem[{{Tacchella} {et~al.}(2016){Tacchella}, {Dekel}, {Carollo},
  {Ceverino}, {DeGraf}, {Lapiner}, {Mandelker}, \& {Primack
  Joel}}]{2016MNRAS.457.2790T}
{Tacchella}, S., {Dekel}, A., {Carollo}, C.~M., {et~al.} 2016, \mnras, 457,
  2790, \dodoi{10.1093/mnras/stw131}

\bibitem[{{Tacchella} {et~al.}(2020){Tacchella}, {Forbes}, \&
  {Caplar}}]{2020MNRAS.497..698T}
{Tacchella}, S., {Forbes}, J.~C., \& {Caplar}, N. 2020, \mnras, 497, 698,
  \dodoi{10.1093/mnras/staa1838}

\bibitem[{{Tacchella} {et~al.}(2022{\natexlab{a}}){Tacchella}, {Smith},
  {Kannan}, {Marinacci}, {Hernquist}, {Vogelsberger}, {Torrey}, {Sales}, \&
  {Li}}]{2022MNRAS.513.2904T}
{Tacchella}, S., {Smith}, A., {Kannan}, R., {et~al.} 2022{\natexlab{a}},
  \mnras, 513, 2904, \dodoi{10.1093/mnras/stac818}

\bibitem[{{Tacchella} {et~al.}(2022{\natexlab{b}}){Tacchella}, {Finkelstein},
  {Bagley}, {Dickinson}, {Ferguson}, {Giavalisco}, {Graziani}, {Grogin},
  {Hathi}, {Hutchison}, {Jung}, {Koekemoer}, {Larson}, {Papovich}, {Pirzkal},
  {Rojas-Ruiz}, {Song}, {Schneider}, {Somerville}, {Wilkins}, \&
  {Yung}}]{2022ApJ...927..170T}
{Tacchella}, S., {Finkelstein}, S.~L., {Bagley}, M., {et~al.}
  2022{\natexlab{b}}, \apj, 927, 170, \dodoi{10.3847/1538-4357/ac4cad}

\bibitem[{{Tacchella} {et~al.}(2022{\natexlab{c}}){Tacchella}, {Johnson},
  {Robertson}, {Carniani}, {D'Eugenio}, {Kumar}, {Maiolino}, {Nelson}, {Suess},
  {{\"U}bler}, {Williams}, {Adebusola}, {Alberts}, {Arribas}, {Bhatawdekar},
  {Bonaventura}, {Bowler}, {Bunker}, {Cameron}, {Curti}, {Egami}, {Eisenstein},
  {Frye}, {Hainline}, {Helton}, {Ji}, {Looser}, {Lyu}, {Perna}, {Rawle},
  {Rieke}, {Rieke}, {Saxena}, {Sandles}, {Shivaei}, {Simmonds}, {Sun},
  {Willmer}, {Willott}, \& {Witstok}}]{2022arXiv220803281T}
{Tacchella}, S., {Johnson}, B.~D., {Robertson}, B.~E., {et~al.}
  2022{\natexlab{c}}, arXiv e-prints, arXiv:2208.03281.
\newblock \doarXiv{2208.03281}

\bibitem[{{Tacconi} {et~al.}(2020){Tacconi}, {Genzel}, \&
  {Sternberg}}]{2020ARA&A..58..157T}
{Tacconi}, L.~J., {Genzel}, R., \& {Sternberg}, A. 2020, \araa, 58, 157,
  \dodoi{10.1146/annurev-astro-082812-141034}

\bibitem[{{Teimoorinia} {et~al.}(2022){Teimoorinia}, {Archinuk}, {Woo},
  {Shishehchi}, \& {Bluck}}]{2022AJ....163...71T}
{Teimoorinia}, H., {Archinuk}, F., {Woo}, J., {Shishehchi}, S., \& {Bluck}, A.
  F.~L. 2022, \aj, 163, 71, \dodoi{10.3847/1538-3881/ac4039}

\bibitem[{{Thorne} {et~al.}(2021){Thorne}, {Robotham}, {Davies}, {Bellstedt},
  {Driver}, {Bravo}, {Bremer}, {Holwerda}, {Hopkins}, {Lagos}, {Phillipps},
  {Siudek}, {Taylor}, \& {Wright}}]{2021MNRAS.505..540T}
{Thorne}, J.~E., {Robotham}, A. S.~G., {Davies}, L. J.~M., {et~al.} 2021,
  \mnras, 505, 540, \dodoi{10.1093/mnras/stab1294}

\bibitem[{Tojeiro {et~al.}(2007)Tojeiro, Heavens, Jimenez, \& Panter}]{vespa}
Tojeiro, R., Heavens, A.~F., Jimenez, R., \& Panter, B. 2007, Monthly Notices
  of the Royal Astronomical Society, 381, 1252

\bibitem[{Tremonti {et~al.}(2004)Tremonti, Heckman, Kauffmann, Brinchmann,
  Charlot, White, Seibert, Peng, Schlegel, Uomoto,
  {et~al.}}]{tremonti2004origin}
Tremonti, C.~A., Heckman, T.~M., Kauffmann, G., {et~al.} 2004, The
  Astrophysical Journal, 613, 898

\bibitem[{{Villaescusa-Navarro} {et~al.}(2021){Villaescusa-Navarro},
  {Angl{\'e}s-Alc{\'a}zar}, {Genel}, {Spergel}, {Somerville}, {Dave},
  {Pillepich}, {Hernquist}, {Nelson}, {Torrey}, {Narayanan}, {Li}, {Philcox},
  {La Torre}, {Maria Delgado}, {Ho}, {Hassan}, {Burkhart}, {Wadekar},
  {Battaglia}, {Contardo}, \& {Bryan}}]{2021ApJ...915...71V}
{Villaescusa-Navarro}, F., {Angl{\'e}s-Alc{\'a}zar}, D., {Genel}, S., {et~al.}
  2021, \apj, 915, 71, \dodoi{10.3847/1538-4357/abf7ba}

\bibitem[{{Villaescusa-Navarro} {et~al.}(2022){Villaescusa-Navarro}, {Genel},
  {Angl{\'e}s-Alc{\'a}zar}, {Perez}, {Villanueva-Domingo}, {Wadekar}, {Shao},
  {Mohammad}, {Hassan}, {Moser}, {Lau}, {Machado Poletti Valle}, {Nicola},
  {Thiele}, {Jo}, {Philcox}, {Oppenheimer}, {Tillman}, {Hahn}, {Kaushal},
  {Pisani}, {Gebhardt}, {Delgado}, {Caliendo}, {Kreisch}, {Wong}, {Coulton},
  {Eickenberg}, {Parimbelli}, {Ni}, {Steinwandel}, {La Torre}, {Dave},
  {Battaglia}, {Nagai}, {Spergel}, {Hernquist}, {Burkhart}, {Narayanan},
  {Wandelt}, {Somerville}, {Bryan}, {Viel}, {Li}, {Irsic}, {Kraljic}, \&
  {Vogelsberger}}]{2022arXiv220101300V}
{Villaescusa-Navarro}, F., {Genel}, S., {Angl{\'e}s-Alc{\'a}zar}, D., {et~al.}
  2022, arXiv e-prints, arXiv:2201.01300.
\newblock \doarXiv{2201.01300}

\bibitem[{Virtanen {et~al.}(2020)Virtanen, Gommers, Oliphant, Haberland, Reddy,
  Cournapeau, Burovski, Peterson, Weckesser, Bright,
  {et~al.}}]{virtanen2020scipy}
Virtanen, P., Gommers, R., Oliphant, T.~E., {et~al.} 2020, Nature methods, 1

\bibitem[{Walt {et~al.}(2011)Walt, Colbert, \& Varoquaux}]{walt2011numpy}
Walt, S. v.~d., Colbert, S.~C., \& Varoquaux, G. 2011, Computing in Science \&
  Engineering, 13, 22

\bibitem[{{Wang} \& {Lilly}(2020{\natexlab{a}})}]{2020ApJ...895...25W}
{Wang}, E., \& {Lilly}, S.~J. 2020{\natexlab{a}}, \apj, 895, 25,
  \dodoi{10.3847/1538-4357/ab8b5e}

\bibitem[{{Wang} \& {Lilly}(2020{\natexlab{b}})}]{2020ApJ...892...87W}
---. 2020{\natexlab{b}}, \apj, 892, 87, \dodoi{10.3847/1538-4357/ab7b7d}

\bibitem[{{Weaver} {et~al.}(2022){Weaver}, {Kauffmann}, {Ilbert}, {McCracken},
  {Moneti}, {Toft}, {Brammer}, {Shuntov}, {Davidzon}, {Hsieh}, {Laigle},
  {Anastasiou}, {Jespersen}, {Vinther}, {Capak}, {Casey}, {McPartland},
  {Milvang-Jensen}, {Mobasher}, {Sanders}, {Zalesky}, {Arnouts}, {Aussel},
  {Dunlop}, {Faisst}, {Franx}, {Furtak}, {Fynbo}, {Gould}, {Greve}, {Gwyn},
  {Kartaltepe}, {Kashino}, {Koekemoer}, {Kokorev}, {Le F{\`e}vre}, {Lilly},
  {Masters}, {Magdis}, {Mehta}, {Peng}, {Riechers}, {Salvato}, {Sawicki},
  {Scarlata}, {Scoville}, {Shirley}, {Silverman}, {Sneppen}, {Smolc̆i{\'c}},
  {Steinhardt}, {Stern}, {Tanaka}, {Taniguchi}, {Teplitz}, {Vaccari}, {Wang},
  \& {Zamorani}}]{2022ApJS..258...11W}
{Weaver}, J.~R., {Kauffmann}, O.~B., {Ilbert}, O., {et~al.} 2022, \apjs, 258,
  11, \dodoi{10.3847/1538-4365/ac3078}

\bibitem[{{Wechsler} \& {Tinker}(2018)}]{2018ARA&A..56..435W}
{Wechsler}, R.~H., \& {Tinker}, J.~L. 2018, \araa, 56, 435,
  \dodoi{10.1146/annurev-astro-081817-051756}

\bibitem[{{Weisz} {et~al.}(2014){Weisz}, {Dolphin}, {Skillman}, {Holtzman},
  {Gilbert}, {Dalcanton}, \& {Williams}}]{2014ApJ...789..147W}
{Weisz}, D.~R., {Dolphin}, A.~E., {Skillman}, E.~D., {et~al.} 2014, \apj, 789,
  147, \dodoi{10.1088/0004-637X/789/2/147}

\bibitem[{{Whitaker} {et~al.}(2014){Whitaker}, {Rigby}, {Brammer}, {Gladders},
  {Sharon}, {Teng}, \& {Wuyts}}]{2014ApJ...790..143W}
{Whitaker}, K.~E., {Rigby}, J.~R., {Brammer}, G.~B., {et~al.} 2014, \apj, 790,
  143, \dodoi{10.1088/0004-637X/790/2/143}

\bibitem[{{Whitler} {et~al.}(2022){Whitler}, {Stark}, {Endsley}, {Leja},
  {Charlot}, \& {Chevallard}}]{2022arXiv220605315W}
{Whitler}, L., {Stark}, D.~P., {Endsley}, R., {et~al.} 2022, arXiv e-prints,
  arXiv:2206.05315.
\newblock \doarXiv{2206.05315}

\bibitem[{Wiener(1930)}]{wiener1930generalized}
Wiener, N. 1930, Acta mathematica, 55, 117

\bibitem[{{Wild} {et~al.}(2020){Wild}, {Taj Aldeen}, {Carnall}, {Maltby},
  {Almaini}, {Werle}, {Wilkinson}, {Rowlands}, {Bolzonella}, {Castellano},
  {Gargiulo}, {McLure}, {Pentericci}, \& {Pozzetti}}]{2020MNRAS.494..529W}
{Wild}, V., {Taj Aldeen}, L., {Carnall}, A., {et~al.} 2020, \mnras, 494, 529,
  \dodoi{10.1093/mnras/staa674}

\bibitem[{{Williams} {et~al.}(2009){Williams}, {Quadri}, {Franx}, {van Dokkum},
  \& {Labb{\'e}}}]{2009ApJ...691.1879W}
{Williams}, R.~J., {Quadri}, R.~F., {Franx}, M., {van Dokkum}, P., \&
  {Labb{\'e}}, I. 2009, \apj, 691, 1879, \dodoi{10.1088/0004-637X/691/2/1879}

\bibitem[{{Willott} {et~al.}(2022){Willott}, {Doyon}, {Albert}, {Brammer},
  {Dixon}, {Muzic}, {Ravindranath}, {Scholz}, {Abraham}, {Artigau},
  {Brada{\v{c}}}, {Goudfrooij}, {Hutchings}, {Iyer}, {Jayawardhana}, {LaMassa},
  {Martis}, {Meyer}, {Morishita}, {Mowla}, {Muzzin}, {Noirot}, {Pacifici},
  {Rowlands}, {Sarrouh}, {Sawicki}, {Taylor}, {Volk}, \&
  {Zabl}}]{2022PASP..134b5002W}
{Willott}, C.~J., {Doyon}, R., {Albert}, L., {et~al.} 2022, \pasp, 134, 025002,
  \dodoi{10.1088/1538-3873/ac5158}

\bibitem[{{Worthey} \& {Ottaviani}(1997)}]{1997ApJS..111..377W}
{Worthey}, G., \& {Ottaviani}, D.~L. 1997, \apjs, 111, 377,
  \dodoi{10.1086/313021}

\bibitem[{{Wuyts} {et~al.}(2007){Wuyts}, {Labb{\'e}}, {Franx}, {Rudnick}, {van
  Dokkum}, {Fazio}, {F{\"o}rster Schreiber}, {Huang}, {Moorwood}, {Rix},
  {R{\"o}ttgering}, \& {van der Werf}}]{2007ApJ...655...51W}
{Wuyts}, S., {Labb{\'e}}, I., {Franx}, M., {et~al.} 2007, \apj, 655, 51,
  \dodoi{10.1086/509708}

\bibitem[{{York} {et~al.}(2000){York}, {Adelman}, {Anderson}, {Anderson},
  {Annis}, {Bahcall}, {Bakken}, {Barkhouser}, {Bastian}, {Berman}, {Boroski},
  {Bracker}, {Briegel}, {Briggs}, {Brinkmann}, {Brunner}, {Burles}, {Carey},
  {Carr}, {Castander}, {Chen}, {Colestock}, {Connolly}, {Crocker}, {Csabai},
  {Czarapata}, {Davis}, {Doi}, {Dombeck}, {Eisenstein}, {Ellman}, {Elms},
  {Evans}, {Fan}, {Federwitz}, {Fiscelli}, {Friedman}, {Frieman}, {Fukugita},
  {Gillespie}, {Gunn}, {Gurbani}, {de Haas}, {Haldeman}, {Harris}, {Hayes},
  {Heckman}, {Hennessy}, {Hindsley}, {Holm}, {Holmgren}, {Huang}, {Hull},
  {Husby}, {Ichikawa}, {Ichikawa}, {Ivezi{\'c}}, {Kent}, {Kim}, {Kinney},
  {Klaene}, {Kleinman}, {Kleinman}, {Knapp}, {Korienek}, {Kron}, {Kunszt},
  {Lamb}, {Lee}, {Leger}, {Limmongkol}, {Lindenmeyer}, {Long}, {Loomis},
  {Loveday}, {Lucinio}, {Lupton}, {MacKinnon}, {Mannery}, {Mantsch}, {Margon},
  {McGehee}, {McKay}, {Meiksin}, {Merelli}, {Monet}, {Munn}, {Narayanan},
  {Nash}, {Neilsen}, {Neswold}, {Newberg}, {Nichol}, {Nicinski}, {Nonino},
  {Okada}, {Okamura}, {Ostriker}, {Owen}, {Pauls}, {Peoples}, {Peterson},
  {Petravick}, {Pier}, {Pope}, {Pordes}, {Prosapio}, {Rechenmacher}, {Quinn},
  {Richards}, {Richmond}, {Rivetta}, {Rockosi}, {Ruthmansdorfer}, {Sandford},
  {Schlegel}, {Schneider}, {Sekiguchi}, {Sergey}, {Shimasaku}, {Siegmund},
  {Smee}, {Smith}, {Snedden}, {Stone}, {Stoughton}, {Strauss}, {Stubbs},
  {SubbaRao}, {Szalay}, {Szapudi}, {Szokoly}, {Thakar}, {Tremonti}, {Tucker},
  {Uomoto}, {Vanden Berk}, {Vogeley}, {Waddell}, {Wang}, {Watanabe},
  {Weinberg}, {Yanny}, {Yasuda}, \& {SDSS Collaboration}}]{2000AJ....120.1579Y}
{York}, D.~G., {Adelman}, J., {Anderson}, John~E., J., {et~al.} 2000, \aj, 120,
  1579, \dodoi{10.1086/301513}

\bibitem[{{Zhou} {et~al.}(2022){Zhou}, {Merrifield}, \&
  {Arag{\'o}n-Salamanca}}]{2022MNRAS.513.5446Z}
{Zhou}, S., {Merrifield}, M., \& {Arag{\'o}n-Salamanca}, A. 2022, \mnras, 513,
  5446, \dodoi{10.1093/mnras/stac1279}

\bibitem[{{Zhu} \& {M{\'e}nard}(2013)}]{2013ApJ...773...16Z}
{Zhu}, G., \& {M{\'e}nard}, B. 2013, \apj, 773, 16,
  \dodoi{10.1088/0004-637X/773/1/16}

\end{thebibliography}
\bibliographystyle{aasjournal}

\appendix

\section{Connecting the PSD and ACF in the Extended Regulator Model} \label{ap:model}

We start by building physical intuition of how different physical processes related to galaxies can affect stochasticity and correlations in the SFRs of individual galaxies across cosmic time, summarized through their power spectral densities (PSDs) and associated auto-covariance functions (ACFs). We focus in particular on derivations of the ACF, which can in some cases be easier to interpret than the PSD. In \S\ref{subsec:psd_acf}, we provide a brief set of definitions for the PSD and ACF, their relationship with each other, and useful associated quantities. In \S\ref{subsec:white_noise}, we outline the expected behavior for completely uncorrelated SFRs. In \S\ref{subsec:damped_walk}, we outline the relationship between a damped random walk process, the PSD, and the ACF. In \S\ref{subsec:regulator} and \S\ref{subsec:regulator_ext}, we derive results for the Regulator and Extended Regulator models presented in \citetalias{2020MNRAS.497..698T}.

\subsection{Overview of Formalism}
\label{subsec:psd_acf}

We start by informally defining a \textit{stochastic process} as something that can generate infinite realizations of a \textit{time series}
$\{ x_1, x_2, \dots x_n \} \equiv \{x_t\}_{1}^{n} \equiv \mathbf{x}_{n}$ at any times $t=1,\dots,n$ (i.e. the $x_t$ values change every time we simulate from the process). The collection of $\mathbf{x}_{n}$ values will then follow some joint probability distribution $P(\mathbf{x}_{n})$ which is defined by the stochastic process.

We can use the time-dependent \textit{mean}
\begin{equation}
    \mu(t) \equiv \int_{-\infty}^{+\infty} x_t \, P(x_t) \,{\rm d}x_t
\end{equation}
as a simple summary statistic to describe how this process evolves over time, given the marginal distribution $P(x_t)$ of $x_t$ defined by our process. Many stochastic processes are defined with $\mu(t) = 0$, so modifying them to follow some non-zero mean is as simple as adding in a chosen mean function to the generated data $\mathbf{x}_{n}$.

The simplest way to explore the correlation structure in a given stochastic process is to compute the \textit{auto-covariance function} (ACF)\footnote{The prefix ``auto-'' is often used to emphasize that the calculation is done at two different times for the same process, rather than between two different processes.}
\begin{equation}
    \mathcal{C}(t, t') = \int_{-\infty}^{+\infty} \left[x_t - \mu(t)\right] \left[x_{t'} - \mu({t'})\right] \, P(x_t, x_{t'}) \, {\rm d}x_t {\rm d}x_{t'}
\end{equation}
between $x_t$ and $x_{t'}$ at two different times $t$ and $t'$. As with the mean, $P(x_t, x_{t'})$ is the joint distribution of $x_t$ and $x_{t'}$ defined by the process.
% Note that these two summary statistics -- the mean $\mu(t)$ and auto-covariance $\mathcal{C}(t, t')$ -- are by no means sufficient to describe many stochastic stochastic processes which may have more complex, higher-order correlation structures, but are sufficient to fully describe the processes we will be exploring in this work.

Assuming our process is \textit{stationary} such that the auto-covariance function only depends on the separation (i.e. time lag) between any two given times $\tau \equiv t - t'$ rather than the individual times $t$ and $t'$ themselves, we can instead write the auto-covariance function as
\begin{equation}
\boxed{
    \mathcal{C}(\tau) = \int_{-\infty}^{+\infty} \left[x_t - \mu(t)\right] \left[x_{t + \tau} - \mu({t + \tau})\right] \, P(x_t, x_{t + \tau}) \, {\rm d}x_t {\rm d}x_{t + \tau}
    }
\end{equation}
We can use the auto-covariance function $\mathcal{C}(\tau)$ to also define the \textit{auto-correlation function} as
\begin{equation}
\boxed{
    \rho(\tau) \equiv \mathcal{C}(\tau) / \mathcal{C}(0) \equiv \mathcal{C}(\tau) / \sigma^2
    }
\end{equation}
which is normalized to be between $1$ and $-1$. Note that at $\tau=0$ the auto-correlation function is always $1$ since it's normalized by the variance
\begin{equation}
\boxed{
    \sigma^2 \equiv \mathcal{C}(\tau=0)
    }
\end{equation}

It can also be useful to define a timescale over which a stochastic process is correlated. One definition is the \textit{auto-correlation time} $\tau_A$, which tries to account for contributions from correlations across all possible time lags $\tau$. This can be computed via
\begin{equation}
\boxed{
    \tau_A \equiv \int_{-\infty}^{+\infty} \rho(\tau) \,{\rm d}\tau
}
\end{equation}

In addition to defining and investigating correlation structure as a function of time $t$, we can also do the same as a function of frequency $f$. Defining a ``windowed'' version of $x(t)$
\begin{equation}
    x_T(t) \equiv x_t w_T(t) =
    \begin{cases}
    x_t & t - \frac{T}{2} < t < t + \frac{T}{2} \\
    0 & {\rm otherwise}
    \end{cases}
\end{equation}
for a window function $w_T(t)$ with some width (duration) $T$ centered around $t$, the average \textit{power} of a signal can be computed via
\begin{equation}
    \mathcal{P} = \lim_{T \rightarrow \infty}\frac{1}{T}\int_{-\infty}^{+\infty} |x_T(t)|^2 \, {\rm d}t
\end{equation}
where we take the limit $T \rightarrow \infty$ assuming the stochastic process is not localized in time.
Using Parseval's theorem, which states that power is conserved if we move from describing our process in the time domain to the frequency domain, we can rewrite this expression in terms of the frequency $f$ as
\begin{equation}
    \mathcal{P} = \lim_{T \rightarrow \infty}\frac{1}{T}\int_{-\infty}^{+\infty} |\hat{x}_T(f)|^2 \, {\rm d}f
\end{equation}
where
\begin{equation}
    \hat{x}_T(f) = \int_{-\infty}^{+\infty} x_T(t) \, e^{-2\pi i f t} \, {\rm d}t
\end{equation}
is the \textit{Fourier transform} of $x_T(t)$.
We now define the \textit{power spectral density} (PSD) as the integrand of the above expression, i.e.
\begin{equation}
\boxed{
    \mathcal{S}(f) \equiv \lim_{T \rightarrow \infty}\frac{1}{T} |\hat{x}_T(f)|^2
    }
\end{equation}
We can interpret the PSD as the relative amount of power as a function of frequency, where larger values indicate stronger correlations across particular frequencies.

While the ACF and PSD can be computed directly from a given stochastic process, they can also be directly computed from each other. Based the Wiener-Khinchin theorem, which states that in the continuous-time limit the PSD $\mathcal{S}(f)$ and ACF $\mathcal{C}(\tau)$ are Fourier pairs, we can convert between the two using the following relations:
\begin{equation}
\boxed{
    \mathcal{S}(f) = \int_{-\infty}^{+\infty} \mathcal{C}(\tau) \, e^{-2\pi i f \tau} \, {\rm d}\tau \quad\Longleftrightarrow\quad
    \mathcal{C}(\tau) = \int_{-\infty}^{+\infty} \mathcal{S}(f) \, e^{+2\pi i \tau f} \, {\rm d}f
    }
\end{equation}
This property is extremely useful, as many stochastic processes (such as the ones discussed below) can be much easier to describe in frequency rather than in time (and vice versa).

Note that due to the nature of the Fourier transform, there will be offsets in the overall normalization depending on how the PSD and ACF are parameterized. In other words, an maximum value of $\sigma^2$ in the PSD may not correspond to a maximum value of $\sigma^2$ in the ACF. Since these normalizations are often arbitrary, we will used the variables $s^2$ and $\sigma^2$ to refer to the variance/overall amplitude in the ACF/PSD, respectively.

In the following subsections, we will include expressions for $\mathcal{S}(f)$, $\mathcal{C}(\tau)$, $\sigma^2$, $\rho(\tau)$, and $\tau_A$ for all cases under consideration.

\subsection{White Noise} \label{subsec:white_noise}

The simplest stochastic process is white noise. This process has equal power at all frequencies and is defined by the PSD
\begin{equation}
    \mathcal{S}_{\rm WN}(f) = s^2
\end{equation}
where $\sigma^2$ is a constant that defines the intrinsic stochasticity. While the direct Fourier transform is ill-defined, taking the expanding window limit gives an ACF of
\begin{equation}
    \mathcal{C}_{\rm WN}(\tau) = \sigma^2 \times \delta_1(\tau) =
    \begin{cases}
    \sigma^2 & \tau = 0 \\
    0 & {\rm otherwise}
    \end{cases}
\end{equation}
where $\delta_1(\tau)$ is the Kronecker delta function that gives $1$ when $\tau=0$ and $0$ otherwise. This corresponds to a variance, auto-correlation function, and auto-correlation time of
\begin{align}
    \sigma_{\rm WN}^2 &= \sigma^2 \\
    \rho_{\rm WN}(\tau) &= \delta_1(\tau) \\
    \tau_{A, \rm WN} &= 0
\end{align}
This makes sense for a white noise process with no intrinsic correlations -- the auto-correlation time is 0 and the auto-correlation function is only non-zero for $\tau = 0$.

\subsection{Damped Random Walk} \label{subsec:damped_walk}

Many natural processes have some characteristic timescale $\tau_{\rm decor}$ where for $\tau < \tau_{\rm decor}$ it is strongly correlated and for $\tau_{\rm decor}$ it becomes uncorrelated (i.e. it loses its ``memory'' of the previous values and behaves like the white noise process described in \S\ref{subsec:white_noise}). One way to describe such a process is by defining a \textit{damped random walk} with a broken power-law PSD of
\begin{equation}
    \mathcal{S}_{\rm DRW}(f) = \frac{s^2}{1 + (2\pi\tau_{\rm dec})^2 f^2}
\end{equation}
% s^2 = \sigma^2 \times 2 \tau_{\rm eq}
This is damped as a function of $f^2$ with a characteristic damping scale of $2 \pi \tau_{\rm decor}$.

As shown in \citetalias{2020MNRAS.497..698T}, assuming that the gas mass is directly related to the SFR of the galaxy, that the conversion from gas mass to SFR follows a stochastic process with an equilibrium timescale $\tau_{\rm eq}$, and the gas inflow rate is a white noise process, the galaxy's SFR will follow a damped random walk with $\tau_{\rm dec} = \tau_{\rm eq}$. The normalization $\sigma$ is the ``long-term'' variability and is directly related to the stochasticity of the inflow rate. This gives (see also \citetalias{2019MNRAS.487.3845C} and \citetalias{2020MNRAS.497..698T}):
\begin{equation}
    \mathcal{C}_{\rm DRW}(\tau) = \sigma^2 \times e^{-|\tau|/\tau_{\rm eq}}
\end{equation}
The corresponding variance, auto-correlation function, and auto-correlation time are
\begin{align}
    \sigma_{\rm DRW}^2 &= \sigma^2 \\
    \rho_{\rm DRW}(\tau) &= e^{-|\tau|/\tau_{\rm eq}} \\
    \tau_{A, \rm DRW} &= 2\tau_{\rm eq}
\end{align}

\subsection{Regulator Model} \label{subsec:regulator}

In \citetalias{2020MNRAS.497..698T}, the ``Regulator Model'' is defined as the case where the gas inflow rate onto the galaxy is also a stochastic process (i.e. it's not just white noise). Assuming that this process also follows a damped random walk with an inflow timescale $\tau_{\rm in}$, the combined PSD will be the product of the two PSDs:
\begin{align}
    \mathcal{S}_{\rm Reg}(f) &= \mathcal{S}_{\rm eq}(f) \times \mathcal{S}_{\rm in}(f) \nonumber \\
    &= \frac{s_{\rm eq}^2}{1 + (2\pi\tau_{\rm eq})^2 f^2} \times \frac{s_{\rm in}^2}{1 + (2\pi\tau_{\rm in})^2 f^2} \nonumber \\
    &\equiv \frac{s_{\rm gas}^2}{1 + ((2\pi\tau_{\rm eq})^2 + (2\pi\tau_{\rm in})^2) f^2 + (2\pi\tau_{\rm eq})^2 (2\pi\tau_{\rm in})^2 f^4}
\end{align}
% s_{\rm gas} = \sigma_{\rm gas}^2 \times (\tau_{\rm in} + \tau_{\rm eq})
This now includes two damping terms: one that scales with $f^2$ and one that scales with $f^4$. Since the suppress at large $f$ (short timescales), they lead to even longer correlations.

Using the Wiener-Khinchin theorem, this PSD corresponds to an ACF of
\begin{equation}
    \mathcal{C}_{\rm Reg}(\tau) = \sigma_{\rm gas}^2 \times \frac{\tau_{\rm in} \, e^{-|\tau|/\tau_{\rm in}} - \tau_{\rm eq} \, e^{-|\tau|/\tau_{\rm eq}}}{\tau_{\rm in} - \tau_{\rm eq}}
\end{equation}
The corresponding variance, auto-correlation function, and auto-correlation time are
\begin{align}
    \sigma_{\rm Reg}^2 &= \sigma_{\rm gas}^2 \\
    \rho_{\rm Reg}(\tau) &= \frac{\tau_{\rm in} \, e^{-|\tau|/\tau_{\rm in}} - \tau_{\rm eq} \, e^{-|\tau|/\tau_{\rm eq}}}{\tau_{\rm in} - \tau_{\rm eq}} \\
    \tau_{A, \rm Reg} &= 2(\tau_{\rm in} + \tau_{\rm eq})
\end{align}

\subsubsection{$\tau_{\rm in} = \tau_{\rm eq}$ and the Matern32 Connection}
\label{subsubsec:matern32}

In the limit where $\tau_{\rm in} = \tau_{\rm eq}$, the ACF from \S\ref{subsec:regulator} becomes undefined even though the PSD is simply
\begin{equation}
    \mathcal{S}_{\rm Reg}(f) = \frac{s_{\rm gas}^2}{1 + 2(2\pi\tau_{\rm eq})^2f^2 + (2\pi\tau_{\rm eq})^4 f^4}
\end{equation}
However, simply recomputing the ACF from the above PSD (or taking the limit as $\tau_{\rm in} \rightarrow \tau_{\rm eq}$) gives the well-defined expression
\begin{equation}
    \mathcal{C}_{\rm Reg}(\tau) = \sigma_{\rm gas}^2 \times \left(1 + \frac{|\tau|}{\tau_{\rm eq}}\right) \, e^{-|\tau| / \tau_{\rm eq}}
\end{equation}
which corresponds to
\begin{align}
    \sigma_{0, {\rm Reg}}^2 &= \frac{\sigma_{\rm gas}^2}{2\tau_{\rm eq}} \\
    \rho_{\rm Reg}(\tau) &= \left(1 + \frac{|\tau|}{\tau_{\rm eq}}\right) \, e^{-|\tau| / \tau_{\rm eq}} \\
    \tau_{A, \rm Reg} &= 4\tau_{\rm eq}
\end{align}
This parallels the original damped random walk case in \S\ref{subsec:damped_walk} closely except the prefactor has change from $1 \rightarrow 1 + \tau/\tau_{\rm eq}$, which doubles the auto-correlation time.

The ACF for this special case can be shown to reduce \textit{exactly} to that of the Matern32 kernel, a common choice of ACF when modelling a range of stochastic processes. In particular, \citet{2019ApJ...879..116I} found the Matern32 kernel to best reproduce observed SFR correlation structure from simulations compared to several alternatives. As a result, we should interpret the Regulator Model with $\tau_{\rm in} \neq \tau_{\rm eq}$ to be a direct generalization of that work.

\subsection{Extended Regulator Model} \label{subsec:regulator_ext}

\citetalias{2020MNRAS.497..698T} introduced the Extended Regulator model, which -- in addition to gas inflow physics -- also includes a prescription for star formation within giant molecular clouds (GMCs). The formation and disruption of GMCs introduce additional stochasticity and a new correlation timescale in the system $\tau_{\rm dyn}$. This arises because the SFR of the galaxy is correlated over the timescale of the star formation processes. While this is originally linked to the lifetime of GMCs, we expand the definition to include the effects of dynamical processes such as spiral arms and bars affect local star formation in galaxies \citep{2015MNRAS.453..739K, 2019MNRAS.487.3581F, 2021ApJ...918...13S} and thus call it $\tau_{\rm dyn}$ as opposed to $\tau_L$ as in \citetalias{2020MNRAS.497..698T}. \citetalias{2020MNRAS.497..698T} show that the formation and disruption of GMCs follow a damped random walk such that the PSD and ACF are
\begin{align}
    \mathcal{S}_{\rm dyn}(f) &= \frac{s_{\rm dyn}^2}{1 + (2\pi\tau_{\rm dyn})^2 f^2} \\
    \mathcal{C}_{\rm dyn}(\tau) &= \sigma_{\rm dyn}^2 \times e^{-|\tau|/\tau_{\rm dyn}}
\end{align}
% s_{\rm dyn}^2 = \sigma_{\rm dyn}^2 \times 2\tau_{\rm dyn}
with timescale $\tau_{\rm dyn}$ and scatter $\sigma_{\rm dyn}$. Following \citetalias{2020MNRAS.497..698T} and assuming that this star-formation processes is decoupled from the processes related to gas cycling (converting gas into stars; $\tau_{\rm eq}$ and inflows (bringing in gas; $\tau_{\rm in}$), the PSD of the Extended Regulator model is the sum of the two PSDs:
\begin{align}
    \mathcal{S}&_{\rm ExReg}(f) = \mathcal{S}_{\rm Reg}(f) + \mathcal{S}_{\rm dyn}(f) \nonumber \\
    &= \frac{s_{\rm gas}^2}{1 + ((2\pi\tau_{\rm eq})^2 + (2\pi\tau_{\rm in})^2) f^2 + (2\pi\tau_{\rm eq})^2 (2\pi\tau_{\rm in})^2 f^4} + \frac{s_{\rm dyn}^2}{1 + (2\pi\tau_{\rm dyn})^2 f^2}
\end{align}
The corresponding ACF is then likewise the sum of the two ACFs:
\begin{equation}
\boxed{
    \mathcal{C}_{\rm ExReg}(\tau) = \sigma_{\rm gas}^2 \times \frac{\tau_{\rm in} \, e^{-|\tau|/\tau_{\rm in}} - \tau_{\rm eq} \, e^{-|\tau|/\tau_{\rm eq}}}{\tau_{\rm in} - \tau_{\rm eq}} + \sigma_{\rm dyn}^2 \times e^{-|\tau|/\tau_{\rm dyn}}
    }
\end{equation}
The corresponding variance, auto-correlation function, and auto-correlation time are
\begin{align}
    \sigma_{\rm ExReg}^2 &= \sigma_{\rm gas}^2 + \sigma_{\rm dyn}^2 \\
    \rho_{\rm ExReg}(\tau) &= \frac{\sigma_{\rm gas}^2}{\sigma_{\rm ExReg}^2} \times \frac{\tau_{\rm in} \, e^{-|\tau|/\tau_{\rm in}} - \tau_{\rm eq} \, e^{-|\tau|/\tau_{\rm eq}}}{\tau_{\rm in} - \tau_{\rm eq}} + \frac{\sigma_{\rm dyn}^2}{\sigma_{\rm ExReg}^2} \times e^{-|\tau|/\tau_{\rm dyn}} \\
    \tau_{A, {\rm ExReg}} &= 2 \times \left(\frac{\sigma_{\rm gas}^2}{\sigma_{\rm ExReg}^2} \times (\tau_{\rm in} + \tau_{\rm eq}) + \frac{\sigma_{\rm dyn}^2}{\sigma_{\rm ExReg}^2} \times \tau_{\rm dyn}\right)
\end{align}

\section{Response curves for individual spectral features}
\label{app:resp_curves}

\begin{figure}
    \centering
    \includegraphics[width=0.81\textwidth]{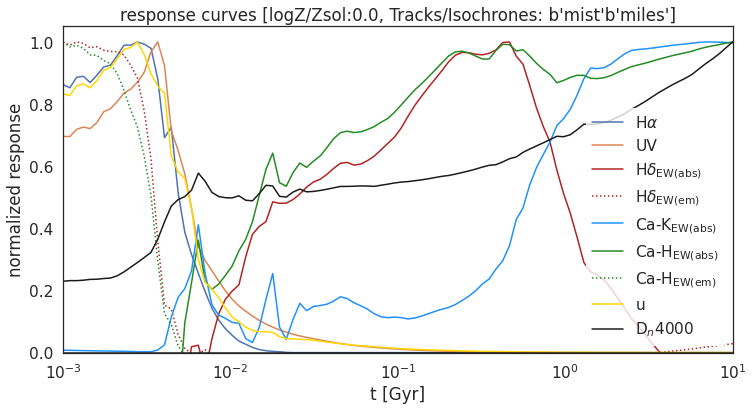}
    \caption{Response curves for spectral features we consider, ranging from H$\alpha$ and rest-UV flux that trace star formation on short timescales, to the D$_n$(4000)\AA{} break that traces the median age of the stellar population.}
    \label{fig:respcurves}
\end{figure}

To better understand the timescales over which the spectral features we consider in this work are sensitive to star formation, we evaluate the relative strength of each spectral feature used in Section \ref{sec:single} using a simple stellar population (SSP) at various ages using the same assumptions as Table \ref{tab:spec_generation}, and plot the normalized results in Figure. \ref{fig:respcurves}.

\section{Varying the base SFHs}
\label{app:varbase}

\begin{figure}
    \centering
    \includegraphics[width=\textwidth]{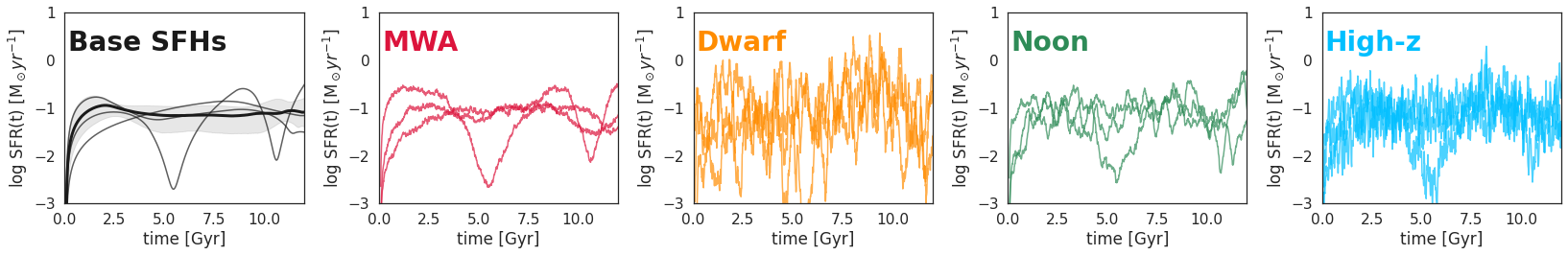}
    \includegraphics[width=\textwidth]{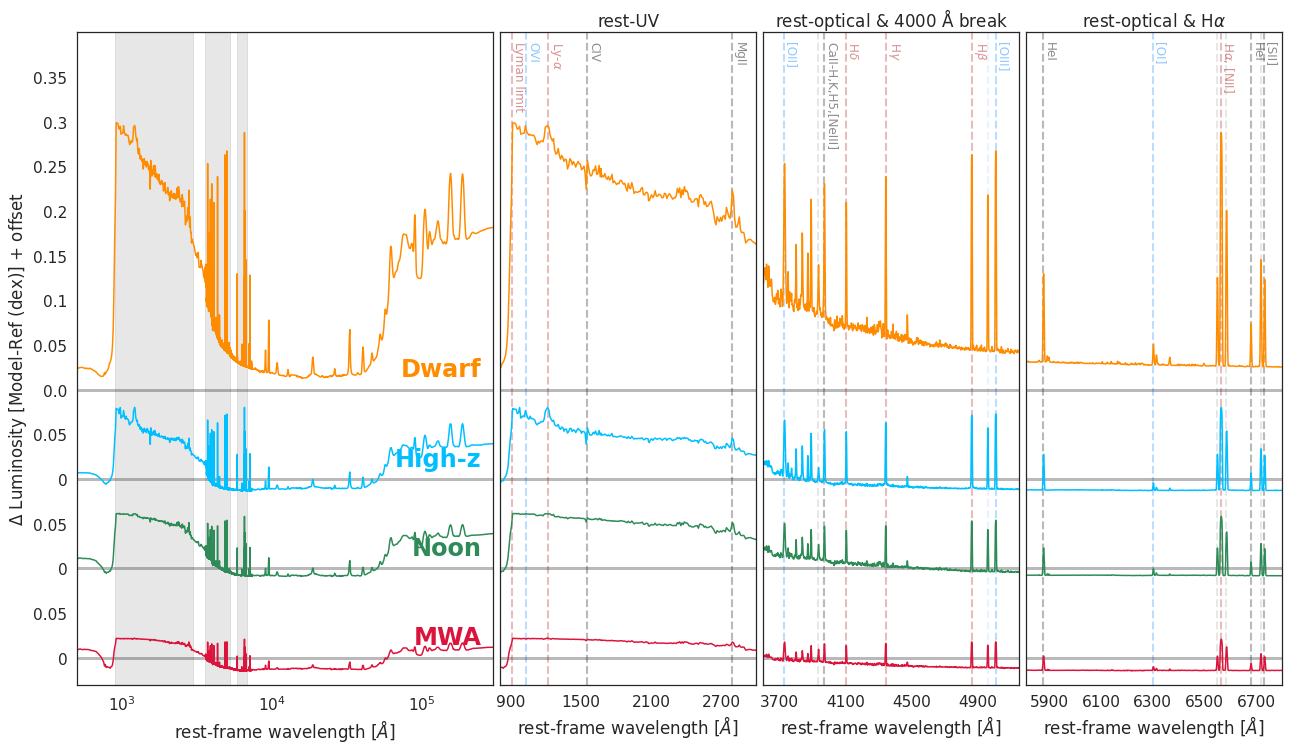}
    \caption{Implementing the ExReg kernel with the realistic SFHs described in \citet{2019ApJ...879..116I}. The top panel shows draws from a Dirichlet distribution, with perturbations from the various toy models added on to the base SFHs. The bottom panels follow the same format as Figure \ref{fig:extended_regulator_spec_cases1}, and show that despite the increased scatter due to SFH variations, the differences between the models are still distinguishable.}
    \label{fig:varsfh_specfeat}
\end{figure}

\begin{figure}
    \centering
    \includegraphics[width=\textwidth]{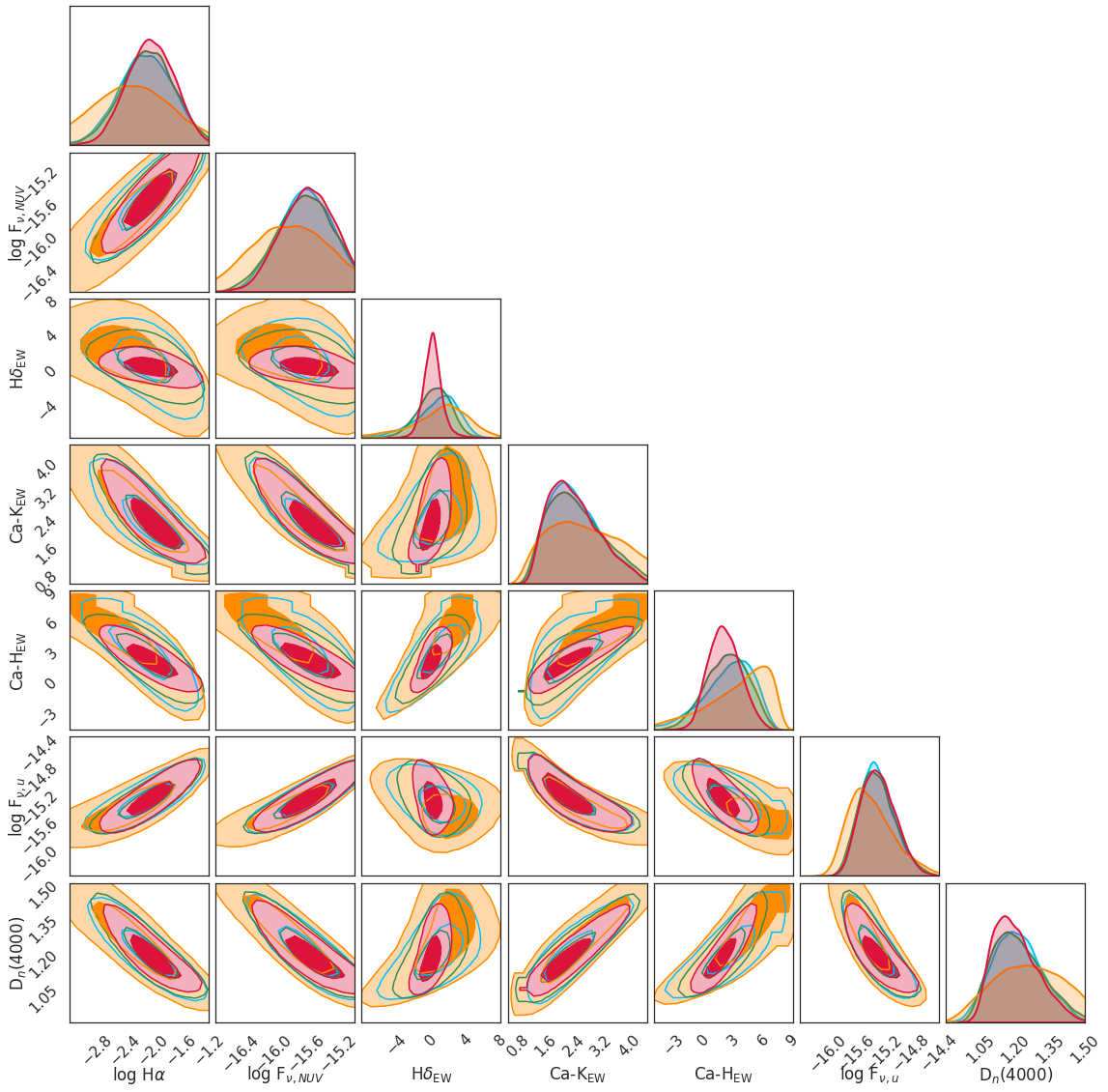}
    \caption{Implementing the ExReg kernel with the realistic SFHs described in \citet{2019ApJ...879..116I}. The top panel shows draws from a Dirichlet distribution, with perturbations from the various toy models added on to the base SFHs. The bottom panels follow the same format as Figure \ref{fig:extended_regulator_spec_cases1_corner}, and show that despite the increased scatter due to SFH variations, the differences between the models are still distinguishable.}
    \label{fig:varsfh_corner}
\end{figure}

The results in Section \ref{sec:single} mainly deal with the case where each underlying population is given \textit{identical} base SFHs. Here, we additionally consider the
% slightly more realistic
case where each individual base SFH
% $\ln {\rm SFR}_{\rm base}(t)$
is itself drawn around some mean SFH, $\ln {\rm SFR}_{\rm pop}(t)$. This type of doubly-stochastic process is generally known as a \textit{(log-)Cox process}. We construct our SFHs based on the framework in \citet{2019ApJ...879..116I}, which models the SFR as a smooth interpolation (using a GP) over points in time where galaxies formed evenly-spaced quantiles of their total mass (e.g., $t_{\rm 25}$, $t_{50}$, and $t_{\rm 75}$ being the times when the galaxy formed 25\%, 50\%, and 75\% of its total stellar mass) in addition to the present-day SFR and a particular formation time $t_0$\footnote{Following \citet{2019ApJ...879..116I}, $t_0$ here is set by the age of the universe at a given redshift.}. Following \citet{2019ApJ...879..116I}, the times $t_{\rm 25}$, $t_{50}$, and $t_{\rm 75}$ are drawn from a Dirichlet distribution with $\alpha = 5.0$, which has been shown to agree well with SFHs from cosmological simulations. For simplicity, the stellar masses of the base SFHs are fixed at $10^9$ M$_\odot$ and SFRs at the time of observation are drawn from a normal distribution designed to mimic a portion of observed $\sim 0.3\,{\rm dex}$ scatter in the observed SFR-M$^*$ correlation. Based on this model, we then construct a base population SFH $\ln {\rm SFR}_{\rm pop}(t)$ that is relatively constant across several Gyr, as seen in the top-left panel of Figure \ref{fig:varsfh_specfeat}.

The SFH realizations generated using this sample of varying SFHs is shown in Figure \ref{fig:varsfh_specfeat}, with the three panels on the top right showing perturbations to the base SFH realisations in the top-left. Noticeably, even in this case where the underlying population itself possesses some intrinsic variability in their base SFHs around some population mean, we still observe some differences in Figure \ref{fig:varsfh_corner} in the joint distribution of observables, noticeably H$\alpha$ and H$\delta$. The former difference comes from the varying sSFR distributions due to the perturbations of the different stochasticity archetypes, and the latter due to the varying amount of SFH burstiness in the last $\sim 1-2$ Gyr, with the median of the distribution rising with decreasing $\tau_{\rm gas}$.

\section{Verifying the covariance for binned SFHs}
\label{app:binnedsfh_validation}

In practice, many non-parametric SFH codes use SFHs binned in roughly logarithmic bins in lookback time with varying priors on stochasticity. In this Section, we verify that directly sampling from the covariance function using bin-centers is equivalent to sampling high-resolution SFHs and degrading them to the same coarse time-bins.

\begin{figure}
    \centering
    \includegraphics[width=\textwidth]{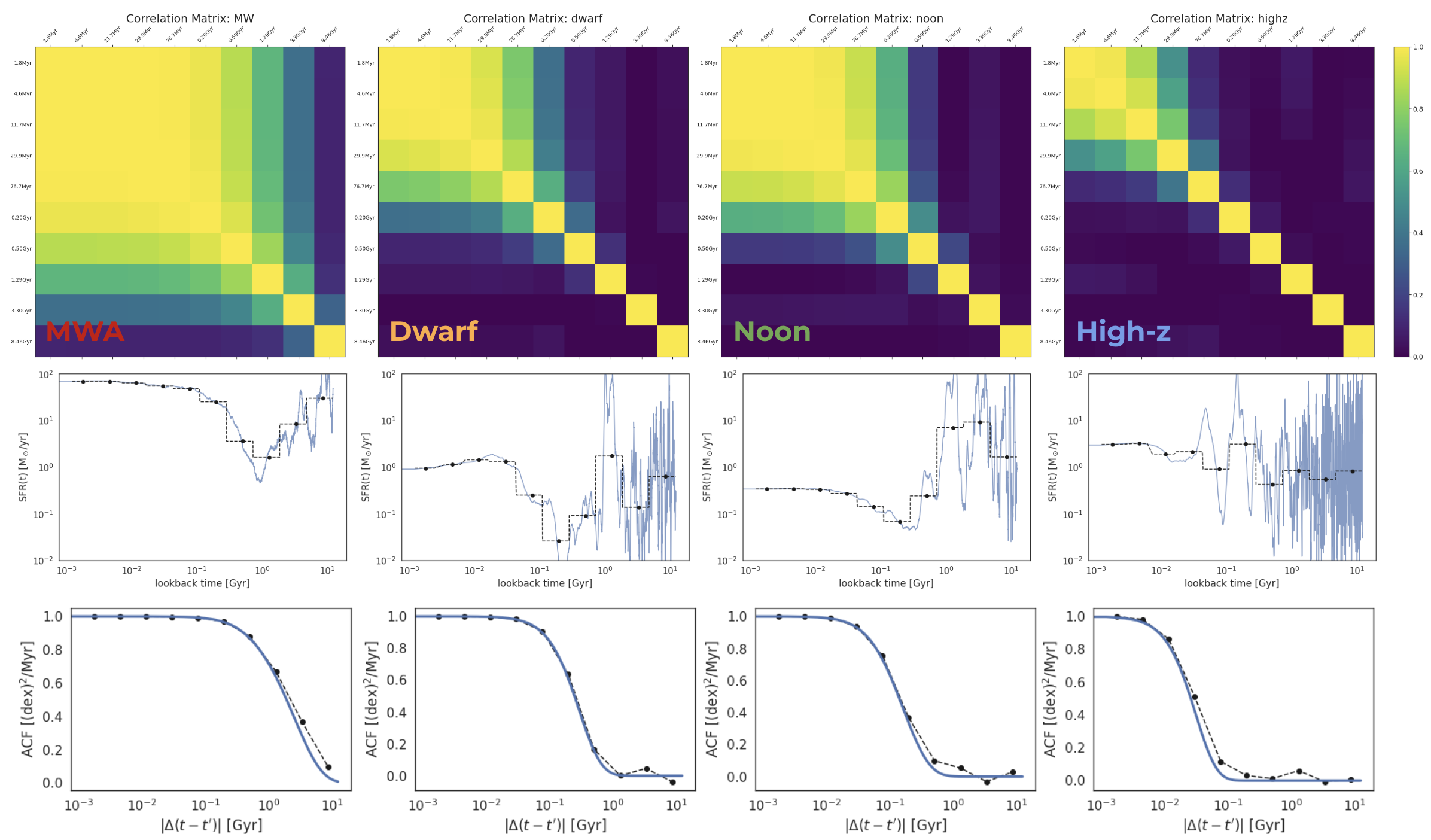}
    \caption{\textbf{Top:} Covariance function calculated for logarithmically spaced time bins for the four cases considered in this paper. \textbf{Middle: }Individual high-resolution SFHs degraded to the 10 log-spaced bins in lookback time. \textbf{Bottom:} Comparison of the computed correlation function to the analytic ACF for the four cases. Differences are due to shot-noise from a limited sample size and finite length for the time-series. }
    \label{fig:bin_covs_validation}
\end{figure}

We start with 1000 samples of high-resolution SFHs from a GP corresponding to each of the four galaxy regimes described in \S\ref{sec:model}. We then degrade them to binned SFHs with 10 equally spaced bins in log lookback time such that the SFR in a given bin is the average of the SFR in that interval, as shown in the middle panels of Figure \ref{fig:bin_covs_validation}. As expected, the covariance matrix computed from the binned SFHs matches the analytical estimate, with small differences due to spectral leakage from the finite length of the time-series. Repeating this analysis now drawing from the SFHs with the coarse time array corresponding to the bin centers with the same kernels confirms that the coarse SFHs are consistent with the distribution obtained by coarsening the high-resolution SFHs.
This also results in a significantly faster GP that may be more suitable for forward-modeling large ensembles of observations that require repeated computation of the covariance matrix.

\end{document}